\documentclass[preprint,5p,times,twocolumn]{elsarticle}
\usepackage{amssymb}
\usepackage{amsmath}
\usepackage{tabularx}
\usepackage[utf8]{inputenc}
\usepackage{longtable}
\usepackage{siunitx}
\usepackage{float}
\usepackage{upgreek}
\usepackage{algorithm}
\usepackage{algorithmicx}
\usepackage{soul}
\usepackage{array}
\usepackage{algpseudocode}
\usepackage[none]{hyphenat}
\usepackage{microtype}
\usepackage{stfloats} 
\usepackage{bm}
\usepackage{bbm}
\usepackage[table]{xcolor}
\usepackage{booktabs}
\usepackage{pdflscape}
\usepackage{enumitem}
\usepackage{caption}
\usepackage{subcaption}
\PassOptionsToPackage{hyphens}{url}
\usepackage{hyperref}
\usepackage{xurl}
\usepackage{todonotes}

\hypersetup{colorlinks}

\usepackage[table]{xcolor}

\definecolor{slategrey}{RGB}{112,128,144}
\definecolor{midnightblue}{rgb}{0.16, 0.26, 0.40}
\definecolor{firebrick}{rgb}{0.69, 0.17, 0.15}

\newcolumntype{L}[1]{>{\raggedright\arraybackslash}p{#1}} 
\newcolumntype{C}[1]{>{\centering\arraybackslash}p{#1}}   
\newcolumntype{R}[1]{>{\raggedleft\arraybackslash}p{#1}}

\usepackage{xcolor}

\journal{Applied Energy}

\begin{document}

\begin{frontmatter}

\titlebeforetext{FLOAT: Fatigue-Aware Design Optimization of Floating Offshore Wind Turbine Towers}

%\title{A Lightweight Fatigue Estimation Method with Multiphysics Validation: \\ Redesign of the IEA 22MW Floating Offshore Reference Tower for Fatigue}
\title{FLOAT: Fatigue-Aware Design Optimization of Floating Offshore Wind Turbine Towers}

% \title{FLOAT: Fatigue-Aware Lightweight Optimization and Assessment for Towers}

%\title{Fatigue Design for Floating Offshore Wind Turbine Upscaling: \\ Multiphysics Simulation and AI-Driven Redesign of the IEA 22MW Reference Tower}

\affiliation[inst1]{organization={Department of Mechanical Engineering, Massachusetts Institute of Technology},%Department and Organization
            city={Cambridge},
            state={MA},
            country={USA}}

\affiliation[inst2]{organization={LAETA-INEGI, Faculty of Engineering, University of Porto},
            city={Porto},
            country={Portugal}}

\affiliation[inst4]{organization={TEMA - Centre for Mechanical Technology and Automation, University of Aveiro},
            city={Aveiro},
            country={Portugal}}           

\affiliation[inst7]{organization={CONSTRUCT, Faculty of Engineering, University of Porto},
            city={Porto},
            country={Portugal}}
            
\affiliation[inst3]{organization={Faculty of Mechanical Engineering, Delft University of Technology},
            city={Delft},
            country={Netherlands}}

\affiliation[inst6]{organization={School of Engineering, Brown University},
            city={Providence},
            state={RI},
            country={USA}}

\cortext[cor1]{Corresponding author.}
 
\author[inst1,inst2,inst4]{João Alves Ribeiro\corref{cor1}}\ead{jpar@mit.edu}
\author[inst7]{Francisco Pimenta}
\author[inst3,inst6]{Bruno Alves Ribeiro}
\author[inst4]{Sérgio M. O. Tavares}

\author[inst1]{Faez Ahmed}

%% Abstract
\begin{abstract}
Upscaling is central to offshore wind’s cost-reduction strategy, with increasingly large rotors and nacelles requiring taller and stronger towers. In Floating Offshore Wind Turbines (FOWTs), this trend amplifies fatigue loads due to coupled wind--wave dynamics and platform motion. Conventional fatigue evaluation requires millions of high-fidelity simulations, creating prohibitive computational costs and slowing design innovation.
This paper presents \textbf{FLOAT} (Fatigue-aware Lightweight Optimization and Analysis for Towers), a framework that accelerates fatigue-aware tower design. It integrates three key contributions: a lightweight fatigue estimation method that enables efficient optimization, a Monte Carlo–based probabilistic wind–wave sampling approach that reduces required simulations, and enhanced high-fidelity modeling through pitch/heave–platform calibration and High-Performance Computing (HPC) execution. 
The framework is applied to the IEA 22 MW FOWT tower, delivering, to the authors’ knowledge, the first fatigue-oriented redesign of this benchmark model: FLOAT~22\,MW FOWT tower. Validation against 6{,}468 simulations demonstrates that the optimized tower extends the estimated fatigue life from $\sim$9 months to 25 years while avoiding resonance, and that the lightweight fatigue estimator provides conservative predictions with a mean relative error of $-8.6$\%. Achieving this lifetime requires increased tower mass, and the final design represents the lowest-mass fatigue-compliant configuration within the selected design space. 
All results and the reported lifetime extension are obtained within the considered fatigue scope, namely DLC 1.2 under aligned wind–wave conditions for the selected site distributions. By reducing simulation requirements by orders of magnitude, \textbf{FLOAT} provides a computationally efficient pathway for reliable and scalable tower design in next-generation FOWTs, bridging industrial needs and academic research while generating high-fidelity datasets that can support data-driven and AI-assisted design methodologies. All \textbf{FLOAT} resources, including the framework and the FLOAT 22 MW tower, are openly available at \url{https://github.com/Joao97ribeiro/FLOAT} and \url{https://github.com/Joao97ribeiro/FLOAT-22-280-RWT-Semi}.
\end{abstract}

%%Graphical abstract
\begin{graphicalabstract}
\includegraphics[width=1\textwidth]{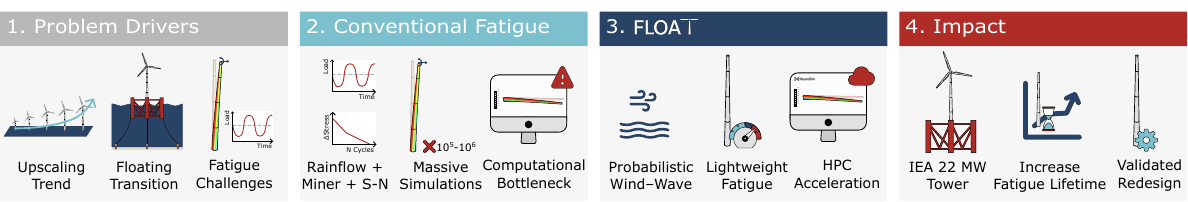}
\end{graphicalabstract}

\pagebreak

%%Research highlights
\onecolumn
\begin{highlights}

\item Introduces a lightweight fatigue estimation method to accelerate tower design optimization.

\item Reduces the number of offshore fatigue simulations via the proposed probabilistic wind–wave sampling method.

\item Demonstrates the method through fatigue-aware design optimization of the IEA 22 MW floating tower.

\item Performs pitch/heave–platform calibration and leverages HPC to enhance and scale floating turbine simulations.

\item Provides open access to the \textbf{FLOAT} framework and the redesigned 22 MW floating tower.

\end{highlights}
\twocolumn

%% Keywords

\end{frontmatter}
%% main text
%%

\section{Introduction}

\textbf{\textit{Problem: Floating and Upscaling Amplify Fatigue.}}  
The offshore wind industry is transitioning from fixed-bottom to floating to reach deeper waters with stronger and more stable winds, unlocking new sites and enabling capacity growth. Floating wind has progressed from pioneering 2–3~MW demonstrators in the early 2010s (Hywind Demo~\cite{HywindDemo}, WindFloat 1~\cite{WindFloat1}) through 6–9.5~MW commercial deployments in the 2020s (Hywind Scotland~\cite{hywind-scotland}, WindFloat Atlantic~\cite{WindFloatAtlantic}, Kincardine~\cite{Kincardine}, Hywind Tampen~\cite{hywind-tampen}) to current 15–20~MW designs~\cite{15mw_new,20mw_new}. Research models mirror this path, from NREL's 5~MW~\cite{osti_921803} in 2007 to the IEA 22~MW~\cite{1cd3e417f8854b808c5372670588d3d0} in 2024. While floating deployment and upscaling improve energy capture and lower the Levelized Cost of Energy (LCoE), they also place new demands on the tower, which must grow taller to support longer blades, maintain a continuous load path from the Rotor–Nacelle Assembly (RNA) to the floating platform, and resist greater loads from heavier nacelles, bigger rotors and platform motions, making fatigue one of the dominant design constraints, as highlighted in the recent comprehensive review of offshore wind tower design and optimization~\cite{ALVESRIBEIRO2025126294}.

\textbf{\textit{Bottleneck: Conventional Fatigue Evaluation.}}
Standard fatigue assessment relies on time-domain simulations with rainflow counting, S-N curves, and Palmgren–Miner’s rule. For fixed-bottom turbines, this already requires on the order of $10^5$ load cases~\cite{IEC61400-3-1}, and estimates rise toward $10^6$ for floating systems where additional platform degrees of freedom and Aero-Hydro-Servo-Elastic (AHSE) couplings expand the design space~\cite{IEC61400-3-2}. Even with modern High-Performance Computing (HPC), such demands make iterative redesign prohibitively expensive, establishing fatigue evaluation as the central computational bottleneck that limits rapid exploration of new tower configurations and slows innovation in floating wind design~\cite{en16010002}.

\pagebreak

\textbf{\textit{Solution: Toward an Efficient Fatigue-Aware Design.}}  
Several strategies have been proposed to alleviate the computational burden of fatigue evaluation in FOWTs, including simulation reduction~\cite{Stewart16,Papi_2022} and AI-based surrogate modeling~\cite{LI2020570,LIU2024120238}. While these methods show promising results, they primarily focus on fatigue assessment rather than fatigue-aware design optimization. Heuristic reduction techniques can provide efficiency gains but risk incomplete environmental coverage, whereas probabilistic reduction strategies offer better representativeness by ensuring that rare yet influential load cases are included according to their probability of occurrence. Surrogate models have also demonstrated potential, however their deployment is often constrained by large training datasets, site-specific data availability, and limited integration with the underlying physics. Moreover, no open-source framework currently provides an integrated solution that unites probabilistic sampling, lightweight fatigue estimation, and tower redesign. This gap highlights the need for a physics-informed framework that both accelerates fatigue evaluation and enables fatigue-oriented design optimization, while at the same time generating high-fidelity datasets that can support the next generation of AI-driven methods.

\textbf{\textit{Impact: The IEA 22 MW Tower as a Benchmark.}}  
The IEA 22~MW reference turbine~\cite{1cd3e417f8854b808c5372670588d3d0}, while widely adopted as a benchmark, was conceived for fixed-bottom conditions without explicit fatigue design. Its inadequacy under floating operation exemplifies the critical bottleneck facing the industry: the inability to rapidly iterate and optimize tower designs under fatigue constraints is fundamentally limiting the development of next-generation floating wind systems. Accelerating fatigue-aware design is essential for unlocking the full potential of upscaled floating turbines and achieving the cost reductions necessary for commercial viability.

\begin{figure*}[!b]
\vspace{0.5cm}
\centering
\includegraphics[width=1.0\linewidth]{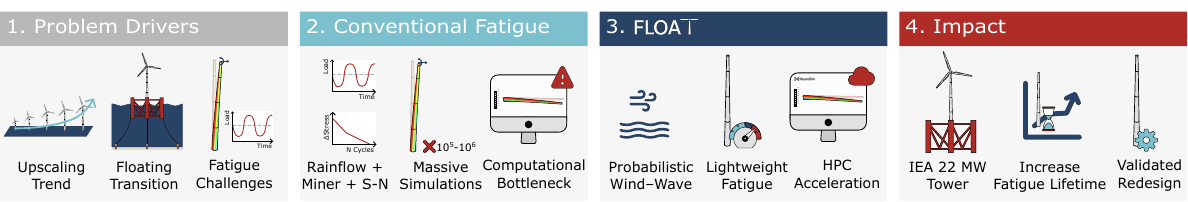}
\caption{\textbf{FLOAT} overcomes the computational bottleneck of fatigue evaluation in floating wind towers by replacing $\sim\!10^{6}$ IEC simulations with probabilistic sampling, lightweight fatigue estimation, and HPC-enabled execution, delivering a validated fatigue-oriented redesign of the IEA 22~MW floating reference, resulting in the FLOAT 22~MW tower.}

\label{fig:abstract}
\end{figure*}

\subsection{Objectives and Contributions} 

The objective of this paper is to address the computational challenges of fatigue-aware tower design in FOWTs. To this end, \textbf{FLOAT} (Fatigue-aware Lightweight Optimization and Analysis for Towers) is introduced as a framework that integrates probabilistic wind–wave sampling, lightweight fatigue estimation, and HPC-based simulation with pitch/heave–platform calibration to enable scalable tower redesign under fatigue constraints.

The main contributions of this work are: 
\begin{itemize}
    \item \textit{\textbf{Lightweight Fatigue Estimation Method}}: enables fast, iterative tower design optimization by calibrating an analytical fatigue response model against reference simulations, avoiding repeated high-fidelity runs.
    
    \item \textit{\textbf{Probabilistic Wind–Wave Sampling}}: introduces a Monte Carlo–based strategy that reduces the number of high-fidelity simulations needed for offshore fatigue assessment.  

    \item \textit{\textbf{First Fatigue-Oriented Redesign of the IEA 22 MW Tower}}: delivers the FLOAT 22 MW tower, a validated design based on 6,468 simulations, achieving a 25-year fatigue life under DLC~1.2 aligned wind--wave conditions and demonstrating the practical applicability of the proposed framework.

    \item \textit{\textbf{HPC and Pitch/Heave–Platform Calibration}}: integrates HPC execution and platform stabilization to enhance the scalability and fidelity of floating turbine simulations.

\end{itemize}

These contributions, together with the trajectory of addressing fatigue in FOWT towers from problem to solution, are synthesized in~\autoref{fig:abstract}, which highlights how \textbf{FLOAT} overcomes the computational bottleneck of conventional fatigue evaluation and enables the first validated 25-year redesign of the IEA 22 MW floating tower, resulting in the FLOAT 22 MW floating tower.

\subsection{Work Outline}
The paper is structured as follows.
\autoref{sec:background} reviews the background and related work on fatigue-aware design of FOWT towers.
\autoref{sec:methodology} presents the proposed \textbf{FLOAT} framework and its core modules.
\autoref{sec:case_study} applies the methodology in a case study on the fatigue-oriented redesign of the IEA 22 MW floating tower.
\autoref{sec:results} reports the main findings, including performance comparisons between reference and optimized designs and validation of the optimized tower. 
\autoref{sec:limitations} discusses the main limitations of the proposed approach and their mitigation.
Finally, \autoref{sec:conclusion} synthesizes the key contributions and highlights directions for further research.

\section{Background and Related Work} \label{sec:background}

This section presents the technical background for fatigue-aware design optimization of FOWT towers. It includes fatigue mechanisms and lifetime evaluation (\autoref{sec:damage_bk}), numerical simulation requirements and reduction strategies (\autoref{sec:back_sim}), reference turbine models with a focus on the IEA 22 MW tower and its design limitations (\autoref{sec:models_back}), software tools with their fatigue-related constraints (\autoref{sec:tools_back}), and a comparison between conventional fatigue evaluation and the proposed \textbf{FLOAT} framework (\autoref{sec:conv_vs_prop}).

\subsection{Fatigue in Offshore Wind Turbine Towers} \label{sec:damage_bk}
\begin{figure*}[b!]
  \centering
  \begin{subfigure}[b]{0.49\linewidth}
    \centering
    \includegraphics[width=\linewidth]{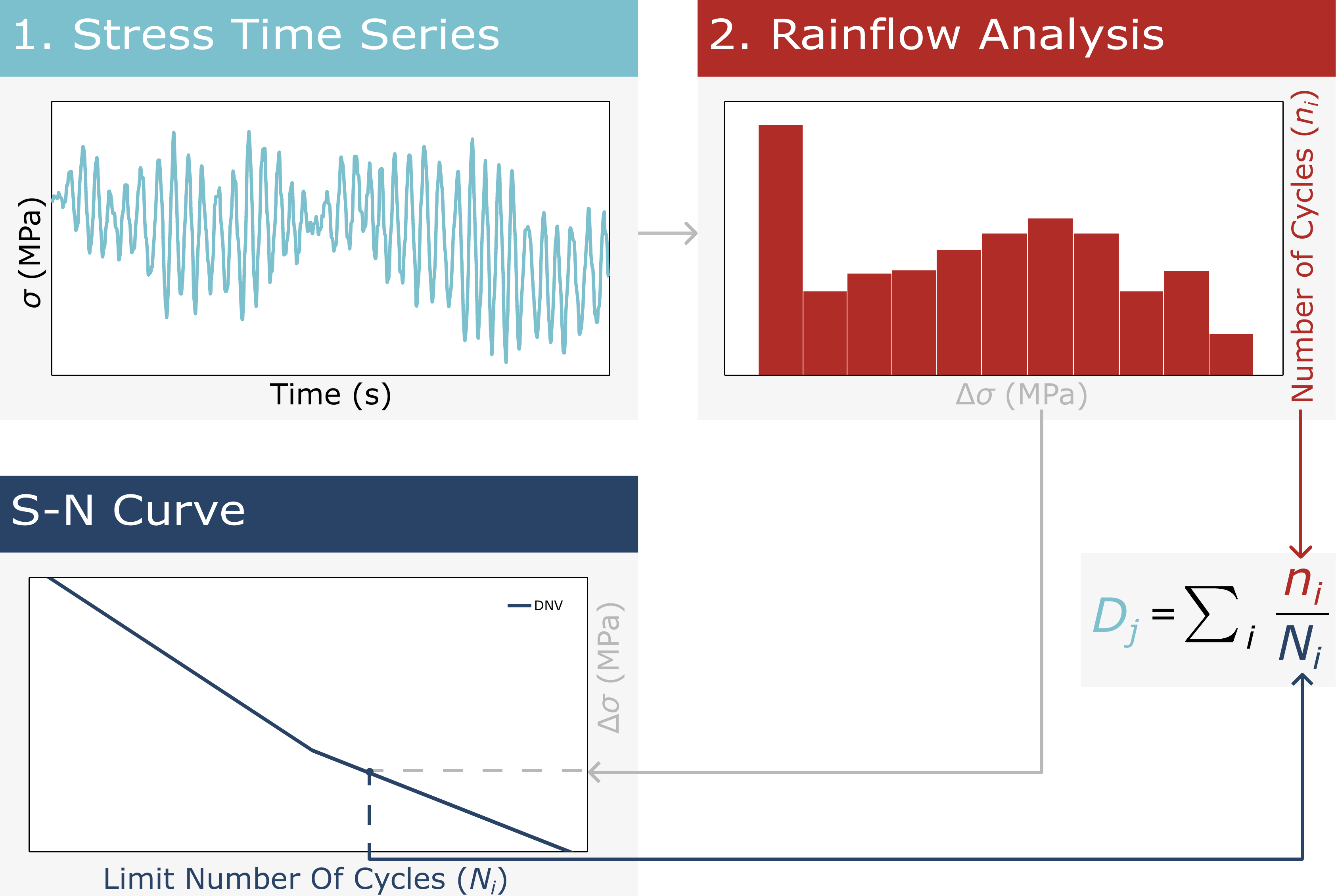}
    \caption{Event Fatigue Evaluation}
    \label{fig:fatigue_a}
  \end{subfigure}
  \hfill
  \begin{subfigure}[b]{0.49\linewidth}
    \centering
    \includegraphics[width=\linewidth]{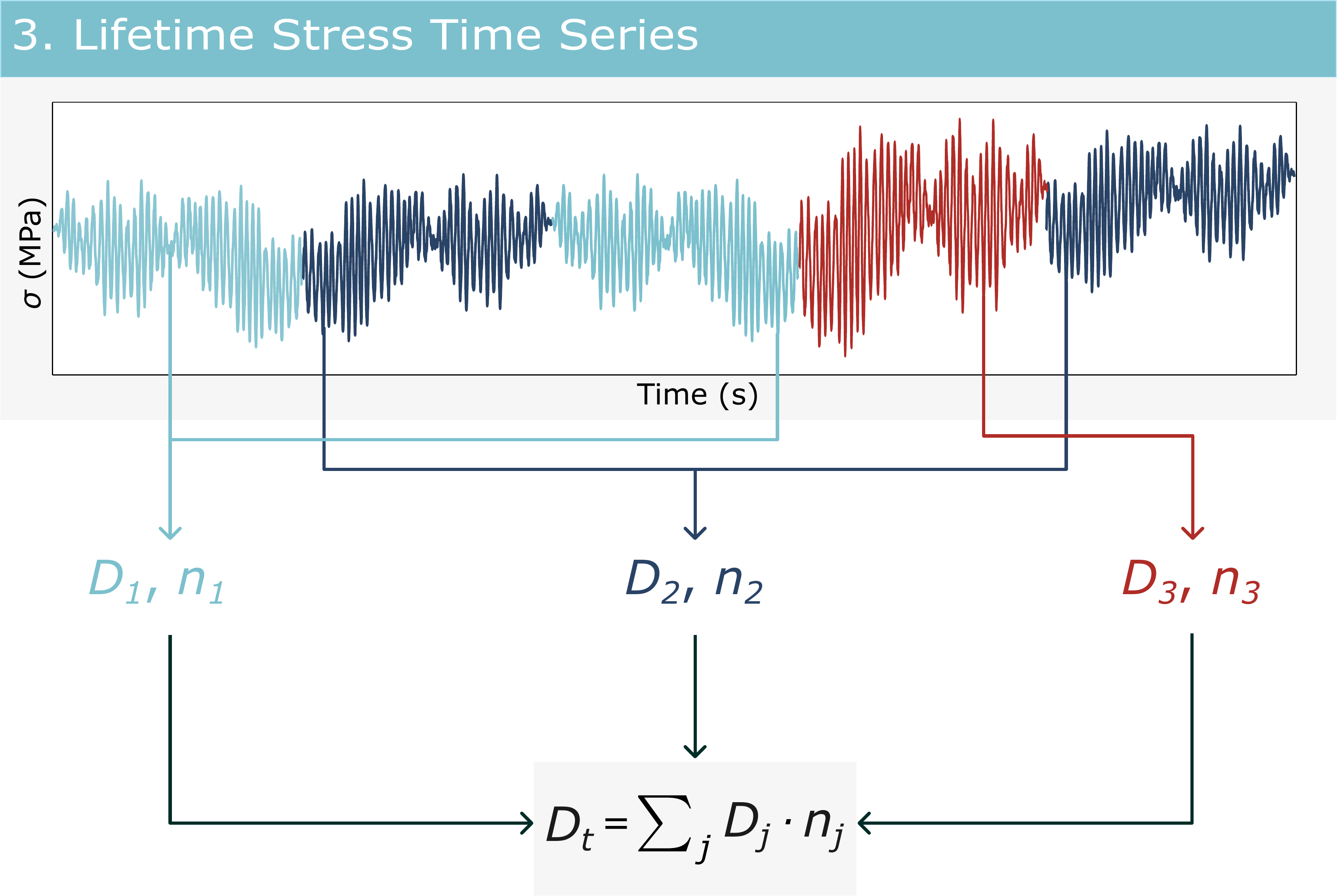}
    \caption{Lifetime Fatigue Estimation}
    \label{fig:fatigue_b}
  \end{subfigure}
  \caption{Fatigue estimation workflow for OWT towers.}
  \label{fig:fatigue_pipeline}
\end{figure*}

Fatigue is a critical failure mechanism in OWT towers, driven by continuous stress cycles from wind and wave loading that can initiate cracks and cause progressive structural degradation~\cite{MCMORLAND2022112499, en14092484}. In floating configurations, platform motion amplifies fatigue effects~\cite{YANG2023119111,YUAN2024119667}, requiring accurate damage estimation for safe and cost-effective tower design. Fatigue assessment follows two steps (\autoref{fig:fatigue_pipeline}): Event Fatigue Evaluation (damage per event type); and Lifetime Fatigue Estimation (combining event damage with expected occurrence counts).

\subsubsection{Event Fatigue Evaluation} \label{sec:event_fatigue}

Each event type is characterized by a stress time series obtained from monitoring data or numerical simulations. This series is processed using the rainflow counting method~\citep{matsuishi_1968,dowling_1972}, which decomposes it into individual stress cycles indexed by $i$, each characterized by a cycle count $n_i$ and a stress range $\Delta\sigma_i$. The number of cycles to failure for each range, $N_i(\Delta\sigma_i)$, is defined by S-N curves from DNV-RP-C203~\cite{DNV-RP-C203}, which, assuming unitary partial safety factors, provide:
\begin{equation}
    N_i(\Delta\sigma_i) = \bar{a} \left[ \Delta \sigma_i \left( \frac{t}{t_{\text{ref}}} \right)^k \right]^{-m}
    \label{eq:ni_explicit}
\end{equation}
\noindent where the term $\left({t}/{t_{\text{ref}}}\right)^k$ adjusts the fatigue strength for a plate with thickness $t$ greater than the reference thickness $t_{\text{ref}}$. The parameters $\bar{a}$, $m$, and $k$ are material constants defined by the S-N curve. The damage associated with a given event $j$ is then estimated using the Palmgren-Miner linear accumulation rule:
\begin{equation}
D_j = \sum_i \dfrac{n_i(\Delta\sigma_i)}{N_i(\Delta\sigma_i)}
\label{eq:damage}
\end{equation}

\subsubsection{Lifetime Fatigue Estimation}
Once each event damage $D_j$ is determined, the total lifetime damage $D_t$ is calculated by weighting $D_j$ by its expected number of occurrences $n_j$ over the design lifetime:

\begin{equation}
    D_t = \sum_j D_j\cdot n_j
    \label{eq:damage_t}
\end{equation}

\noindent The expected count $n_j$ for each event type is estimated as:  
\begin{equation}
    n_j = \frac{LT}{t_j}\cdot p_j
     \label{eq:nj_t}
\end{equation}
where $LT$ is the design lifetime, $t_j$ the event duration, and $p_j$ its probability of occurrence ($\sum_j p_j = 1$). Events are typically 10-minute intervals, consistent with SCADA data (operational data) and standards~\cite{IEC61400-1}. The total fatigue damage must satisfy the Fatigue Limit State \(D_t \leq{1}/{FDF}\), where the fatigue design factor (FDF) accounts for inspection accessibility and the structural consequences of failure~\cite{DNV-RP-C203,wang2005fatigue,ZHAO2021106075}.

While accurate, this process is computationally intensive and often dependent on extensive monitoring or large-scale simulations, a challenge further amplified in FOWTs.

\subsection{Numerical Simulations for FOWT Tower Fatigue} \label{sec:back_sim}
Numerical simulations are essential for evaluating the structural behavior of FOWT towers under coupled wind–wave loading, especially in early design stages when experimental or monitoring data are unavailable. They provide stress time series across a range of operating conditions, serving as input for fatigue evaluation and design validation. For accurate fatigue evaluation, simulations must represent realistic environmental conditions, including wind conditions and wave conditions, the latter defined by the sea state.

\subsubsection{Design Load Cases for FOWT Fatigue Analysis}

IEC 61400 \cite{IEC61400-1, IEC61400-3-1, IEC61400-3-2} defines the wind field properties and sea-state conditions that should be evaluated to ensure safety requirements are met.

For the wind field, \citet{IEC61400-1} defines normal and extreme wind conditions, but for fatigue purposes only the former is of interest. In this case, the wind speed profile is described by the Normal Wind Profile model (NWP) and the Normal Turbulence Model (NTM), which govern the mean evolution over height and the turbulent fluctuations, respectively. 

In terms of sea-state conditions, three families are considered for OWT design:
\begin{itemize}
    \item \textbf{Normal Sea State (NSS):} typical combinations of significant wave height ($H_s$), peak wave period ($T_p$), and mean wave direction ($M_{ww}$) during power production.
    \item \textbf{Severe Sea State (SSS):} 50-year return period conditioned on wind speed.
    \item \textbf{Extreme Sea State (ESS):} concurrent wind and wave extremes with a 50-year return period.
\end{itemize}
Once more, for fatigue analysis, only the NSS is considered.

Having defined the environmental conditions to be evaluated, design standards also prescribe a set of Design Load Cases (DLCs) for ultimate limit state and fatigue analyses. The latter, which are the focus of this work, require consideration of the following scenarios:
\begin{itemize}
    \item \textbf{DLC 1.2:} Normal power production.
	\item \textbf{DLC 2.4:} Power production plus occurrence of fault.
	\item \textbf{DLC 3.1:} Generator start up.
	\item \textbf{DLC 4.1:} Generator shut down.
	\item \textbf{DLC 6.4:} Parked (standing still or idling).
    \item \textbf{DLC 7.2:} Parked with fault conditions.
    \item \textbf{DLC 8.3:} Transport, assembly, maintenance and repair.
\end{itemize}
For each DLC, a joint distribution of environmental conditions is also defined. Since the wind field generation is stochastic in nature, IEC 61400 requires that a minimum of 6 independent realizations should be evaluated for a given set of conditions. Since DLC 1.2 corresponds to most of the operational lifetime and typically contributes most significantly to total fatigue damage \cite{PACHECO2023115913}, this work focuses on that particular DLC, although the same procedure can be applied to the remaining ones.

\subsubsection{Number of Simulations for FOWT Fatigue Analysis}
Fatigue under NTM and NSS is quantified using numerical simulations based on discrete combinations of key environmental parameters ($U$, $H_s$, $T_p$, and $M_{ww}$), as prescribed by IEC 61400-3 \cite{IEC61400-3-1, IEC61400-3-2}, where $U$ denotes the mean wind speed. This discretization results in 266112 distinct environmental conditions, detailed in \autoref{table:bin_ranges_iec} (see \autoref{A:num_sim}), each requiring an individual simulation. To capture stochastic variability, six seeds per wind speed–turbulence bin are recommended, yielding over 1.5 million simulations and motivating reduction strategies for computational feasibility.

\subsubsection{Simulation Reduction for FOWT Fatigue Analysis} \label{sec:red_sim}
To reduce the high computational cost of fatigue simulations in FOWTs, \citet{Stewart16} proposed two strategies: increasing the bin widths of $H_s$, $T_p$, and $M_{ww}$ by a factor of four, reducing the total cases by $4^3$; and excluding bins with very low probability of occurrence, given their negligible impact on fatigue, by applying a 90\% cumulative distribution function (CDF) threshold to retain representative conditions. \citet{Papi_2022} showed that combining both strategies reduces the simulations to 251 per seed (1506 for six seeds). Additional savings can be achieved by omitting wind–wave misalignment, though most rare bins are already excluded. Despite these reductions, all conditions must still be generated before filtering. Moreover, \citet{Vlachogiannis_2025} showed that binning systematically overestimates fatigue and investigated an exhaustive HPC framework to recover accuracy. While more precise, its prohibitive cost for design optimisation underscores the need for strategies that directly select representative cases.

\subsection{Floating Offshore Reference Models} \label{sec:models_back}
For simulation, benchmarking, and standardized design of offshore wind systems, reference wind turbine models are essential. Models from NREL, DTU, and IEA are widely adopted in research and industry. For FOWTs, \autoref{table:fowt_re} (see \autoref{A:fowt_models}) lists key official models, including the NREL 5 MW~\cite{osti_921803}, IEA 15 MW~\cite{osti_1660012}, and IEA 22 MW~\cite{1cd3e417f8854b808c5372670588d3d0}. Beyond these, adaptations have been proposed, such as coupling the DTU 10 MW model~\cite{bak2013dtu} with spar support structures~\cite{Borg_2016}.

\subsubsection{IEA 22 MW Reference Model} The most recent FOWT reference model is the IEA 22 MW, which includes a semi-submersible support structure. In response to the growing need for upscaling and the transition from fixed-bottom to floating foundations, it is expected to become a central benchmark in floating wind research, as demonstrated in recent studies \cite{Zahle_2024, DANGI2025122248, Collier_2024}.

\subsubsection{IEA 22 MW Tower Design Limitations} 
The IEA 22 MW tower was originally designed for fixed-bottom conditions and is used without modification in the floating configuration~\cite{1cd3e417f8854b808c5372670588d3d0}. As noted in~\cite{osti_2409185}, it is too light for the increased dynamic loads and was not designed with fatigue considerations, even in fixed-bottom use. These limitations reduce fatigue life and structural reliability. In floating operation, the first tower dominated natural frequency lies close to the third rotor harmonic, risking resonance and amplifying fatigue loads. This highlights the need for redesign to meet the fatigue demands of floating offshore applications.

\subsection{Software Tools for Tower Simulation and Design} \label{sec:tools_back}
Software tools for FOWT tower analysis fall into two main categories. Simulation tools, such as OpenFAST~\cite{Jonkman_2013}, FAST.Farm~\cite{92172f5db61f409db7a831d414c019bd}, HAWC2~\cite{18aac95355e641309b54a6830618c5ca}, and Bladed~\cite{Bladed}, perform high-fidelity AHSE modeling to capture the dynamic response of the system. Design and Optimization tools, including WISDEM~\cite{osti_2345922} and WEIS~\cite{10.1115/IOWTC2021-3533}, integrate structural and cost models to support early-stage decision-making. Both commercial and open-source platforms are in use, with a growing emphasis on transparency and reproducibility, supported by initiatives such as NREL’s open-source development. \autoref{tab:software_overview} (see~\autoref{A:software_tools}) summarizes the main tools in each category.

\subsubsection{Software Tools Limitations for Fatigue}
Fatigue is critical in OWT design and optimization, although existing tools face limitations. WISDEM, built on OpenMDAO, relies on steady-state models and is unsuitable for fatigue analysis requiring dynamic stress series. WEIS extends WISDEM by coupling with OpenFAST for dynamic simulations, but remains computationally expensive due to the large number of load cases and iterative evaluations. Although HPC solutions such as Inductiva~\cite{Sarmento2024} and Amazon Web Services (AWS)~\cite{aws} support parallel execution of OpenFAST~\cite{InductivaOpenFAST, AWSOpenFAST}, repeated fatigue evaluations remain time-consuming. These limitations highlight the need for simplified fatigue methods that reduce computational cost while maintaining predictive accuracy.

\subsection{From Conventional Fatigue Evaluation to FLOAT} \label{sec:conv_vs_prop}

Conventional fatigue evaluation of OWT towers requires high-fidelity simulations for all DLCs specified by IEC 61400-3. While accurate, this results in high computational cost, slow iteration cycles, and fragmented workflows across multiple tools. The proposed \textbf{FLOAT} methodology integrates probabilistic wind–wave sampling with a lightweight fatigue estimator in a unified automated workflow, reducing simulations and runtime while preserving predictive accuracy. \autoref{tab:fatigue_comparison} contrasts both approaches and motivates the methodology presented in~\autoref{sec:methodology}.

\begin{table}[h!]
\centering
\footnotesize
\caption{Conventional fatigue evaluation versus the \textbf{FLOAT} framework.}
\label{tab:fatigue_comparison}
\begin{tabular}{@{}p{1.4cm}p{3.1cm}p{3.1cm}@{}}
\toprule
\textbf{} 
& \cellcolor{slategrey!20}\textbf{Conventional} 
& \cellcolor{firebrick!20}\textbf{FLOAT} \\
\midrule
Fatigue Evaluation 
& \cellcolor{slategrey!5}High-fidelity per iteration (rainflow, S-N, Miner) 
& \cellcolor{firebrick!5}{Lightweight estimator + final high-fidelity check} \\
\addlinespace[0.1cm]
Simulations
& \cellcolor{slategrey!5}Full IEC binning ($\sim$10$^{5}$–10$^{6}$) 
& \cellcolor{firebrick!5}{Probabilistic sampling ($\sim$10$^{3}$–10$^{4}$)} \\
\addlinespace[0.1cm]
Models 
& \cellcolor{slategrey!5}{IEA 22~MW fixed-bottom reused for floating (too light, no fatigue-driven design)}
& \cellcolor{firebrick!5}{FLOAT 22~MW: fatigue-driven floating redesign of IEA 22~MW} \\
\addlinespace[0.1cm]
Tools
& \cellcolor{slategrey!5}Multi-tool, manual 
& \cellcolor{firebrick!5}{Unified, automated; HPC} \\
\addlinespace[0.1cm]
Efficiency
& \cellcolor{slategrey!5}High cost, slow 
& \cellcolor{firebrick!5}{Lower cost, fast} \\
\bottomrule
\end{tabular}
\end{table}

\section{Methodology} \label{sec:methodology}
\textbf{FLOAT} (\textbf{F}atigue-aware \textbf{L}ightweight \textbf{O}ptimization and \textbf{A}nalysis for \textbf{T}owers)\footnote{The \textbf{FLOAT} framework is openly available at \url{https://github.com/Joao97ribeiro/FLOAT}.} accelerates the iterative design of FOWT towers under fatigue constraints. While high-fidelity simulations are required to assess the reference and to validate optimized design candidates, \textbf{FLOAT} avoids rerunning them at every intermediate step of the optimization by introducing a lightweight fatigue model that reduces computational cost without compromising accuracy. Efficiency is further enhanced by a wind–wave sampling strategy that selects a reduced yet representative set of environmental conditions, enabling accurate fatigue predictions with fewer simulations. An HPC framework and pitch/heave–platform calibration is also integrated to improve simulation realism and computational performance. Together, these elements enable a fast and cost-effective fatigue-aware tower optimization. Although developed for floating platforms, the proposed methodology can be readily extended to fixed-bottom tower configurations.

\begin{figure*}[t!]
\centering
\includegraphics[width=1\linewidth]{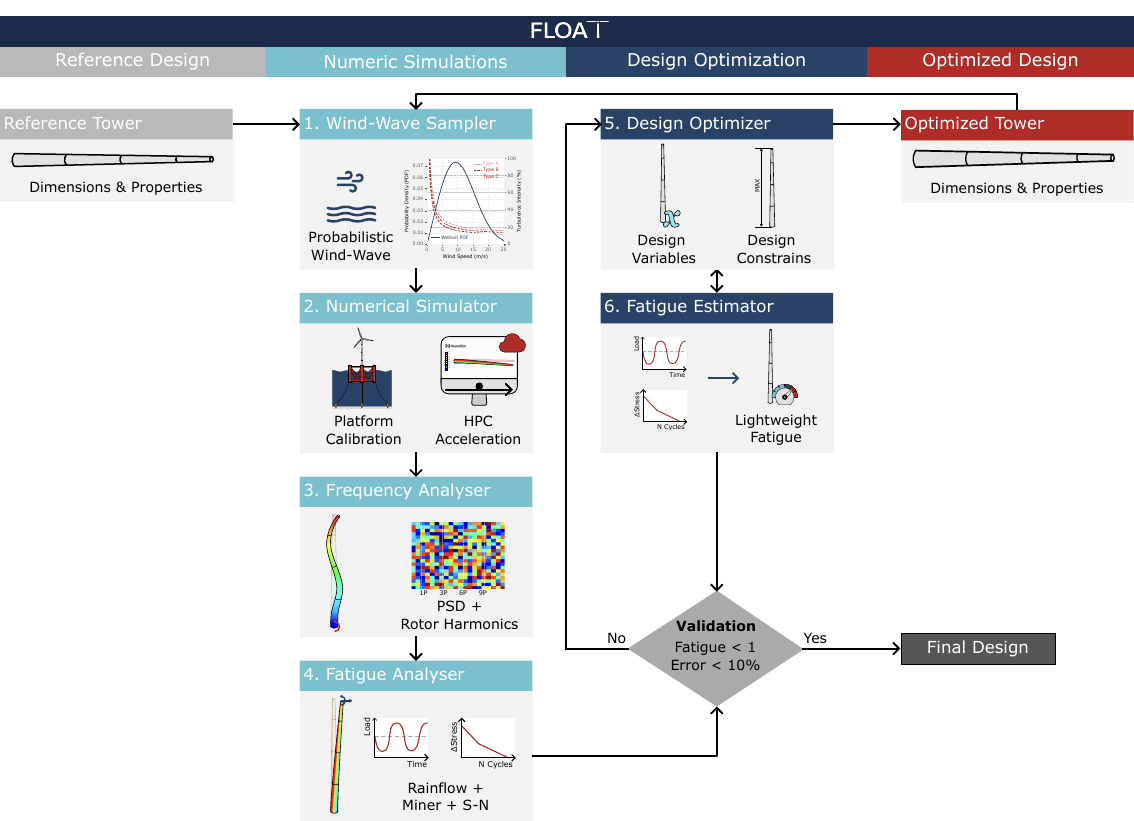}
\caption{\textbf{FLOAT} workflow. The framework consists of six blocks: the \textbf{\textit{Wind–Wave Sampler}} selects representative wind–wave conditions; the \textbf{\textit{Numerical Simulator}} runs large-scale HPC simulations and applies platform calibration; the \textbf{\textit{Frequency Response Analyser}} and \textbf{\textit{Fatigue Analyser}} compute frequency responses and fatigue damage; the \textbf{\textit{Design Optimizer}} iteratively updates geometry; and the \textbf{\textit{Fatigue Estimator}} predicts damage without rerunning full simulations. The process starts from the reference design and, through \textbf{FLOAT}, performs design optimization with validation of each optimized design, repeating additional cycles if needed until converging to a validated final design.}

\label{fig:frameworkdamage}
\end{figure*}

\subsection{FLOAT Workflow}

The following subsections describe the two main components of the \textbf{FLOAT} workflow. First, the workflow itself is detailed, outlining the numerical simulation and design optimization stages. Then, the validation process is presented, highlighting the convergence criteria and re-simulation steps.

\subsubsection{Workflow Description}
The \textbf{FLOAT} framework follows a two-stage process comprising numerical simulation and design optimization, as shown in \autoref{fig:frameworkdamage}. It begins with the numerical simulation stage, where the \textbf{\textit{Wind–Wave Sampler}} selects a representative subset of environmental wind–wave conditions. These scenarios are simulated for the reference tower in OpenFAST~\cite{Jonkman_2013} within the \textbf{\textit{Numerical Simulator}}, which applies pitch/heave-platform calibration to capture floating dynamics and leverages HPC through Inductiva’s platform~\cite{InductivaOpenFAST} for large-scale analyses. The resulting outputs are processed in the \textbf{\textit{Frequency Response Analyser}} and the \textbf{\textit{Fatigue Analyser}} to compute the frequency responses and fatigue damage of the reference tower. The frequency response is used to define the frequency constraint, while the fatigue damage provides the calibration required for the fatigue estimation model during the design optimization stage. The workflow then proceeds to the \textbf{\textit{Design Optimizer}}, where the tower geometry is iteratively updated under fatigue constraints, with damage predicted at each step by the \textbf{\textit{Fatigue Estimator}}, a lightweight analytical model that avoids rerunning high-fidelity simulations. After convergence, the same wind–wave conditions are re-simulated to evaluate the optimized tower and verify its performance against the reference design.

\subsubsection{Workflow Validation} \label{sec:val_work}

Workflow validation relies on two criteria: (i) the lifetime fatigue damage in any tower section must not exceed the admissible limit (unity), and (ii) the relative error between \textbf{FLOAT} and high-fidelity OpenFAST section-wise lifetime fatigue damage must remain below 10\%. If either criterion is violated, additional optimization cycles are performed. The first criterion enforces compliance with fatigue safety requirements, while the second evaluates the accuracy of the lightweight fatigue estimator embedded in \textbf{FLOAT}. Together, they ensure that the optimized tower design satisfies fatigue constraints and that fatigue damage predictions remain consistent with high-fidelity simulations. Between optimization cycles, only the fatigue damage calibration is updated based on the most recent design. Each cycle starts from the original reference tower geometry and uses the same mode shapes, which preliminary tests confirmed do not affect the fatigue response. The optimization continues until an accepted solution is achieved.

Each component of the \textbf{FLOAT} workflow is briefly summarized in the following subsections, 
while the complete procedures and corresponding algorithms are detailed in \autoref{A:worflows}.

\subsection{Wind–Wave Sampler}\label{sec:windwave_samp}
The \textbf{\textit{Wind–Wave Sampler}} defines the environmental conditions used to simulate and estimate tower fatigue in both the reference and optimized designs. Instead of simulating thousands of low-impact bins or applying heuristic filtering strategies such as those proposed by \citet{Stewart16} (see~\autoref{sec:red_sim}), it adopts a direct probabilistic sampling approach inspired by the joint modeling of offshore wind–wave parameters. In this work, the distributions proposed by \citet{Papi_2022} are considered, although these can be readily adapted to different locations. Within this probabilistic framework, the sampler formulates a Monte Carlo–based estimate of expected fatigue damage and selects a fixed number of representative environmental states accordingly. This strategy reduces the number of required high-fidelity simulations, lowering computational cost and accelerating optimization. Rather than seeking a minimum simulation set or claiming formal convergence, the objective is to construct a computationally tractable and representative set of environmental states for reliable lifetime fatigue assessment. Residual sampling uncertainty is addressed through subsequent high-fidelity re-simulation of the optimized design for validation purposes. The complete procedure is detailed in \autoref{alg:FLOWTS} (see~\ref{Aa:FLOWTS}).

\subsubsection{Analytical Formulation of the Wind–Wave Sampler} 
The combined behavior of wind–wave conditions is defined by a joint probability model of the mean wind speed ($U$), significant wave height ($H_s$), wave peak period ($T_p$), and wind–wave misalignment angle ($M_{ww}$), with the associated probability density function (PDF) being given by:
\begin{equation}
\begin{aligned}
    f_{U,H_s,T_p,M_{ww}}(U,H_s,T_p,M_{ww})  &= f_U(U) \cdot f_{H_s}(H_s\mid U) \\ 
    &\quad \cdot f_{T_p}(T_p\mid H_s) \cdot f_{M_{ww}}(M_{ww}\mid U)
\end{aligned}
\end{equation}
\noindent where $f(\cdot)$ are the conditional PDFs of the environmental parameters, whose analytical forms are detailed in \autoref{A:fx}, based on \citet{Papi_2022}.

Building on this foundation, the present work extends the formulation to account for fatigue damage over the turbine’s lifetime, $LT$. Let $D(U, H_s, T_p, M_{ww})$ denote the fatigue damage per unit time under a given environmental state $(U, H_s, T_p, M_{ww})$. The total expected lifetime damage is then defined as:
\begin{equation}
    D_t = LT \cdot \mathbb{E}_{U,H_s,T_p,M_{ww}}[D(U, H_s, T_p, M_{ww})]
    \label{eq:d_prob}
\end{equation}
where the expectation is over the joint distribution of the environmental parameters:
\begin{equation}
\begin{aligned}
\mathbb{E}_{U,H_s,T_p,M_{ww}}[D(U, H_s, T_p, M_{ww})] 
&= \int D(U,H_s,T_p,M_{ww}) \\
&\quad \cdot f_{U,H_s,T_p,M_{ww}}(U,H_s,T_p,M_{ww}) \\
&\quad \cdot dU \, dH_s \, dT_p \, dM_{ww}
\end{aligned}
\end{equation}

The integral can be approximated via a weighted Monte Carlo method using $N$ sampled states:
\begin{equation}
    D_t \approx LT \cdot \sum_{j=1}^{N} w_j \cdot D(U_j, H_{s,j}, T_{p,j}, M_{ww,j})
    \label{eq:monte}
\end{equation}
where $D(U_j, H_{s,j}, T_{p,j}, M_{ww,j})$ is the fatigue damage per unit time for environmental state $j$, and $w_j$ are normalized weights ($\sum_j w_j = 1$). Normalization is required because the wind-speed bins are retained explicitly, while the sampling is applied only to the remaining parameters. In \autoref{alg:FLOWTS}, $U$ is retained in discrete bins and $(H_s,T_p,M_{ww})$ are sampled from the corresponding conditional distributions, so $w_j$ represents the associated probability mass; the relation between these weights and the occurrence probability of each state is detailed in~\ref{A:monte_carlo_val}.

\subsection{Numerical Simulator}\label{sec:numsim}
To simulate the environmental conditions sampled by the \textbf{\textit{Wind–Wave Sampler}}, the \textbf{\textit{Numerical Simulator}} performs dynamic analyses of FOWTs using a physics-based pitch/heave–platform calibration, which adjusts the platform’s rotation (pitch) and vertical position (heave) to achieve static equilibrium under mean environmental loading. The simulations are executed in OpenFAST~\cite{Jonkman_2013} and parallelized on cloud-based infrastructure via Inductiva’s HPC platform~\cite{InductivaOpenFAST}, significantly accelerating computation. Both steady-state and turbulent inflows generated with TurbSim~\cite{turbsim} are supported, enabling detailed assessment of deterministic and stochastic structural responses. The full implementation steps are summarized in \autoref{alg:FastFOWT} (see~\ref{sec:FastFOWT}).

\subsubsection{Analytical Formulation of the Pitch–Platform Calibration}
\begin{figure}[h!]
    \centering
    \includegraphics[width=1\columnwidth]{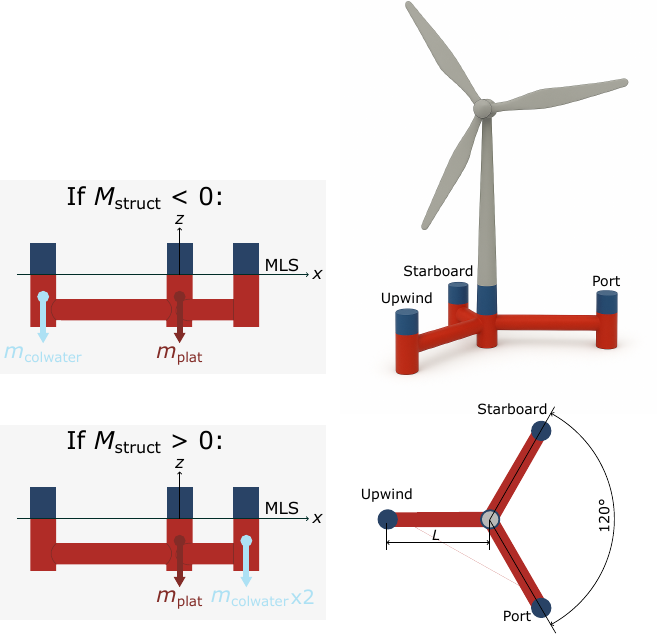}
    \caption{Pitch–platform calibration strategy implemented in \textbf{FLOAT}, based on the structural moment \( M_\text{struct} \).}

    \label{fig:x}
\end{figure}

The pitch–platform calibration stabilizes the floating platform’s rotation by adjusting the water ballast in the semi-submersible columns. The ballast induces a compensating moment, \( M_\text{colwater} \), that balances the total structural moment, \( M_\text{struct} = M_\text{weight} + M_\text{aero} \), where \( M_\text{weight} \) and \( M_\text{aero} \) are the gravitational and aerodynamic contributions. Equilibrium is achieved when \( M_\text{colwater} + M_\text{struct} = 0 \). Ballast is added to the upwind column when \( M_\text{struct} < 0 \), and distributed equally to the port and starboard columns when \( M_\text{struct} > 0 \), as illustrated in \autoref{fig:x}. Further implementation details, including the explicit formulation of the aerodynamic moment and the computation of ballast height and platform mass adjustments, are provided in~\ref{A:pitch_calib}.

%%%%%%%%%%%%%%%%%%%%%%%%%%%%%%%%%%%%%%%%%%%%%%%%%%%%%%%%%%%%%%%%%%%%%%%%%%%%%%%%%%%%%%%%%%%%%%%%%%%%%%
%%%%%%%%%%%%%%%%%%%%%%%%%%%%%%%%%%%%%%%%%%%%%%%%%%%%%%%%%%%%%%%%%%%%%%%%%%%%%%%%%%%%%%%%%%%%%%%%%%%%%%
%%%%%%%%%%%%%%%%%%%%%%%%%%%%%%%%%%%%%%%%%%%%%%%%%%%%%%%%%%%%%%%%%%%%%%%%%%%%%%%%%%%%%%%%%%%%%%%%%%%%%%
%%%%%%%%%%%%%%%%%%%%%%%%%%%%%%%%%%%%%%%%%%%%%%%%%%%%%%%%%%%%%%%%%%%%%%%%%%%%%%%%%%%%%%%%%%%%%%%%%%%%%%

\subsubsection{Analytical Formulation of the Heave–Platform Calibration}
The heave–platform calibration stabilizes the floating platform’s vertical equilibrium by adjusting its total mass. This adjustment is only required when the tower geometry changes and needs to be simulated for validation of the optimized tower. The mass variation of the tower is compensated by an equal and opposite adjustment of the platform mass, ensuring static balance: $\Delta m_\text{tower} + \Delta m_\text{plat} = 0$. An increase in tower mass is balanced by reducing the platform mass, whereas a reduction in tower mass is compensated by increasing the platform mass.

\subsection{Frequency Response Analyser} \label{sec:freq_fado}
The \textbf{\textit{Frequency Response Analyser}} quantifies rotor–aeroelastic excitations across frequency and wind speed by estimating the Power Spectral Density (PSD) for a given output of the \textbf{\textit{Numerical Simulator}}. By default, the analysis is performed on the tower-base fore–aft moment signal, as it concentrates the highest bending loads, governs structural design and fatigue assessment, and emphasizes the low-frequency range where dynamic amplification and fatigue damage accumulate. The resulting heatmap, displayed on a $\log_{10}$ scale, shows frequency along the rows and the representative mean wind speed of each bin along the columns, revealing resonance bands and rotor-harmonic contributions (1P, 3P, 6P, and 9P). The full procedure summarized in \autoref{alg:FRATower} (see~\ref{sec:FRA}).

\subsubsection{Analytical Formulation of the Frequency Response Analyser}
For each simulation $i$, the one-sided PSD of the tower-base fore–aft moment, $\widehat{S}_{xx}^{(i)}(f)$, was estimated using Welch’s method~\cite{1161901}. The method was applied using a prescribed sampling frequency $f_s$ and a Hann window of length $L$ with 50\% overlap. Each PSD was then paired with the mean wind speed ${U}_i$ computed from the corresponding time series.

When multiple simulations shared the same mean wind speed, their spectra were averaged into a representative PSD at $U_j$:
\begin{equation}
\mathbf{S}(f,U_j) = \frac{1}{|\mathcal{J}(U_j)|}\sum_{i \in \mathcal{J}(U_j)} \widehat{S}_{xx}^{(i)}(f)
\label{eq:sf_psd}
\end{equation}
where $\mathcal{J}(U_j)$ is the set of simulations whose mean velocity falls in group $U_j$, 
$|\mathcal{J}(U_j)|$ is its cardinality (number of simulations in the group), 
and $f$ denotes the evaluated frequencies.

The rotor fundamental frequency and its harmonics were obtained from the time-averaged rotor speed, $\overline{\mathrm{RPM}}$, within each wind speed bin $U_b$:
\begin{equation}
f_{1\mathrm{P}}(U_b) = \frac{\overline{\mathrm{RPM}}(U_b)}{60}, \qquad
f_{n\mathrm{P}}(U_b) = n\,f_{1\mathrm{P}}(U_b), \;\; n\in\{3,6,9\}
\label{eq:f_rpm}
\end{equation}
and overlaid on the heatmap to reveal the alignment between excitation harmonics and structural response.

\subsection{Fatigue Analyser} \label{sec:fatigue_fado}

The \textbf{\textit{Fatigue Analyser}} computes fatigue damage from the simulation outputs generated by the \textbf{\textit{Numerical Simulator}}, evaluating damage at the midpoint of each tower section. These sectional values are aggregated over the expected occurrence of each environmental condition to estimate lifetime fatigue. The fore–aft bending moment is taken as the dominant driver of stress variation and thus of fatigue accumulation. The complete procedure is summarized in \autoref{alg:FAITower} (see~\ref{sec:FAITower}).

\subsubsection{Analytical Formulation of the Fatigue Analyser} \label{sec:fatigue_analyser_sn}
\begin{figure}[!ht]
\centering
\includegraphics[width=1.0\linewidth]{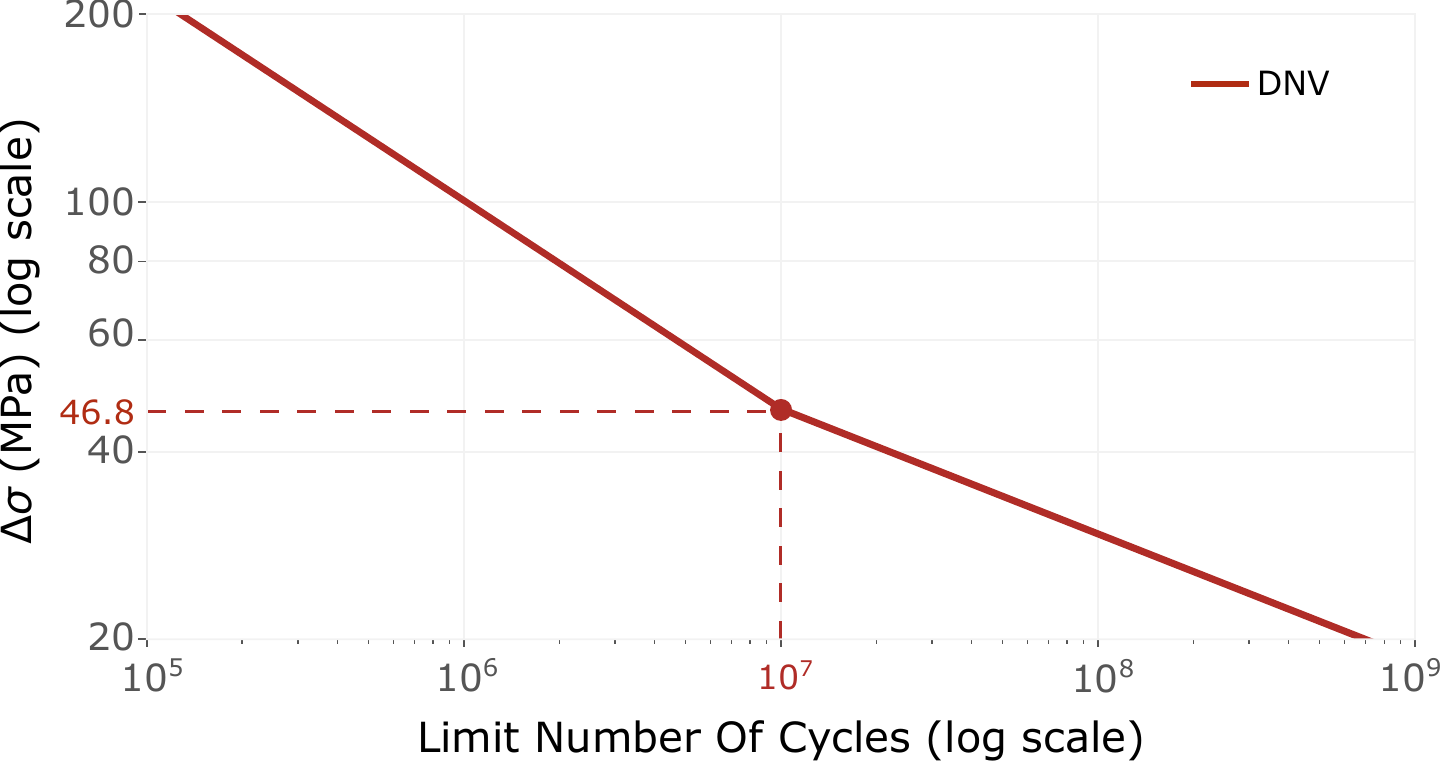}
\caption{S-N curve of type E from the DNV-RP-C203 standard \cite{DNV-RP-C203}.}
\label{fig:fatigue_curve}
\end{figure}

Fatigue damage is computed following the methodology presented in~\autoref{sec:damage_bk}, combining rainflow cycle counting with Miner’s rule. The type~E S--N curve from the DNV-RP-C203 standard~\cite{DNV-RP-C203} is adopted, corresponding to detail category~80 in Eurocode~3: Part~1--9 (EN~1993-1-9)~\cite{Eurocode3-Part1-9}, which is recommended for welded joints. This curve features two linear regimes with slopes of 3 and 5, transitioning at $10^7$ cycles, as shown in~\autoref{fig:fatigue_curve}. A thickness exponent $k$ of 0.20 is considered to account for the influence of plate thickness on fatigue strength, with a reference thickness $t_{\text{ref}}$ of 25~mm.

\subsection{Design Optimizer}\label{sec:desing_opt}

The \textbf{\textit{Design Optimizer}} performs fatigue-aware tower design optimization by minimizing tower mass subject to multiple constraints. The design variables are the outer diameter and wall thickness at discrete tower nodes, with linear tapering between nodes forming conical frustums in each section. Fatigue governs the optimization, while additional constraints enforce structural requirements (stress, buckling, and natural frequency) and geometric consistency (diameter-to-thickness ratio, tapering, and monotonicity). As mass is minimized under multiple constraints, any increase during optimization reflects the additional material required to satisfy all imposed requirements. Fatigue is evaluated at each iteration using the \textbf{\textit{Fatigue Estimator}} introduced in this work (see~\autoref{sec:fatigue_estimator}), thereby avoiding high-fidelity re-simulations.

The optimization is implemented in WISDEM~\cite{osti_2345922} using Sequential Least Squares Programming (SLSQP) with central finite-difference gradients, though other solvers are also available. Tower-top loads are defined from a representative DLC and treated as fixed inputs within the optimization loop, where they are used exclusively for structural consistency checks, while fatigue constraints govern the final tower dimensioning. This assumption is adopted to improve computational efficiency, as fully coupled AHSE re-simulations at each iteration would be prohibitive. To further reduce computational cost, a land-based tower model is employed within the optimization loop, with frequency constraints corrected from the reference floating configuration and applied consistently across iterations. All simplifications are limited to the optimization stage; the final optimized design is subsequently validated through full high-fidelity floating simulations under the sampled environmental conditions, ensuring compliance with the acceptance criteria and capturing any load redistribution induced by geometry changes. The complete procedure is summarized in \autoref{alg:OPTITower} (see~\ref{sec:OPTITower}).

\subsubsection{Analytical Formulation of the Design Optimizer}

\begin{figure}[h!]
	\centering
    \includegraphics[width=3 cm]{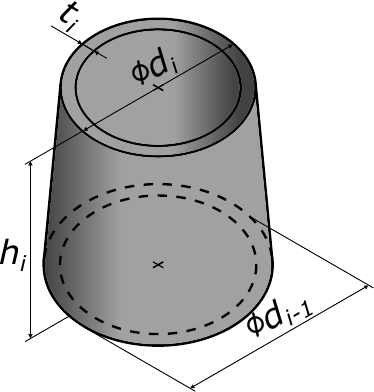}
    \caption{Tower section \(i\) defined by bottom diameter \(d_{i-1}\), top diameter \(d_i\), constant height \(h_i\), and wall thickness \(t_i\).}
    \label{fig:section_tower}
\end{figure}

The goal of the design optimization is to determine the optimal geometry of the wind turbine tower. The tower is divided into \(n\) conical sections, each defined by a bottom diameter \(d_{i-1}\), a top diameter \(d_i\), a constant height \(h_i\), and a uniform wall thickness \(t_i\), as illustrated in \autoref{fig:section_tower}. The section heights \(h_i\) are fixed according to the reference tower.

%%%%%%%%%%%%%%%%%%%%%%%%%%%%%%%%%%%%%%%%%%%%%%%%%%%%%%%%%%%%%%%%%%%%%%%%%%%%%%%%%%%%%%%%%%%%%%%%%%%%%%
%%%%%%%%%%%%%%%%%%%%%%%%%%%%%%%%%%%%%%%%%%%%%%%%%%%%%%%%%%%%%%%%%%%%%%%%%%%%%%%%%%%%%%%%%%%%%%%%%%%%%%
%%%%%%%%%%%%%%%%%%%%%%%%%%%%%%%%%%%%%%%%%%%%%%%%%%%%%%%%%%%%%%%%%%%%%%%%%%%%%%%%%%%%%%%%%%%%%%%%%%%%%%
%%%%%%%%%%%%%%%%%%%%%%%%%%%%%%%%%%%%%%%%%%%%%%%%%%%%%%%%%%%%%%%%%%%%%%%%%%%%%%%%%%%%%%%%%%%%%%%%%%%%%%

The design vector is:
\[
\boldsymbol{x} = [d_0, \dots, d_n, \, t_1, \dots, t_n],
\]
where \(d_i\) and \(t_j\) denote the outer diameters and wall thicknesses, respectively.  
Each variable is subject to the bounds:
\[
d_i \in [d_{\min}, d_{\max}], \quad i = 0, \dots, n,
\]
\[
t_j \in [t_{\min}, t_{\max}], \quad j = 1, \dots, n.
\]

The objective is to minimize the total tower mass:

\begin{equation}
\min_{\boldsymbol{x}} \quad m_{\text{tower}}(\boldsymbol{x})
\label{eq:mass}
\end{equation}

\noindent where the design is subject to the following constraints:

\begin{enumerate}
\item \textbf{{Stress constraint.}} The von Mises stress \(\sigma_{\text{vM}}\) must remain below the material yield stress \(\sigma_y\), scaled by safety factors \(\gamma_f\) (load uncertainty), \(\gamma_m\) (material variability), and \(\gamma_n\) (consequence of failure):

\begin{equation}
\frac{\gamma_f \cdot \gamma_m \cdot \gamma_n \cdot \sigma_{\text{vM}}(\boldsymbol{x})}{\sigma_y} \leq 1.0
\label{eq:stress1}
\end{equation}

\item \textbf{{Buckling constraints.}} Compliance with Eurocode standards is enforced by requiring both Local Shell Buckling (LSF) and Global Buckling (GF) utilization ratios to stay below unity:
\begin{equation}
\text{LSF}(\boldsymbol{x}) \leq 1.0, \quad
\text{GF}(\boldsymbol{x}) \leq 1.0
\label{eq:buck1}
\end{equation}

\item \textbf{{Frequency constraint.}} To avoid resonance phenomena, the first natural frequency of the tower is constrained to lie within a prescribed target range:
\begin{equation}
f_1^{\min} \leq f_1(\boldsymbol{x}) \leq f_1^{\max}
\label{eq:freq1}
\end{equation}

\item \textbf{{Fatigue constraint.}} Fatigue damage at each section, scaled by $\gamma_d$ (fatigue uncertainty), must remain below unity:
\begin{equation}
D_i(\boldsymbol{x}) \cdot \gamma_d \leq 1.0
\label{eq:damage1}
\end{equation}

\item \textbf{{Monotonicity constraints.}} The outer diameter and wall thickness must decrease or remain constant from base to top:
\begin{equation}
d_{i+1} \leq d_i, \quad t_{i+1} \leq t_i
\label{eq:mono1}
\end{equation}

\item \textbf{{Diameter-to-thickness constraint.}} The diameter-to-thickness ratio must remain within specified bounds to ensure structural integrity and manufacturability:
\begin{equation}
\left( \frac{d}{t} \right)_{\text{min}} \leq \frac{d_i}{t_i} \leq \left( \frac{d}{t} \right)_{\text{max}}
\label{eq:d_to_p1}
\end{equation}

\item \textbf{{Taper constraint.}} Enforces a maximum allowable conical frustum taper ratio per section:
\begin{equation}
\text{taper}_{\min} \leq \frac{d_{i+1}}{d_i} \leq \text{taper}_{\max}
\label{eq:taper1}
\end{equation}

\end{enumerate}
As the optimization uses a land-base tower model, the floating reference frequency is converted using the relation proposed in~\cite{PIMENTA2024117367}:
\begin{equation}
f_1^{\text{floating}} = \frac{f_1^{\text{land-base}}(\boldsymbol{x})}{r_I}
\label{eq:float_to_land}
\end{equation}
where $1/r_I\approx 1.57$ for this particular case study.

\subsection{Fatigue Estimator} \label{sec:fatigue_estimator}
The \textbf{\textit{Fatigue Estimator}} is a lightweight analytical model embedded in the \textbf{\textit{Design Optimizer}} to accelerate fatigue-aware tower design optimization. At each iteration, section-level fatigue damage is estimated directly from geometric parameters, S--N curve properties, and a calibrated damage profile, avoiding high-fidelity re-simulations within the optimization loop. The computational speed-up is achieved by treating fatigue as a geometry-driven surrogate, where damage is updated through stress rescaling induced by geometric changes, rather than by recomputing the full coupled AHSE simulations. This approximation is suitable when the structural response is bending-dominated and geometry updates remain close to the calibration tower. As load redistribution effects are not explicitly captured, final design acceptance is always based on high-fidelity re-simulation of the optimized geometry under the sampled environmental conditions. The calibration tower corresponds to the reference design in the first optimization cycle and to the most recent optimized design in subsequent cycles. The complete procedure is summarized in \autoref{alg:FASTower} (see~\ref{sec:FASTower}).

\subsubsection{Analytical Formulation of the Fatigue Estimator}
Based on the formulation detailed in ~\autoref{sec:event_fatigue}, the damage associated with a given event $j$ can be expressed as:
\begin{equation}
    D_j =  \frac{1}{\bar{a}} \sum_i n_i \left[ \Delta \sigma_i \left( \frac{t}{t_{\text{ref}}} \right)^k \right]^m
    \label{eq:palmgren_miner_brackets}
\end{equation}

\noindent Thus, damage scales with stress range and thickness factor:
\begin{equation}
    D_j \propto \left[\Delta\sigma_i \left(\frac{t}{t_{\text{ref}}} \right)^k\right]^m
    \label{eq:damage_scaling}
\end{equation}

On the other hand, for a given tower cross section, and assuming the stresses variations are induced by bending moments $M$ alone, the stresses are given by $\sigma = {M}/{I_y}\, r = {M}/{\omega_y}$, where $\sigma$ is the axial stress, $I_y$ the second moment of area and $\omega_y$ the section modulus. Further assuming a thin wall circular cross section, as typical of wind turbine towers, the expression can be simplified to $\sigma\approx {M}/{\left(\pi r^2 t\right)}$, where $r$ is the outer radius and $t$ the wall thickness. Hence, stress scales with radius and thickness as:

\begin{equation}
    \sigma \propto r^{-2} t^{-1}
    \label{eq:stress_scaling}
\end{equation}

Combining \autoref{eq:damage_scaling} and \autoref{eq:stress_scaling}, with a fixed slope $m$, the fatigue damage at a tower cross section scales as:
\begin{equation}
    D = C r^{-2m} t^{-m}\left(\frac{t}{t_{\text{ref}}} \right)^{k\cdot m}
    \label{eq:optimal_rt}
\end{equation}
\noindent $m$ is set by default to 4, the average slope of the two S-N curve regimes (see~\autoref{sec:fatigue_analyser_sn}), and $C$ is a constant recalibrated from the geometry and fatigue damage of the calibration tower.

%\subsection{FLOAT Tool}
%\todo[inline]{The integration with NREL tools mentioned here is suggested as a possible future collaboration.}
%\textcolor{red}{\textbf{FLOAT} is planned to be made publicly available, with potential integration into NREL tools either as a complete framework or as standalone modules. For instance, the \textbf{\textit{Wind–Wave Sampler}} could be added to the WEIS\footnote{\url{https://github.com/WISDEM/WEIS}} with minimal modification, requiring only specification of the number of samples for each environmental parameter in the \texttt{metocean\_conditions} block. Likewise, the \textbf{\textit{Fatigue Estimator}} could be incorporated into the WISDEM\footnote{\url{https://github.com/WISDEM/WISDEM}} by extending the generic modeling options file to include SN curve parameters, reference tower geometry, and corresponding fatigue damage values in the \texttt{TowerSE} section.}

\section{Case Study} \label{sec:case_study}
To demonstrate its practical relevance, the \textbf{FLOAT} method was applied to a large-scale case study: the redesign of the IEA 22 MW FOWT tower. The original reference model employs a semi-submersible support structure, but its tower was neither tailored for floating conditions nor constrained by fatigue requirements. Using the proposed method, a fatigue-oriented redesign was performed by combining rapid fatigue estimation with high-fidelity simulations. The resulting geometry is, to the authors’ knowledge, the first publicly available 22~MW FOWT tower optimized for fatigue, referred to as the FLOAT 22~MW tower. This case study highlights \textbf{FLOAT}’s effectiveness in addressing real-world structural challenges in next-generation OWTs.

\subsection{IEA 22 MW Floating Offshore Reference Tower}

The case study focuses on the IEA 22 MW reference FOWT. Its main specifications are listed in \autoref{tab:22_mw_reference}.
%%%%%%%%%%%%%%%%%%%%%%%%%%%%%%%%%%%%%%%%%%%%%%%%%%%%%%%%%%%%%%%%%%%%%%%%%%%%%%%%%%%%%%%%%%%%%%%%%%%%%%
\begin{table}[h!]
\footnotesize
\centering
\caption{Main parameters of the IEA 22 MW reference wind turbine \cite{1cd3e417f8854b808c5372670588d3d0}.}
\begin{tabular}{@{}p{2.75cm}p{1.0cm}p{2.75cm}p{1.0cm}@{}}
\toprule
\textbf{Parameter} & \textbf{Value} & \textbf{Parameter} & \textbf{Value} \\ \midrule
Rated Power (MW) & 22.0 & Min Rotor Speed (rpm) & 1.807 \\
Blades Number & 3 & Max Rotor Speed (rpm) & 7.061 \\
Rotor Diameter (m) & 284 & Rotor-Nacelle Mass (t) & 1218.685 \\
Cut-in Wind Speed (m/s) & 3.0 & Hub Height (m) & 170 \\
Rated Wind Speed (m/s) & 25.0 & Blade Length (m) & 137.8 \\
Tower Mass (t) & 1574 & Blade Mass (t) & 82.301 \\
\bottomrule
\end{tabular}
\label{tab:22_mw_reference}
\end{table}
%%%%%%%%%%%%%%%%%%%%%%%%%%%%%%%%%%%%%%%%%%%%%%%%%%%%%%%%%%%%%%%%%%%%%%%%%%%%%%%%%%%%%%%%%%%%%%%%%%%%%%

\subsubsection{Tower Geometry} 
\begin{figure*}[h!]
	\centering
    \includegraphics[width=\textwidth]{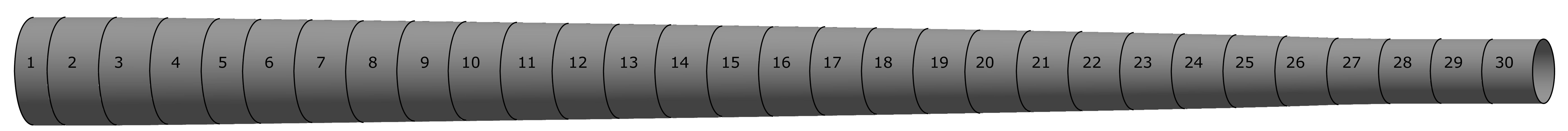}
    \caption{Tower geometry divided into 30 sections (indexed from bottom to top).}
    \label{fig:tower_geom}
\end{figure*}

The tower is modeled with 30 tapered cylindrical sections (\autoref{fig:tower_geom}), each defined by bottom diameter $d_{i-1}$, top diameter $d_i$, height $h_i$, and wall thickness $t_i$ (\autoref{fig:section_tower}). Full dimensions are provided in \autoref{tab:geom_tower} (see~\ref{A:table_ref}).

%%%%%%%%%%%%%%%%%%%%%%%%%%%%%%%%%%%%%%%%%%%%%%%%%%%%%%%%%%%%%%%%%%%%%%%%%%%%%%%%%%%%%%%%%%%%%%%%%%%%%%
%%%%%%%%%%%%%%%%%%%%%%%%%%%%%%%%%%%%%%%%%%%%%%%%%%%%%%%%%%%%%%%%%%%%%%%%%%%%%%%%%%%%%%%%%%%%%%%%%%%%%%
%%%%%%%%%%%%%%%%%%%%%%%%%%%%%%%%%%%%%%%%%%%%%%%%%%%%%%%%%%%%%%%%%%%%%%%%%%%%%%%%%%%%%%%%%%%%%%%%%%%%%%
%%%%%%%%%%%%%%%%%%%%%%%%%%%%%%%%%%%%%%%%%%%%%%%%%%%%%%%%%%%%%%%%%%%%%%%%%%%%%%%%%%%%%%%%%%%%%%%%%%%%%%

\subsubsection{Tower Material} The tower material is steel, with properties in \autoref{tab:steel_properties}.

\begin{table}[H]
\footnotesize
\centering
\caption{Steel properties used in the IEA 22 MW tower \cite{1cd3e417f8854b808c5372670588d3d0}.}
\begin{tabular}{@{}l r l r@{}}
\toprule
\textbf{Property} & \textbf{Value} & \textbf{Property} & \textbf{Value} \\
\midrule
Density & 7850~kg\,m$^{-3}$ & Young’s Modulus & 200~GPa \\
Shear Stiffness & 79.3~GPa & Poisson’s Ratio & 0.265 \\
Yield Strength & 345~MPa & Tensile/Compressive Strength & 450~MPa \\
\bottomrule
\end{tabular}
\label{tab:steel_properties}
\end{table}

\subsubsection{Tower Loads} 
The redesign adopts the ultimate tower-top loads defined in the IEA~22~MW reference model and does not recompute these loads for each updated tower geometry. This choice is consistent with the scope of this work, which focuses on fatigue-driven tower redesign for FOWTs. Tower-top loads are used exclusively to enforce structural consistency constraints during optimization (stress, buckling, and frequency), while fatigue constraints govern the final tower dimensioning. The applied force and moment components are summarized in \autoref{tab:applied_loads}.

\begin{table}[h!]
\footnotesize
\centering
\caption{Forces and moments applied at the top of the IEA 22 MW tower \cite{1cd3e417f8854b808c5372670588d3d0} ($x$: downwind, $y$: lateral, $z$: vertical).}
\begin{tabular}{@{}p{0.5cm}C{1.5cm}p{1cm}C{1.7cm}@{}}
\toprule
\textbf{Force} & \textbf{Value (MN)} & \textbf{Moment} & \textbf{Value (MN\,m)} \\
\midrule
$F_x$ & 5.7 & $M_x$ & $-1.6$ \\
$F_y$ & 0.09 & $M_y$ & $-37.6$ \\
$F_z$ & $-11.3$ & $M_z$ & 10.7 \\
\bottomrule
\end{tabular}
\label{tab:applied_loads}
\end{table}

\subsection{Numerical Simulations of the IEA 22 MW Tower}
Numerical simulations were conducted to calculate the initial fatigue damage of the reference tower and to validate the optimized design generated with \textbf{FLOAT}, by evaluating its fatigue performance against the reference tower.

\subsubsection{Wind–Wave Sampling}  \label{sec:sampl-windwave}
\begin{figure}[h!]
    \centering
\includegraphics[width=0.925\columnwidth]{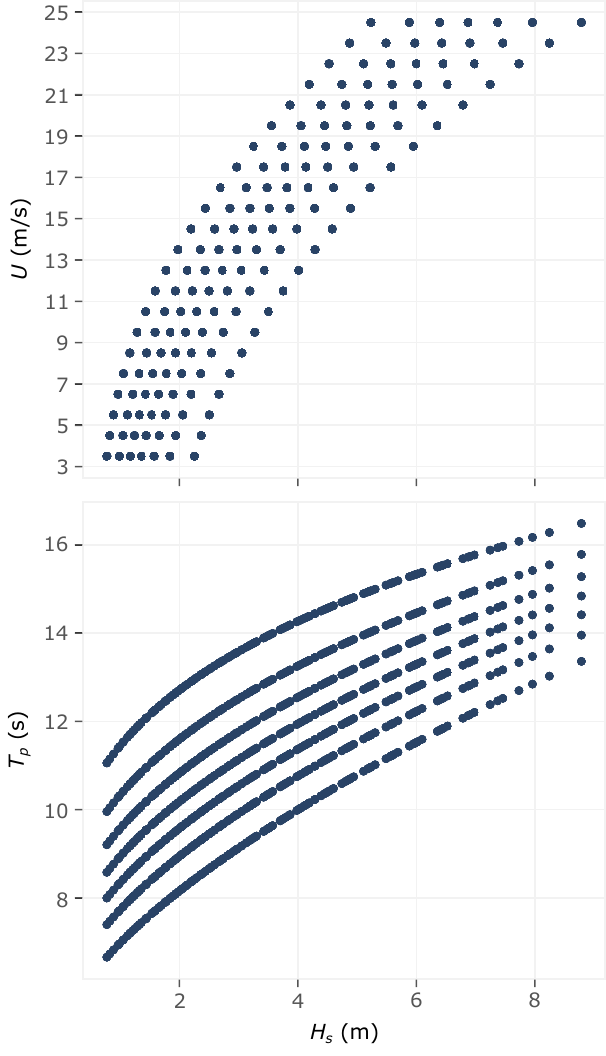}
\caption{Sampled environmental conditions generated by the \textbf{FLOAT} \textbf{\textit{Wind–Wave Sampler}} for the case study: 
wind speed $U$ vs.\ significant wave height $H_s$ and $H_s$ vs.\ peak wave period $T_p$, 
resulting in 6{,}468 environmental cases per design.}
    \label{fig:sampling}
\end{figure}

Wind–wave sampling was performed using the approach proposed in this paper to define the environmental conditions for FOWT fatigue assessment, implemented through the \textbf{\textit{Wind–Wave Sampler}} module of \textbf{FLOAT} (see~\autoref{sec:windwave_samp}). Following IEC Class~A specifications, 22 turbulent wind speeds were considered, each evaluated with six independent random seeds to represent stochastic variability. For each wind speed, seven significant wave heights were sampled and, for each wave height, seven peak wave periods were selected, resulting in 49 $(H_s, T_p)$ wave conditions per wind speed and a total of 6{,}468 ($22 \times 6 \times 49$) environmental simulations per design. The number of simulations was chosen to balance computational cost and runtime rather than to identify a minimum set for formal convergence. Fatigue damage contributions from the sampled states are aggregated using occurrence-probability weights derived from the adopted environmental distributions. In this work, wind direction variability and wind--wave misalignment are not considered to limit the number of required simulations, as is common at the design and pre-design stages. The results are therefore reported for aligned wind--wave conditions, although the proposed approach is readily extensible to other configurations. \autoref{fig:sampling} presents the complete set of sampled conditions, and an example of the wind--wave sampling procedure for a wind speed of 12.5~m/s is provided in \autoref{A:wave_samples}.

\subsubsection{Numerical Simulator Setup}  
The \textbf{\textit{Numerical Simulator}} of \textbf{FLOAT} (see~\autoref{sec:numsim}) was used to perform high-fidelity dynamic simulations with pitch/heave–platform calibration. Its accuracy was verified by benchmarking the IEA 22~MW simulation results against those published in the reference study~\cite{1cd3e417f8854b808c5372670588d3d0}, as detailed in~\autoref{A:benchmarking_fado_iea}. All 6,468 sampled cases were simulated in OpenFAST, each with a total duration of 1,000~s, of which the first 400~s were discarded to remove transients, resulting in 10~minutes of effective simulation time. For intermediate calibration runs, shorter 200-second simulations were used, discarding the first 100 seconds to remove transients. \autoref{A:pitch_fado} analyzes the dynamic improvements achieved through pitch/heave–platform calibration, showing that both platform pitch and heave responses are substantially reduced and driven close to zero after calibration. Simulations were distributed across 250 cloud instances provided by Inductiva~\cite{InductivaOpenFAST}, each of type \texttt{c2d-highcpu-2} (2~vCPUs, 4~GB memory). This setup completed all 6,468 cases in 12.6~h at a total cost of $\sim$\$47, compared to 124 days if executed sequentially on a single instance, corresponding to an approximate 250$\times$ speed-up. Further details on the HPC benchmarking of the \textbf{FLOAT} \textbf{\textit{Numerical Simulator}} are provided in \autoref{A:benchmark_hpc}.

\subsection{Evaluation of the IEA 22 MW FOWT Tower} \label{sec:res_original}
Frequency response and fatigue analyses of the reference tower were conducted through numerical simulations executed within the \textbf{FLOAT} framework.

\subsubsection{Frequency Response Analysis} \label{case:freq}
\begin{figure}[h!]
    \centering
    \includegraphics[width=1\columnwidth]{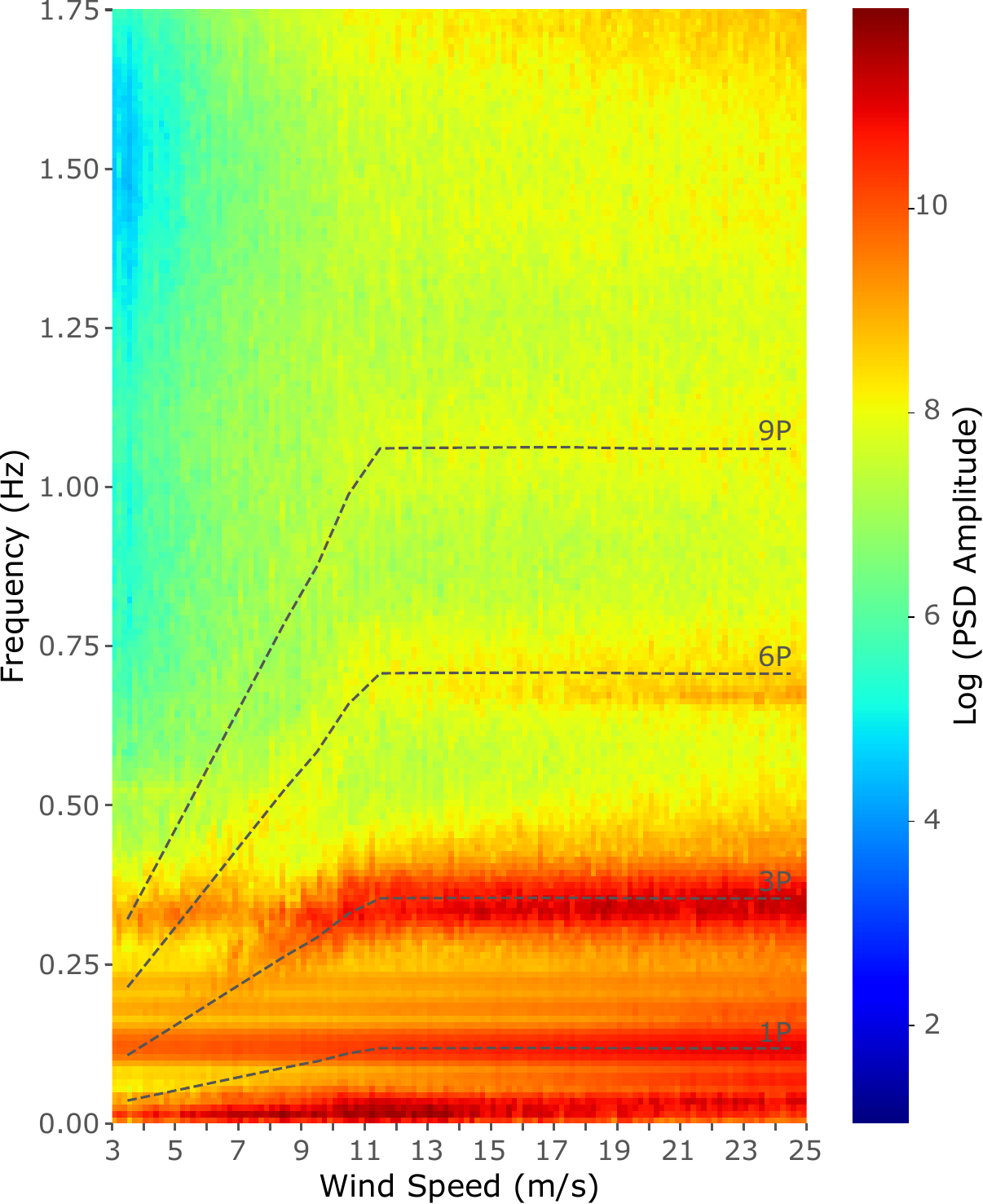}
\caption{PSD heatmap of tower-base fore–aft moments for the IEA~22~MW FOWT \cite{1cd3e417f8854b808c5372670588d3d0} across wind speeds, obtained using \textbf{FLOAT}. Dashed lines indicate rotor harmonics (1P, 3P, 6P, and 9P).}
    \label{fig:psd_heatmap}
\end{figure}

The \textbf{\textit{Frequency Response Analyser}} of \textbf{FLOAT} (see~\autoref{sec:freq_fado}) was applied to assess the tower’s dynamic behavior by generating a PSD heatmap of the tower-base fore–aft moment across wind speeds, as shown in \autoref{fig:psd_heatmap}. The analysis was restricted to the operational range between cut-in and rated wind speeds, 3–25 m/s, and to frequencies up to 1.75 Hz. A pronounced energy concentration near the 3P harmonic, around 0.35~Hz, closely matches the tower’s first natural frequency in the floating configuration ($\approx$0.34~Hz). This value is obtained by converting the land-based frequency of 0.214~Hz using \autoref{eq:float_to_land}. The proximity between the 3P excitation and the floating natural frequency indicates a potential resonance risk under specific operating conditions. Consequently, a stiff–stiff configuration was adopted for the optimized tower, ensuring that the first natural frequency is shifted above the 3P excitation range.

\subsubsection{Fatigue Analysis}   \label{case:damage}
\begin{figure}[h!]
    \centering
    \includegraphics[width=1.0\columnwidth]{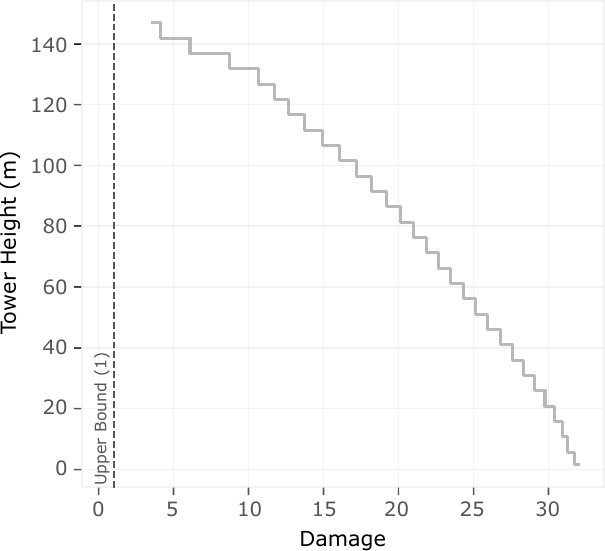}
\caption{Fatigue damage profile of the original IEA 22~MW FOWT tower \cite{1cd3e417f8854b808c5372670588d3d0}, obtained using \textbf{FLOAT}.}
    \label{fig:damage_profile}
\end{figure}

The \textbf{\textit{Fatigue Analyser}} of \textbf{FLOAT} (see~\autoref{sec:fatigue_fado}) 
was used to evaluate the cumulative fatigue damage at the midpoint of each tower section along the tower height, 
as shown in \autoref{fig:damage_profile}. Over a 25-year lifetime, the highest damage occurs near the base, reaching 32.1, which corresponds to a fatigue life of only 9 months. Even at the top, where the minimum value is 3.47, the fatigue life is limited to about 7 years. These results establish the reference and underscore the need for fatigue-aware design optimization of the IEA~22~MW FOWT tower.

\subsection{Optimization of the IEA 22 MW Reference Tower}\label{sec:design_optimization}
The objective is to explicitly incorporate fatigue into the tower design process, with structural mass minimized subject to fatigue constraints. The approach follows the design framework of the IEA 22 MW reference study~\cite{1cd3e417f8854b808c5372670588d3d0}, with targeted modifications aimed at improving fatigue performance and extending the structural lifetime.

\subsubsection{Modifications to the IEA 22 MW Optimization}
The optimization strategy was modified to explicitly account for fatigue-driven constraints. A section-wise fatigue constraint was imposed at the midpoint of each tower section, limiting the cumulative lifetime fatigue damage. In the first cycle, a fatigue uncertainty factor $\gamma_d$ of 1.0 was applied, since the design was still far from convergence and enforcing a stricter criterion at this stage would have resulted in an unnecessarily conservative tower. In subsequent cycles, $\gamma_d$ was set to 1.11, corresponding to an effective damage limit of $D < 0.9$, in order to introduce a robustness margin once the geometry was closer to the final design. To mitigate the high fatigue damage observed at the tower base (see~\autoref{fig:damage_profile}) and keep it within the admissible limit, 
the upper bound of the bottom diameter was increased from 10~m to 12~m, 
matching the dimension of the floater’s central column. Furthermore, the first natural frequency was shifted from the soft–stiff design of the original IEA~22~MW FOWT tower 
to a stiff–stiff configuration, constrained between the 3P ($\sim$0.35~Hz) and 6P ($\sim$0.71~Hz) rotor harmonics 
with a 15\% margin (0.4–0.6~Hz) to avoid resonance, based on the dynamic response analysis shown in \autoref{fig:psd_heatmap}. 
As the optimization was performed using a land-based model, the floating frequency constraint was converted to its land-based equivalent using \autoref{eq:float_to_land}, yielding a target range of 0.25–0.38~Hz.

\subsubsection{New Tower Optimization}  
To achieve a fatigue-aware tower design with minimum structural weight, 
the optimization objective is defined as the minimization of the tower mass. 
The design variables are:

\begin{itemize}
\item \textbf{Outer diameter:} [6, 12] m
\item \textbf{Wall thickness:} [0.0375, 0.15] m
\end{itemize}

The applied design constraints include:
\begin{itemize}
\item \textbf{Diameter-to-thickness:} Diameter-to-thickness ratios were constrained between 80 and 160 to satisfy manufacturability requirements.

\item \textbf{Monotonicity:} Wall thickness was enforced to decrease monotonically along the tower height to avoid unrealistic local increases.

\item \textbf{Stress:} The maximum von Mises stress along the tower was constrained not to exceed the steel yield strength of 345~MPa. Partial safety factors of 1.35 for loads, 1.3 for material strength, and 1.0 for the consequence of failure were applied.

\item \textbf{Frequency:} The first natural frequency was constrained to remain above the 3P and below the 6P rotor harmonics, enforced within 0.25–0.38~Hz in the land-based model (corresponding to 0.4–0.6~Hz in the floating case) to avoid resonance.

\item \textbf{Buckling:} Global and shell buckling constraints were enforced according to DNV-RP-C202 \cite{DNV-RP-C202}.

\item \textbf{Fatigue:} Lifetime fatigue damage at the midpoint of each of the 30 sections was constrained to remain below unity. A fatigue uncertainty factor of 1.0 was applied in the first optimization cycle and increased to 1.11 in subsequent cycles, resulting in an effective damage limit of approximately $0.9$.
\end{itemize}

Within the optimization framework, tower-top loads were kept constant and equal to those of the IEA 22~MW reference model. 
These loads were used exclusively to verify stress, buckling, and frequency constraints, 
while fatigue damage governed the final tower geometry.

No outfitting factor was applied, as auxiliary systems (e.g., platforms, stairs) are non-load-bearing and were therefore excluded.

The optimization was carried out within the \textbf{FLOAT} framework using the \textbf{\textit{Design Optimizer}} (see~\autoref{sec:desing_opt}), with fatigue handled through the \textbf{\textit{Fatigue Estimator}} introduced in this work (see~\autoref{sec:fatigue_estimator}). The optimization problem was solved using the SLSQP algorithm with central finite-difference gradients, an optimality tolerance of $10^{-3}$, a step size of $10^{-4}$, and a maximum of 100 iterations. The procedure was executed through WISDEM, incorporating the \textbf{\textit{Fatigue Estimator}} within the optimization loop.

\section{Results and Discussion} \label{sec:results}
The results obtained with the \textbf{FLOAT} model are organized into three categories. First, \autoref{sec:res_optimization_process} describes the optimization procedure that produced a fatigue-oriented design for a new 22 MW FOWT tower. Next, \autoref{sec:res_comparison} presents a structural performance comparison between the reference and optimized towers, and \autoref{sec:res_validation_new_tower} validates the optimized designs through additional frequency response and fatigue assessments. The final design was achieved after two optimization cycles. 
For clarity, results from the first (Optimized~1) and second (Optimized~2, corresponding to the final FLOAT~22~MW floating tower design) optimization iterations are reported together for each category.

\subsection{Fatigue-Aware Design Optimization of the IEA 22 MW FOWT Tower}\label{sec:res_optimization_process}

This section presents the results of the fatigue-aware design optimization with \textbf{FLOAT}, applied to the IEA 22 MW FOWT tower following the procedure in \autoref{sec:design_optimization}. The objective was to redesign the tower to reduce lifetime fatigue damage while minimizing mass and satisfying fatigue and other design constraints.

\subsubsection{Optimization Process} 

\begin{figure}[!ht]
    \centering
\includegraphics[width=1\columnwidth]{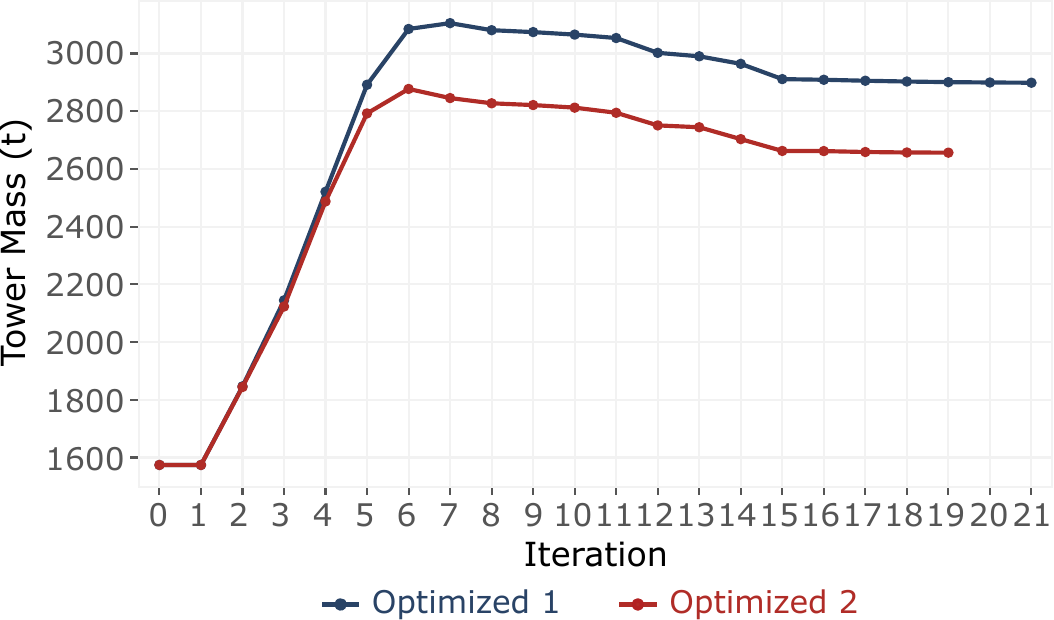}
\caption{Evolution of the objective function (tower mass) during the optimization of the IEA~22~MW FOWT tower using \textbf{FLOAT}.}
    \label{fig:objective_function}
\end{figure}

\autoref{fig:objective_function} shows the evolution of the objective function, defined as tower mass minimization, over the two optimization cycles (Optimized~1 and Optimized~2), while \autoref{tab:costmass} summarizes the corresponding mass and cost values for the reference and optimized designs. The reference tower, with a mass of 1,574~t, was not designed to satisfy a 25-year fatigue lifetime under offshore operating conditions and exhibits an estimated operational lifetime of approximately 9 months. When fatigue resistance and additional structural constraints were imposed, increases in wall thickness and diameter were required to reduce cyclic stresses, resulting in a tower mass of 2,899~t after 21 iterations in the first optimization cycle (+84.2\%). In the second cycle, after 19 iterations, a more efficient redistribution of material reduced the mass to 2,656~t (+68.8\%), corresponding to a 15.4\% reduction relative to the first-cycle design.

\begin{table}[h!]
\footnotesize
\centering
\caption{Comparison of mass and estimated cost between the reference and optimized IEA~22~MW towers obtained with \textbf{FLOAT}.}
\begin{tabular}{@{}lccc@{}}
\toprule
\textbf{Parameter} 
& \cellcolor{slategrey!15}\textbf{Reference} 
& \cellcolor{midnightblue!15}\textbf{Optimized 1} 
& \cellcolor{firebrick!15}\textbf{Optimized 2} \\
\midrule
Tower mass [t]  
& \cellcolor{slategrey!5}1,574 
& \cellcolor{midnightblue!5}\textcolor{midnightblue}{2,899 \,(+84.2\%)} 
& \cellcolor{firebrick!5}\textcolor{firebrick}{\textbf{2,656 \,(+68.8\%)}} \\
Tower cost [USD] 
& \cellcolor{slategrey!5}4.11M 
& \cellcolor{midnightblue!5}\textcolor{midnightblue}{ 7.11M \,(+73.2\%)} 
& \cellcolor{firebrick!5}\textcolor{firebrick}{\textbf{6.57M \,(+60.1\%)}} \\
\bottomrule
\end{tabular}
\label{tab:costmass}
\end{table}

Although heavier than the reference, this configuration represents the minimum-mass solution that satisfies the imposed fatigue, frequency, stress, and buckling constraints within the selected material and design space. Higher-strength steels combined with localized reinforcements may further reduce wall thickness and cyclic stresses, enabling lighter yet fatigue-resistant configurations. Estimated costs followed the same trend, increasing from 4.11~M~USD (reference) to 7.11~M~USD in Optimized~1 (+73.2\%) and then decreasing to 6.57~M~USD in Optimized~2 (+60.1\%). This design evolution extended the tower lifetime to 25 years, confirming fatigue as the governing constraint in the final tower dimensioning. The influence of fatigue as the active constraint is examined in detail in the following subsection through the evolution of the individual design constraints.

\subsubsection{Design Constraints Evolution} 

\begin{figure}[!ht]
    \centering
    \includegraphics[width=1\columnwidth]{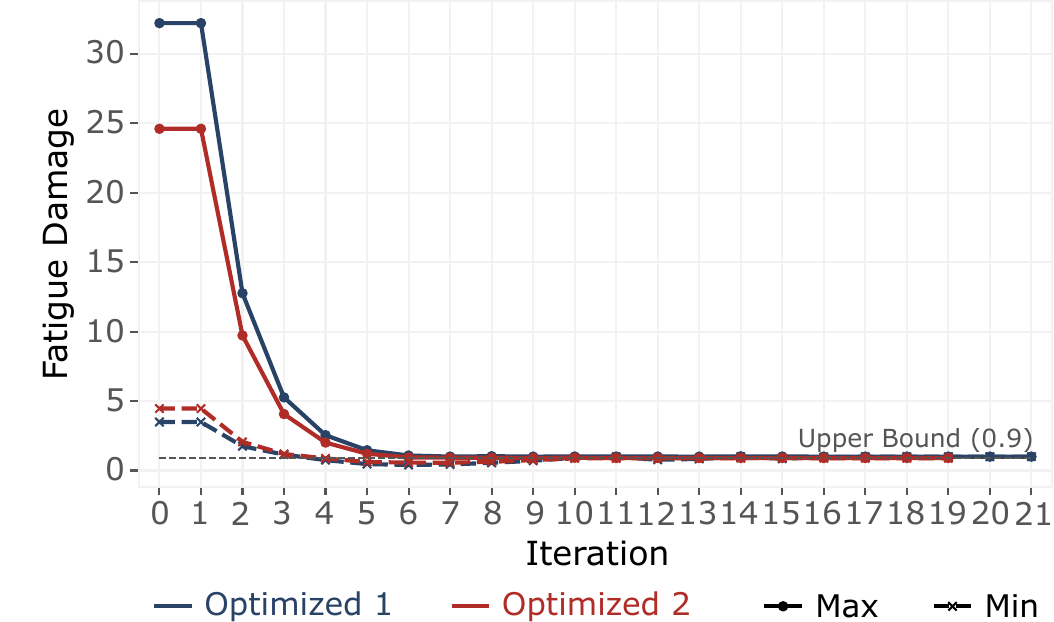}
\caption{Evolution of fatigue damage, the governing design constraint during the optimization of the IEA 22 MW FOWT tower using \textbf{FLOAT}.}
    \label{fig:fatigue_constraint}
\end{figure}

\autoref{fig:fatigue_constraint} highlights the evolution of fatigue damage, showing the minimum and maximum values across all sections. In both cycles (Optimized~1 and Optimized~2), several tower sections initially exceeded the admissible limit, but fatigue damage progressively decreased as the geometry adapted and converged within the allowable range. The convergence trends are similar in both cycles, although fewer iterations were required in the second cycle, as it is closer to the final design. This convergence behavior reflects a clear trade-off between tower mass (see \autoref{fig:objective_function}) and fatigue damage: during the first 6 to 7 iterations, fatigue-driven geometric changes increase wall thickness and mass, leading to a rapid reduction in fatigue damage. Once fatigue compliance is achieved, subsequent iterations focus on redistributing material to improve mass efficiency while maintaining fatigue requirements, yielding the minimum-mass tower configuration within the considered design space.

\begin{figure*}[!t]
    \centering
    \includegraphics[width=1\textwidth]{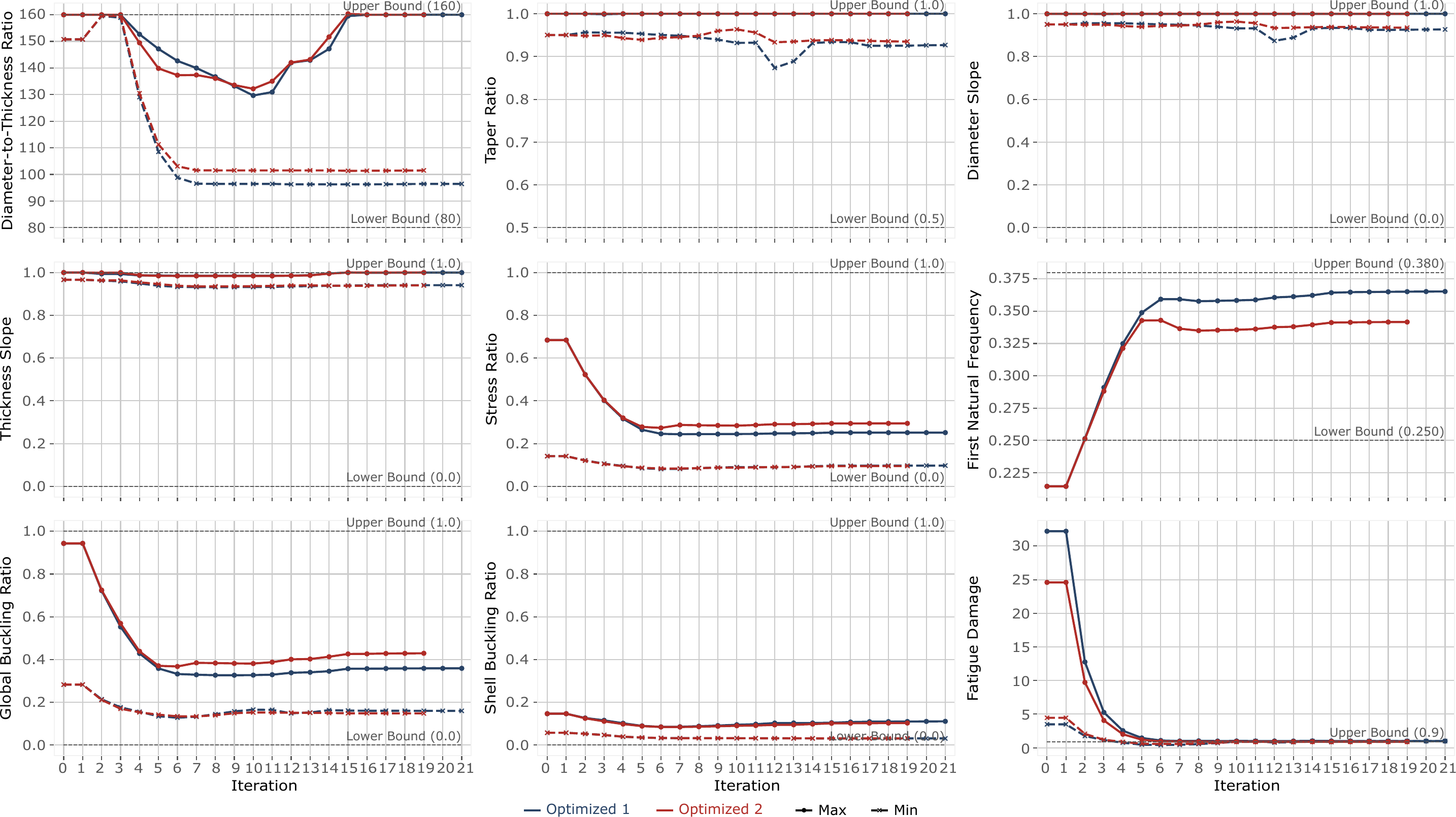}
\caption{Evolution of design constraint satisfaction during the optimization of the IEA 22 MW FOWT tower using \textbf{FLOAT}.}
\label{fig:all_constraints_evolution}
\end{figure*}

\autoref{fig:all_constraints_evolution} compares the fatigue constraint with the remaining structural constraints, namely stress utilization, buckling utilization, and the first natural frequency, as well as geometric constraints related to the diameter-to-thickness ratio, tapering, and monotonicity limits. Throughout the optimization process, stress, buckling, and geometric constraints remained satisfied and therefore acted primarily as safeguarding conditions rather than active drivers of the design. The first natural frequency, initially below the target band, shifted upward into compliance in both optimization cycles, requiring only minor geometric adjustments during the first two iterations. In contrast, fatigue consistently governed the design evolution through geometry-driven changes, which in turn led to mass redistribution. Together with the mass evolution discussed in the previous subsection, these trends confirm that fatigue resistance acts as the governing design constraint shaping the final tower configuration.

\subsubsection{Optimized 22 MW FOWT Tower} 
\begin{figure*}[h!]
    \centering
    \includegraphics[width=0.95\textwidth]{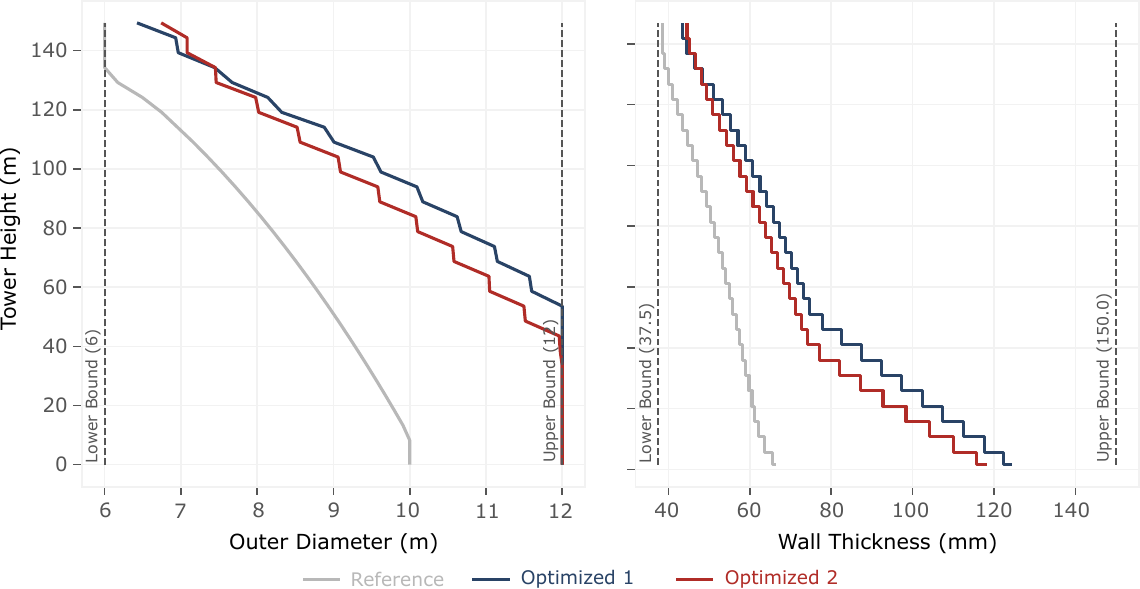}
\caption{Comparison of tower geometries for the original and optimized IEA 22 MW FOWT designs obtained with \textbf{FLOAT}.}
    \label{fig:new_tower}
\end{figure*}

\autoref{fig:new_tower} compares the original and optimized geometries of the IEA~22~MW Semi-Submersible FOWT tower, while \autoref{tab:geometry_bottom_top} summarizes the key geometric parameters, namely tower diameter, wall thickness, and the diameter-to-thickness ratio (D/t). The final optimized design (Optimized~2) increased both diameter and wall thickness, with the base diameter growing from 10.0~m to 12.0~m (+20.0\%) and thickness from 66~mm to 118~mm (+78.2\%). At the top, the diameter rose from 6.0~m to 6.741~m (+12.4\%) and thickness from 38~mm to 45~mm (+15.8\%). The maximum differences relative to the reference reached +30.1\% in diameter at the intermediate section and +78.2\% in thickness at the bottom section. These changes reduced the minimum diameter-to-thickness ratio from 150.8 to 101.5, improving resistance against local buckling. 
The full geometric parameters of the reference tower and the final optimized tower (FLOAT 22~MW Semi-Submersible FOWT tower) are reported in \autoref{A:table_ref} and \autoref{A:table_opt}, respectively.

\begin{table}[h!]
\footnotesize
\centering
\caption{Geometric comparison between the reference and optimized IEA 22 MW towers obtained with \textbf{FLOAT}.}
\label{tab:geometry_bottom_top}
\begin{tabular}{@{}lccc@{}}
\toprule
\textbf{Parameter} 
& \cellcolor{slategrey!20}\textbf{Reference} 
& \cellcolor{midnightblue!20}\textbf{Optimized 1} 
& \cellcolor{firebrick!20}\textbf{Optimized 2} \\
\midrule
\multicolumn{4}{l}{\textbf{(1) Diameter [m] }} \\
\midrule
Bottom
& \cellcolor{slategrey!5}10.000 
& \cellcolor{midnightblue!5}\textcolor{midnightblue}{12.000 \,(20.0\%)} 
& \cellcolor{firebrick!5}\textbf{\textcolor{firebrick}{12.000 \,(20.0\%)}} \\
Top 
& \cellcolor{slategrey!5}6.000  
& \cellcolor{midnightblue!5}\textcolor{midnightblue}{6.424 \,(7.1\%)} 
& \cellcolor{firebrick!5}\textbf{\textcolor{firebrick}{6.741 \,(12.4\%)}} \\
Max Diff  
& \cellcolor{slategrey!5}-- 
& \cellcolor{midnightblue!5}\textcolor{midnightblue}{3.069 (34.4\%)} 
& \cellcolor{firebrick!5}\textbf{\textcolor{firebrick}{2.766 (30.1\%)}} \\
\midrule
\multicolumn{4}{l}{\textbf{(2) Thickness [m] }} \\
\midrule
Bottom
& \cellcolor{slategrey!5}0.066 
& \cellcolor{midnightblue!5}\textcolor{midnightblue}{0.124 \,(87.6\%)} 
& \cellcolor{firebrick!5}\textbf{\textcolor{firebrick}{0.118 \,(78.2\%)}} \\
Top    
& \cellcolor{slategrey!5}0.038 
& \cellcolor{midnightblue!5}\textcolor{midnightblue}{0.043 \,(13.1\%)} 
& \cellcolor{firebrick!5}\textbf{\textcolor{firebrick}{0.045 \,(15.8\%)}} \\
Max Diff
& \cellcolor{slategrey!5}-- 
& \cellcolor{midnightblue!5}\textcolor{midnightblue}{0.058 (87.6\%)} 
& \cellcolor{firebrick!5}\textbf{\textcolor{firebrick}{0.052 (78.2\%)}} \\
\midrule
\multicolumn{4}{l}{\textbf{(3) Diameter-to-Thickness Ratio (D/t) [--]}} \\
\midrule
D/t Ratio 
& \cellcolor{slategrey!5}150.8–160.0 
& \cellcolor{midnightblue!5}\textcolor{midnightblue}{96.4–160.0} 
& \cellcolor{firebrick!5}\textbf{\textcolor{firebrick}{101.5–160.0}} \\
\bottomrule
\end{tabular}
\end{table}

\subsection{Structural Performance Comparison Between Original and Optimized 22 MW FOWT Towers}
\label{sec:res_comparison}

\begin{figure*}[h!]
    \centering
    \subfloat[Fatigue damage]{%
    \label{fig:tower_damage_comparison}
        \includegraphics[width=0.32\textwidth]{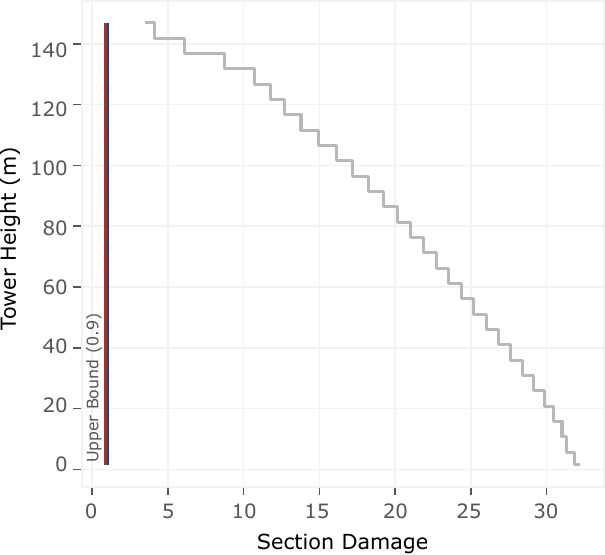}
    }\hfill
    \subfloat[Axial stress]{%
    \label{fig:tower_stress_comparison}
        \includegraphics[width=0.32\textwidth]{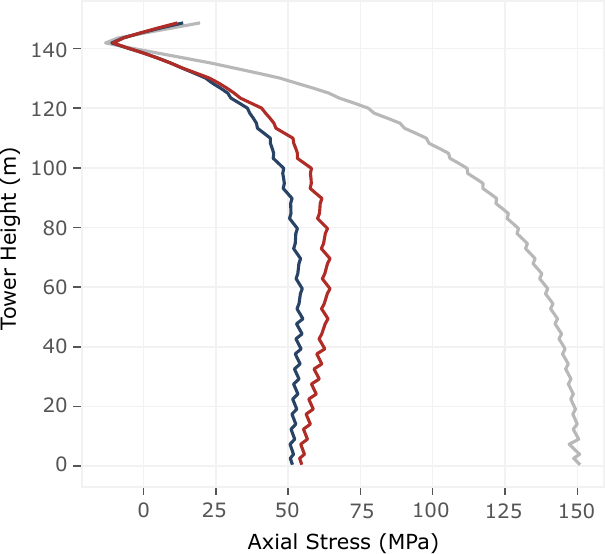}
    }\hfill
    \subfloat[Top deflection]{%
    \label{fig:tower_deflection_comparison}
        \includegraphics[width=0.32\textwidth]{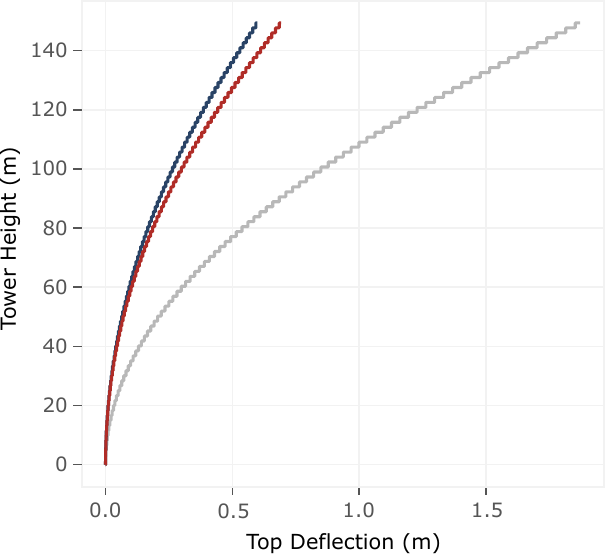}
    }\\[3ex]
    \subfloat[Buckling]{%
    \label{fig:tower_buckling_comparison}
        \includegraphics[width=0.64\textwidth]{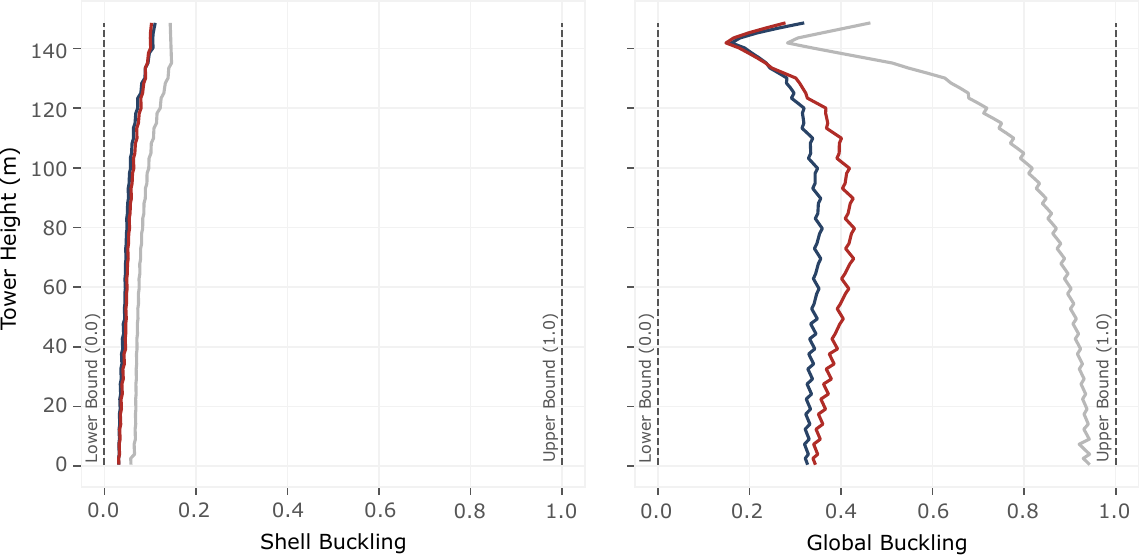}
    }
    \vspace{0.3cm}
    \includegraphics[width=0.64\textwidth]{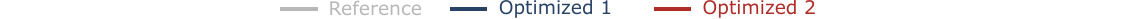}

    \caption{Structural performance of the IEA 22 MW FOWT tower using \textbf{FLOAT}.}
    \label{fig:tower_structural_performance}
\end{figure*}

This section presents a detailed comparison between the original IEA 22~MW FOWT tower (reference) and the optimized configurations obtained from the first optimization cycle (Optimized~1) and the final design (Optimized~2), focusing on key structural performance metrics. The overall results are illustrated in \autoref{fig:tower_structural_performance}, which presents fatigue damage, axial stress, top deflection, and buckling utilization profiles, while the corresponding numerical values are summarized in \autoref{tab:structural_bottom_top_final}.

\begin{table}[h!]
\footnotesize
\centering
\caption{Structural performance comparison between reference and optimized IEA 22 MW towers obtained with \textbf{FLOAT}.}
\label{tab:structural_bottom_top_final}
\begin{tabular}{@{}lccc@{}}
\toprule
\textbf{Metric}
& \cellcolor{slategrey!20}\textbf{Reference}
& \cellcolor{midnightblue!20}\textbf{Optimized 1}
& \cellcolor{firebrick!20}\textbf{Optimized 2} \\
\midrule
\multicolumn{4}{l}{\textbf{(1) Fatigue Damage [--]}} \\
\midrule
Bottom & \cellcolor{slategrey!5}32.128
& \cellcolor{midnightblue!5}\textcolor{midnightblue}{1.000\,($-96.9$\%)} 
& \cellcolor{firebrick!5}\textbf{\textcolor{firebrick}{0.900 \,(\textbf{\boldmath$-97.2$\%}
)}} \\
Top     & \cellcolor{slategrey!5}3.471
& \cellcolor{midnightblue!5}\textcolor{midnightblue}{1.000 \,($-71.2$\%)} 
& \cellcolor{firebrick!5}\textbf{\textcolor{firebrick}{0.900 \,(\textbf{\boldmath$-74.1$\%})}} \\
Max Diff & \cellcolor{slategrey!5}--
& \cellcolor{midnightblue!5}\textcolor{midnightblue}{$-31.128$ ($-96.9$\%)}
& \cellcolor{firebrick!5}\textbf{\textcolor{firebrick}{\textbf{\boldmath$-31.228$ (\boldmath$-97.2$\%)}}} \\
\midrule
\multicolumn{4}{l}{\textbf{(2) Axial Stress [MPa]}} \\
\midrule
Bottom  & \cellcolor{slategrey!5}150.944
& \cellcolor{midnightblue!5}\textcolor{midnightblue}{51.514 \,($-65.9$\%)} 
& \cellcolor{firebrick!5}\textbf{\textcolor{firebrick}{54.761 \,(\textbf{\boldmath$-63.7$\%})}} \\
Top     & \cellcolor{slategrey!5}19.601
& \cellcolor{midnightblue!5}\textcolor{midnightblue}{13.744 \,($-29.9$\%)} 
& \cellcolor{firebrick!5}\textbf{\textcolor{firebrick}{11.790 \,(\textbf{\boldmath$-39.85$\%})}} \\
Max Diff & \cellcolor{slategrey!5}--
& \cellcolor{midnightblue!5}\textcolor{midnightblue}{$-99.430$ ($-65.9$\%)}
& \cellcolor{firebrick!5}\textbf{\textcolor{firebrick}{\textbf{\boldmath$-96.183$ (\boldmath$-63.7$\%)}}} \\
\midrule
\multicolumn{4}{l}{\textbf{(3) Top Deflection [m]}} \\
\midrule
Bottom  & \cellcolor{slategrey!5}0.000
& \cellcolor{midnightblue!5}\textcolor{midnightblue}{0.000 (0.0\%)}
& \cellcolor{firebrick!5}\textbf{\textcolor{firebrick}{0.000 (0.0\%)}} \\
Top     & \cellcolor{slategrey!5}1.870
& \cellcolor{midnightblue!5}\textcolor{midnightblue}{0.599 \,($-68.0$\%)} 
& \cellcolor{firebrick!5}\textbf{\textcolor{firebrick}{0.692 \,(\textbf{\boldmath$-63.0$\%})}} \\
Max Diff & \cellcolor{slategrey!5}--
& \cellcolor{midnightblue!5}\textcolor{midnightblue}{$-1.271$ ($-68.0$\%)}
& \cellcolor{firebrick!5}\textbf{\textcolor{firebrick}{\textbf{\boldmath$-1.178$ (\boldmath$-63.0$\%)}}} \\
\midrule
\multicolumn{4}{l}{\textbf{(4a) Shell Buckling [--]}} \\
\midrule
Bottom  & \cellcolor{slategrey!5}0.057
& \cellcolor{midnightblue!5}\textcolor{midnightblue}{0.031 \,($-45.6$\%)} 
& \cellcolor{firebrick!5}\textbf{\textcolor{firebrick}{0.032 \,(\textbf{\boldmath$-45.6$\%})}} \\
Top     & \cellcolor{slategrey!5}0.144
& \cellcolor{midnightblue!5}\textcolor{midnightblue}{0.111 \,($-22.8$\%)} 
& \cellcolor{firebrick!5}\textbf{\textcolor{firebrick}{0.103 \,(\textbf{\boldmath$-28.4$\%})}} \\
Max Diff & \cellcolor{slategrey!5}--
& \cellcolor{midnightblue!5}\textcolor{midnightblue}{$-0.052$ ($-35.1$\%)}
& \cellcolor{firebrick!5}\textbf{\textcolor{firebrick}{\textbf{\boldmath$-0.053$ (\boldmath$-35.8$\%)}}} \\
\midrule
\multicolumn{4}{l}{\textbf{(4b) Global Buckling [--]}} \\
\midrule
Bottom  & \cellcolor{slategrey!5}0.943 
& \cellcolor{midnightblue!5}\textcolor{midnightblue}{0.327\,($-65.4$\%)} 
& \cellcolor{firebrick!5}\textbf{\textcolor{firebrick}{0.344 \,(\textbf{\boldmath$-63.6$\%})}} \\
Top     & \cellcolor{slategrey!5}0.464
& \cellcolor{midnightblue!5}\textcolor{midnightblue}{0.319 \,($-31.2$\%)} 
& \cellcolor{firebrick!5}\textbf{\textcolor{firebrick}{0.278 \,(\textbf{\boldmath$-40.0$\%})}} \\
Max Diff & \cellcolor{slategrey!5}--
& \cellcolor{midnightblue!5}\textcolor{midnightblue}{$-0.616$ ($-65.4$\%)}
& \cellcolor{firebrick!5}\textbf{\textcolor{firebrick}{\textbf{\boldmath$-0.599$ (\boldmath$-63.6$\%)}}} \\
\midrule
\multicolumn{4}{l}{\textbf{(5) First Natural Frequency (land-based) [Hz]}} \\
\midrule
$f_1$  
& \cellcolor{slategrey!5}0.214
& \cellcolor{midnightblue!5}\textcolor{midnightblue}{0.365 \,($+70.5$\%)} 
& \cellcolor{firebrick!5}\textbf{\textcolor{firebrick}{0.342 \,(\textbf{\boldmath$+59.4$\%})}} \\
\bottomrule
\end{tabular}
\end{table}

\begin{figure*}[b!]
    \centering
    \includegraphics[width=1\textwidth]{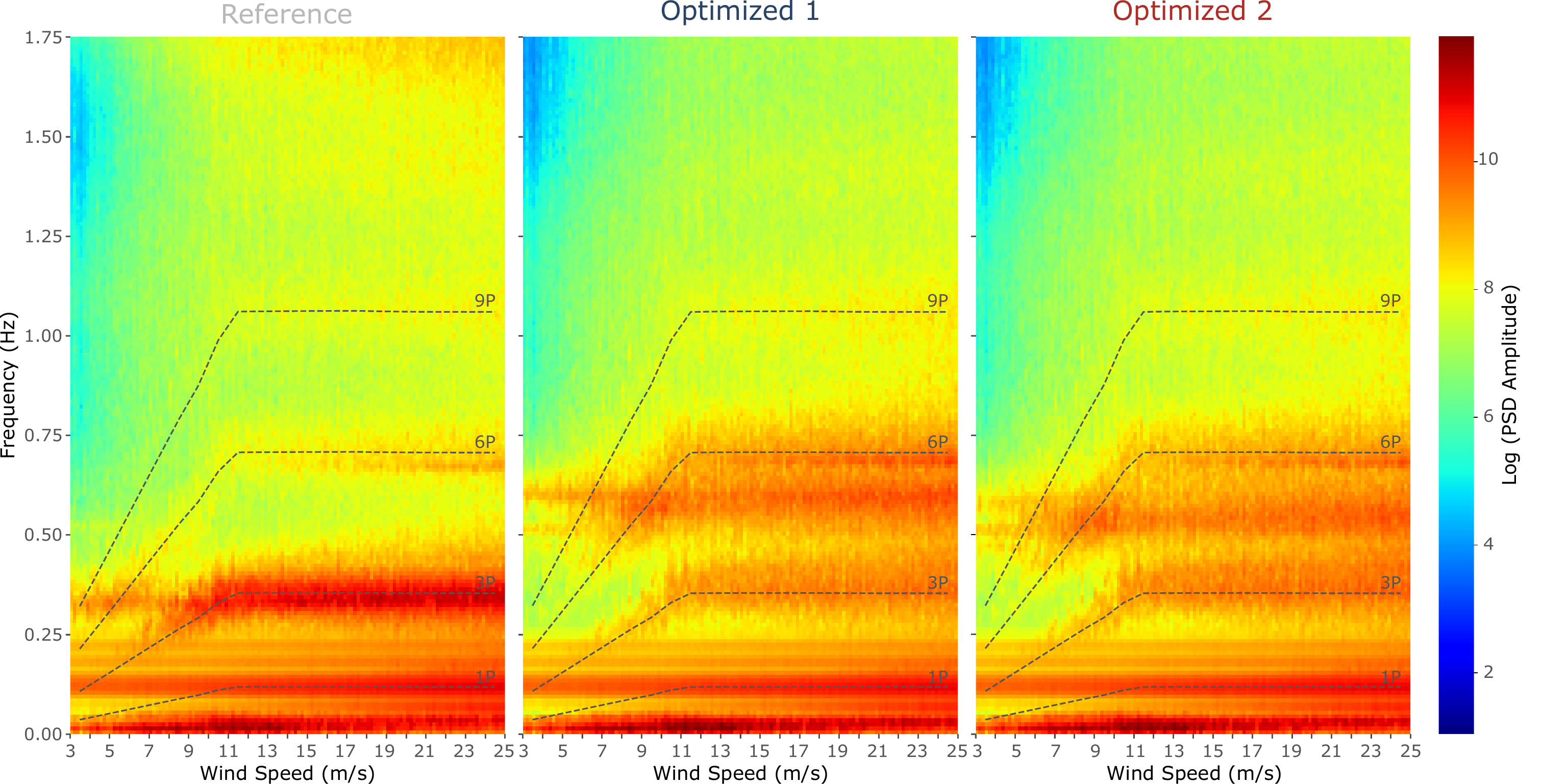}
    \caption{PSD heatmap of tower-base fore–aft bending moments for the IEA 22~MW FOWT across wind speeds, obtained using \textbf{FLOAT}. From left to right: reference tower, first optimization tower, and second optimization tower. Dashed lines mark rotor harmonics (1P, 3P, 6P, 9P).}
    \label{fig:comparation_freq}
\end{figure*}

\subsubsection{Fatigue Damage}
As shown in \autoref{fig:tower_damage_comparison}, fatigue damage reached its maximum value at the bottom of the reference tower (32.128) but was reduced to 0.900 in the final design ($-97.2$\%), representing the location with the largest reduction. At the top, the damage decreased from 3.471 to 0.900 ($-74.2$\%), keeping all locations below the critical threshold of 1.

\subsubsection{Axial Stress}
\autoref{fig:tower_stress_comparison} shows that reference stresses of 150.944~MPa at the bottom decreased to 54.761~MPa ($-63.7$\%) in the final design, representing the section with the maximum reduction. This improvement reflects the higher stiffness of the tower required to meet the fatigue damage constraints.

\subsubsection{Top Deflection}
In \autoref{fig:tower_deflection_comparison}, the top deflection of 1.870~m in the reference tower was reduced to 0.692~m ($-63.0$\%) in the final design, representing the location with the largest reduction in displacement. Once more, this improvement reflects the higher stiffness of the tower required to meet the fatigue damage constraints.

\subsubsection{Buckling}
\autoref{fig:tower_buckling_comparison} shows that shell buckling utilization decreased from the reference to the final design, with similar reductions observed along the tower and a maximum reduction of about 0.053. Global buckling utilization improved more substantially, especially at the bottom, decreasing from 0.943 to 0.344, corresponding to a reduction of 0.599 ($-63.6$\%) and representing the location with the greatest improvement. Once more, this improvement reflects the higher stiffness of the tower required to meet the fatigue damage constraints.

\subsubsection{First Natural Frequency}
The first natural frequency increased from 0.214~Hz in the reference tower to 0.342~Hz in the final design ($+59.4$\%), as shown in \autoref{tab:structural_bottom_top_final}. When converted from the land-based to the floating case using \autoref{eq:float_to_land}, the reference value of 0.214~Hz corresponds to approximately 0.34~Hz, while the final design shifts to about 0.54~Hz. This places the optimized tower within the stiff–stiff configuration, safely constrained between 3P ($\sim$0.35~Hz) and 6P ($\sim$0.71~Hz), ensuring resonance is avoided and reflecting the increased stiffness of the optimized design.

\subsection{Validation of the Optimized 22 MW FOWT Towers} \label{sec:res_validation_new_tower} 
Validation of the optimized 22~MW towers is required to assess whether additional optimization cycles are necessary, as observed in this case where two iterations were required. To this end, additional frequency response and fatigue assessments were performed using the \textbf{\textit{Frequency Response Analyser}} and the \textbf{\textit{Fatigue Analyser}} of \textbf{FLOAT}, respectively. For each optimization cycle, the reference wind--wave conditions (see~\autoref{sec:sampl-windwave}) were re-simulated to evaluate the optimized towers and assess their frequency response and fatigue performance relative to the reference design.

\subsubsection{Frequency Response Validation}

\autoref{fig:comparation_freq} confirms the effectiveness of the optimizations by comparing the PSD heatmaps of tower-base fore–aft moments for the reference and optimized towers. The reference response, already analysed in \autoref{case:freq}, revealed strong amplification near the 3P harmonic ($\sim0.35$ Hz), close to the reference tower’s first floating natural frequency ($\sim0.34$ Hz), indicating a resonance problem. The validation highlights how this resonance is mitigated in the optimized designs: the first optimization shifts part of the energy away from 3P towards frequencies below 6P, while the final design (second optimization) achieves further attenuation both at 3P and near the 6P limit ($\sim0.71$ Hz). These observations are in line with the first natural frequency of the final design (0.54~Hz), which safely places the optimized tower within the stiff–stiff configuration between the 3P and 6P limits. The progressive decrease in harmonic amplification across iterations confirms that the final design provides a more robust frequency–response performance than the reference.

\subsubsection{Fatigue Performance Validation}

\begin{figure*}[t!]
    \centering
    \subfloat[Reference and optimized designs across the tower height]{%
    \label{fig:tower_damage_comparison_all_a}
        \includegraphics[width=1\columnwidth]{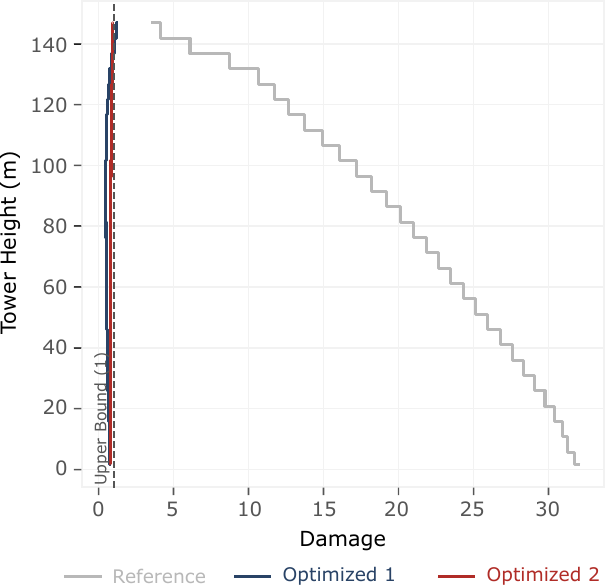}
    }\hfill
    \subfloat[Optimized designs with admissible fatigue bounds indicated]{%
    \label{fig:tower_damage_comparison_all_b}
        \includegraphics[width=1\columnwidth]{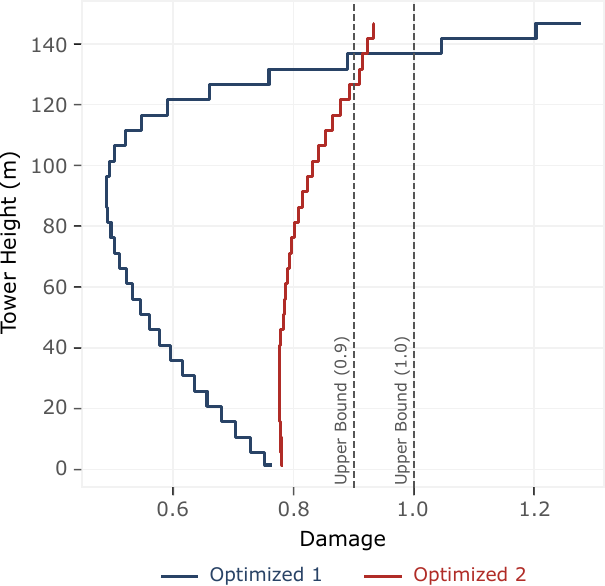}
    }
    \caption{Comparison of fatigue damage profiles for the IEA 22~MW FOWT tower obtained with \textbf{FLOAT}.}
    \label{fig:tower_damage_comparison_all}
\end{figure*}

\autoref{fig:tower_damage_comparison_all_a} compares the fatigue damage of the reference and optimized towers, high-fidelity simulations. The reference largely exceeds admissible levels (see \autoref{case:damage}), reaching 32.1 at the base, while the first optimization substantially reduces fatigue damage but still exceeds the admissible limit at the upper sections, with values of about 1.3, as shown in \autoref{tab:damage_bottom_top_val}. This corresponds to a lifetime extension from 9 months for the reference design to approximately 19 years with the Optimized~1 design, although still below the required 25-year lifetime. As shown in \autoref{fig:tower_damage_comparison_all_b}, the second optimization targeted a more conservative bound of 0.9, resulting in a fatigue damage profile in which all sections remain below unity and close to 0.9, thereby satisfying the 25-year fatigue life requirement and validating the first convergence criterion of the \textbf{FLOAT} workflow (see \autoref{sec:val_work}).

\begin{table}[h!]
\footnotesize
\centering
\caption{Damage comparison between reference and optimized IEA 22~MW towers obtained with \textbf{FLOAT}.}
\label{tab:damage_bottom_top_val}
\begin{tabular}{@{}lccc@{}}
\toprule
\textbf{Metric}
& \cellcolor{slategrey!20}\textbf{Reference}
& \cellcolor{midnightblue!20}\textbf{Optimized 1}
& \cellcolor{firebrick!20}\textbf{Optimized 2} \\
\midrule
Bottom & \cellcolor{slategrey!5}32.128
& \cellcolor{midnightblue!5}\textcolor{midnightblue}{0.764\,($-97.6$\%)} 
& \cellcolor{firebrick!5}\textbf{\textcolor{firebrick}{0.781 \,(\textbf{\boldmath$-97.6$\%}
)}} \\
Top     & \cellcolor{slategrey!5}3.471
& \cellcolor{midnightblue!5}\textcolor{midnightblue}{1.277 \,($-63.2$\%)} 
& \cellcolor{firebrick!5}\textbf{\textcolor{firebrick}{0.932 \,(\textbf{\boldmath$-73.1$\%})}} \\
Max Diff & \cellcolor{slategrey!5}--
& \cellcolor{midnightblue!5}\textcolor{midnightblue}{$-31.364$ ($-97.6$\%)}
& \cellcolor{firebrick!5}\textbf{\textcolor{firebrick}{\textbf{\boldmath$-31.347$ (\boldmath$-97.6$\%)}}} \\
\bottomrule
\end{tabular}
\end{table}

To validate the second convergence criterion, the mean relative error between \textbf{FLOAT} damage estimates and high-fidelity simulations was computed and reported in \autoref{tab:damage_fado_val}. For the final design, the error remains below 10\%, with a mean value of $-8.6$\%, a maximum deviation of 3.9\% at the top sections, and a minimum deviation of $-13.8$\%, indicating that \textbf{FLOAT} slightly underestimates fatigue damage in a conservative manner. These results confirm that both convergence criteria were satisfied and demonstrate the effectiveness of \textbf{FLOAT} in producing a reliable fatigue-resistant design.

\begin{table}[h!]
\footnotesize
\centering
\caption{Damage differences between high-fidelity simulation results and \textbf{FLOAT} estimations for the optimized IEA 22~MW towers.}
\label{tab:damage_fado_val}
\begin{tabular}{@{}lcc@{}}
\toprule
\textbf{Metric}
& \cellcolor{midnightblue!20}\textbf{Optimized 1 (Bound = 1.0)}
& \cellcolor{firebrick!20}\textbf{Optimized 2 (Bound = 0.9)} \\
\midrule
Bottom Diff
& \cellcolor{midnightblue!5}\textcolor{midnightblue}{$-0.236$\,($-23.6$\%)} 
& \cellcolor{firebrick!5}\textbf{\textcolor{firebrick}{\boldmath$-0.119$ \,(\textbf{\boldmath$-13.2$\%})}} \\
Top Diff    
& \cellcolor{midnightblue!5}\textcolor{midnightblue}{0.277 \,($27.7$\%)} 
& \cellcolor{firebrick!5}\textbf{\textcolor{firebrick}{0.032 \,(\textbf{\boldmath$3.5$\%})}} \\
Min Diff 
& \cellcolor{midnightblue!5}\textcolor{midnightblue}{$-0.512$ \,($-51.2$\%)} 
& \cellcolor{firebrick!5}\textbf{\textcolor{firebrick}{\boldmath$-0.124$ \,(\textbf{\boldmath$-13.8$\%})}} \\
Max Diff 
& \cellcolor{midnightblue!5}\textcolor{midnightblue}{$0.277$ ($27.7$\%)} 
& \cellcolor{firebrick!5}\textbf{\textcolor{firebrick}{\textbf{\boldmath$0.035$ (\boldmath$3.9$\%)}}} \\
Mean Diff 
& \cellcolor{midnightblue!5}\textcolor{midnightblue}{$-0.356$\,($-35.6$\%)} 
& \cellcolor{firebrick!5}\textbf{\textcolor{firebrick}{\boldmath$-0.077$ \,(\textbf{\boldmath$-8.6$\%})}} \\
\bottomrule
\end{tabular}
\end{table}

\section{Limitations} \label{sec:limitations}

This study prioritizes computational tractability during the intermediate design stage, which introduces modeling simplifications that should be acknowledged. These simplifications are intentionally confined to the optimization loop and do not affect the final acceptance of the optimized tower. In particular, the design optimization relies on a land-based tower model, with a floating-to-land frequency correction applied to preserve dynamic consistency, and evaluates intermediate structural constraints using prescribed tower-top loads that are not recomputed as the tower geometry evolves. In addition, the lightweight fatigue estimator adopts an analytical geometry-based scaling with an averaged S--N slope. While this formulation enables efficient iterative redesign, it does not explicitly resolve geometry-induced changes in coupled AHSE load redistribution during intermediate optimization iterations, which is a deliberate trade-off to enable rapid convergence during early-stage design exploration. To mitigate these limitations, final design acceptance is performed exclusively through full OpenFAST re-simulation of the same sampled environmental conditions, requiring all section-wise fatigue damages to remain admissible and agreement with high-fidelity results within a prescribed tolerance.

The demonstrated case study also reflects a deliberate scoping of environmental conditions and DLCs. Fatigue assessment and redesign are restricted to DLC~1.2 (normal power production), which typically governs cumulative lifetime fatigue damage and therefore defines the primary sizing envelope for tower design. Other IEC fatigue-relevant DLCs, such as fault, shutdown, and parked conditions, are not explicitly included in the presented optimization.

Likewise, wind direction variability and wind--wave misalignment are not considered in the case study in order to control the number of required simulations. The reported results are therefore specific to the aligned wind--wave conditions considered here, although the proposed probabilistic formulation can readily accommodate these effects. Finally, the wind--wave distributions adopted for sampling are based on published models and may not be representative of all sites; extending the framework to site-calibrated distributions would further strengthen its generalizability.

\section{Conclusion} \label{sec:conclusion}

This study introduced \textbf{FLOAT}, a framework for fatigue-aware design optimization of floating offshore wind turbine towers. The main outcomes are:

\begin{itemize}
    \item \textbf{Probabilistic Wind–wave sampling:} reduced $\sim\!1.5 \times 10^{6}$ IEC load cases for FOWTs to 6,468 ($\sim$99.6\% reduction), while maintaining representation of high-severity sea states that dominate fatigue damage.

    \item \textbf{HPC-enabled simulations:} 6,468 OpenFAST cases completed in 13 h at a cost of \$47 using 250 cloud instances, compared to 124 days sequentially ($\sim$250$\times$ faster). Pitch/heave calibration improved floating dynamics fidelity.  
    
    \item \textbf{Lightweight fatigue estimator:} validated against 6,468 simulations of the IEA 22 MW tower, yielding conservative fatigue predictions with a mean relative error of $-8.6\%$ across the 30 sections and a maximum positive deviation of $3.9\%$, while enabling iterative optimization without repeated high-fidelity runs.

    \item \textbf{Fatigue-oriented redesign of the IEA 22 MW tower:} Resulted in the \textit{FLOAT 22 MW FOWT tower}, extending fatigue life from $\sim$9 months to 25 years and reducing bottom and top damage by 97.6\% and 73\%, respectively. The redesign shifted the first floating natural frequency from 0.34\,Hz (soft--stiff) to 0.54\,Hz (stiff--stiff), eliminating resonance risks, with the associated 69\% mass increase reflecting the material required to meet long-term fatigue requirements within the considered design space. This could be mitigated through the adoption of higher-strength materials or localized reinforcements that reduce cyclic stresses.

\end{itemize}

All quantitative fatigue lifetime results reported in this work should be interpreted within the demonstrated DLC~1.2 aligned wind--wave scope; extension to additional DLCs and misaligned environmental conditions represents a natural next step for future studies using the same framework.

The proposed framework, demonstrated on the IEA 22~MW floating tower, is generic and can be adapted to other tower sizes, site conditions, and support structures, including fixed-bottom and land-based configurations, with appropriate recalibration of the environmental inputs and verification settings. Taken together, these contributions position \textbf{FLOAT} as a computationally efficient workflow for scaling floating offshore wind towers within the demonstrated scope, while also providing benchmark datasets to support future AI-driven design tools.

\section*{CRediT authorship contribution statement}

\textbf{João Alves Ribeiro}: Conceptualization, Methodology, Software, Formal Analysis, Investigation, Data Curation, Writing -- Original Draft, Visualization. 
\textbf{Francisco Pimenta}: Conceptualization, Methodology, Software, Formal Analysis, Investigation, Validation, Writing -- Review \& Editing. 
\textbf{Bruno Alves Ribeiro}: Conceptualization, Methodology, Software, Investigation, Validation, Writing -- Review \& Editing. 
\textbf{Sérgio M. O. Tavares}: Conceptualization, Methodology, Validation, Investigation, Writing -- Review \& Editing, Supervision. 
\textbf{Faez Ahmed}: Conceptualization, Methodology, Validation, Investigation, Writing -- Review \& Editing, Supervision.

\section*{Declaration of competing interest}
The authors declare that they have no known competing financial interests or personal relationships that could have appeared to influence the work reported in this paper.

\section*{Acknowledgements}
\textbf{João Alves Ribeiro} acknowledges funding from the Luso-American Development Foundation (FLAD) and the doctoral grant SFRH/BD/151362/2021 (DOI: \href{https://doi.org/10.54499/SFRH/BD/151364/2021}{10.54499/SFRH/BD/151364/2021} (accessed on November 18, 2024)), financed by the Portuguese Foundation for Science and Technology (FCT), Ministério da Ciência, Tecnologia e Ensino Superior (MCTES), Portugal, with funds from the State Budget (OE), European Social Fund (ESF), and PorNorte under the MIT Portugal Program, and by the Alliance for the Energy Transition (56) co-financed by the Recovery and Resilience Plan (PRR) through the European Union.

\textbf{Bruno Alves Ribeiro} acknowledges financial support from FCT through the doctoral grant 2021/08659/BD.

\textbf{Francisco Pimenta} acknowledges the financial support for project 2022.08120.PTDC, M4WIND (DOI: \href{https://doi.org/10.54499/2022.08120.PTDC}{10.54499/2022.08120.PTDC} (accessed on November 18, 2024)), funded by national funds through FCT/MCTES (PIDDAC), and for UID/ECI/04708/2020-CONSTRUCT-Instituto de I\&D em Estruturas e Construções, also funded by national funds through FCT/MCTES (PIDDAC).

The authors would like to thank \textbf{Garrett Barter}, \textbf{Pietro Bortolotti}, and \textbf{Daniel Zalkind} from the National Renewable Energy Laboratory (NREL) for their valuable discussions and support.

\section*{Data availability}

All \textbf{FLOAT} resources, including the framework and the redesigned IEA~22\,MW tower model (FLOAT~22\,MW FOWT tower), are openly available at \url{https://github.com/Joao97ribeiro/FLOAT} and \url{https://github.com/Joao97ribeiro/FLOAT-22-280-RWT-Semi}, respectively.

\clearpage
\newpage
\appendix

\section{Number of Simulations for Fatigue Analysis of OWT Following IEC Standard}\label{A:num_sim}

The total number of simulations for OWT fatigue assessment is obtained by discretizing the environmental parameters defined in IEC 61400-3 \cite{IEC61400-3-1, IEC61400-3-2}: mean wind speed ($U$), significant wave height ($H_s$), peak wave period ($T_p$), and mean wave direction ($M_{ww}$). \autoref{table:bin_ranges_iec} lists the range, bin width, and resulting bin count for each parameter for one seed per mean wind speed and turbulence intensity. The total cases equal the bin product, scaled by the number of seeds.

%%%%%%%%%%%%%%%%%%%%%%%%%%%%%%%%%%%%%%%%%%%%%%%%%%%%%%%%%%%%%%%%%%%%%%%%%%%%%%%%%%%%%%%%%%%%%%%%%%%%%%
%%%%%%%%%%%%%%%%%%%%%%%%%%%%%%%%%%%%%%%%%%%%%%%%%%%%%%%%%%%%%%%%%%%%%%%%%%%%%%%%%%%%%%%%%%%%%%%%%%%%%%
\begin{table}[h!]
\footnotesize
\centering
\caption{Number of simulations for fatigue analysis of OWT using one seed per mean wind speed and turbulence intensity, following IEC standard bin recommendations (adapted from \cite{Papi_2022}).}
\begin{tabularx}{\columnwidth}{@{}L{2.5cm}C{1.7cm}C{1.7cm}C{1.7cm}@{}}
\toprule
\textbf{Parameter} & \textbf{Range} & \textbf{Bin Width} & \textbf{Bins Number} \\ \midrule
$U$ (m/s)           & 4 -- 26       & 2           & 11  \\ \addlinespace[0.1cm]
$H_S$ (m)           & 0 -- 14       & 0.5         & 28  \\ \addlinespace[0.1cm]
$T_P$ (s)           & 3 -- 21       & 0.5         & 36  \\ \addlinespace[0.1cm]
$M_{ww}$ (°)        & $-180$ -- $180$ & 15        & 24  \\ \addlinespace[0.1cm]
\addlinespace[0.15cm]
\textbf{Total combinations} &  &  & \textbf{266112} \\
\bottomrule
\end{tabularx}
\label{table:bin_ranges_iec}
\end{table}
%%%%%%%%%%%%%%%%%%%%%%%%%%%%%%%%%%%%%%%%%%%%%%%%%%%%%%%%%%%%%%%%%%%%%%%%%%%%%%%%%%%%%%%%%%%%%%%%%%%%%%
%%%%%%%%%%%%%%%%%%%%%%%%%%%%%%%%%%%%%%%%%%%%%%%%%%%%%%%%%%%%%%%%%%%%%%%%%%%%%%%%%%%%%%%%%%%%%%%%%%%%%%

\section{Reference Models for Floating Offshore Wind Turbines} \label{A:fowt_models}

\autoref{table:fowt_re} summarizes the main reference models for FOWTs developed by leading institutions, providing standardized configurations widely used for design, benchmarking, and simulation studies.

%%%%%%%%%%%%%%%%%%%%%%%%%%%%%%%%%%%%%%%%%%%%%%%%%%%%%%%%%%%%%%%%%%%%%%%%%%%%%%%%%%%%%%%%%%%%%%%%%%%%%%
%%%%%%%%%%%%%%%%%%%%%%%%%%%%%%%%%%%%%%%%%%%%%%%%%%%%%%%%%%%%%%%%%%%%%%%%%%%%%%%%%%%%%%%%%%%%%%%%%%%%%%
\begin{table}[h!]
\footnotesize
\centering
\caption{FOWT reference models.}
\begin{tabularx}{\columnwidth}{@{}L{1.6cm}C{2cm}C{2cm}C{2cm}@{}}
\toprule
\textbf{Characteristic} & \textbf{NREL 5 MW} & \textbf{IEA 15 MW} & \textbf{IEA 22 MW} \\ \midrule
\textbf{Entity} & NREL & IEA & IEA and DTU \\
\addlinespace[0.1cm]
\textbf{Year} & 2009 & 2021 & 2023 \\
\addlinespace[0.1cm]
\textbf{Support Structure(s)} & Barge, Spar, Semi-submersible & Semi-submersible & Semi-submersible \\
\addlinespace[0.1cm]
\textbf{Rotor Diameter (m)} & 126 & 240 & 284 \\
\addlinespace[0.1cm]
\textbf{Hub Height (m)} & 90 & 150 & 170 \\
\addlinespace[0.1cm]
\textbf{Tower Mass (ton)} & 347.46 & 860 & 1574 \\
\addlinespace[0.1cm]
\textbf{Refs.} & \cite{osti_921803, osti_979456, osti_1155123} & \cite{osti_1660012} & \cite{1cd3e417f8854b808c5372670588d3d0} \\
\bottomrule
\end{tabularx}
\label{table:fowt_re}
\end{table}
%%%%%%%%%%%%%%%%%%%%%%%%%%%%%%%%%%%%%%%%%%%%%%%%%%%%%%%%%%%%%%%%%%%%%%%%%%%%%%%%%%%%%%%%%%%%%%%%%%%%%%
%%%%%%%%%%%%%%%%%%%%%%%%%%%%%%%%%%%%%%%%%%%%%%%%%%%%%%%%%%%%%%%%%%%%%%%%%%%%%%%%%%%%%%%%%%%%%%%%%%%%%%

\section{Software Tools for Wind Turbine Simulation and Design} \label{A:software_tools}
\autoref{tab:software_overview} summarizes common software tools for wind turbine simulation and design, including their main functions and key references.
%%%%%%%%%%%%%%%%%%%%%%%%%%%%%%%%%%%%%%%%%%%%%%%%%%%%%%%%%%%%%%%%%%%%%%%%%%%%%%%%%%%%%%%%%%%%%%%%%%%%%%
%%%%%%%%%%%%%%%%%%%%%%%%%%%%%%%%%%%%%%%%%%%%%%%%%%%%%%%%%%%%%%%%%%%%%%%%%%%%%%%%%%%%%%%%%%%%%%%%%%%%%%
\begin{table}[h!]
\centering
\caption{Software tools for wind turbine simulation and design.}
\label{tab:software_overview}
\footnotesize
\begin{tabularx}{\columnwidth}{@{}L{1.1cm}C{0.9cm}C{1.2cm}p{3.5cm}C{0.5cm}@{}}
\toprule
\textbf{Software} & \textbf{Developer} & \textbf{Category} & \textbf{Description} & \textbf{Ref.} \\
\midrule
\multicolumn{5}{l}{\textbf{(1) Simulation Tools}} \\
\midrule
\textbf{OpenFAST} & NREL & Open-source & AHSE simulation tool for onshore and offshore turbines. Modular, validated, and widely adopted. & \cite{Jonkman_2013} \\
\addlinespace[0.2cm]
\textbf{FAST.Farm} & NREL & Open-source & Farm-level simulator extending OpenFAST. Models wakes, turbine interactions, and atmospheric conditions. & \cite{92172f5db61f409db7a831d414c019bd} \\
\addlinespace[0.2cm]
\textbf{HAWC2} & DTU & Proprietary (free for academia) & High-fidelity aeroelastic tool for wind turbine certification and academic research. & \cite{18aac95355e641309b54a6830618c5ca} \\
\addlinespace[0.2cm]
\textbf{Bladed} & DNV & Proprietary & Industry-standard commercial tool with GUI, used for design and certification. & \cite{Bladed} \\
\midrule
\multicolumn{5}{l}{\textbf{(2) Design and Optimization Tools}} \\
\midrule
\textbf{WISDEM} & NREL & Open-source & Design and cost modeling framework for wind turbines and plants. Supports multidisciplinary optimization. & \cite{osti_2345922} \\
\addlinespace[0.2cm]
\textbf{WEIS} & NREL & Open-source & Extension of WISDEM focused on floating wind, control co-design, and reliability analysis. & \cite{10.1115/IOWTC2021-3533} \\
\bottomrule
\end{tabularx}
\end{table}
%%%%%%%%%%%%%%%%%%%%%%%%%%%%%%%%%%%%%%%%%%%%%%%%%%%%%%%%%%%%%%%%%%%%%%%%%%%%%%%%%%%%%%%%%%%%%%%%%%%%%%
%%%%%%%%%%%%%%%%%%%%%%%%%%%%%%%%%%%%%%%%%%%%%%%%%%%%%%%%%%%%%%%%%%%%%%%%%%%%%%%%%%%%%%%%%%%%%%%%%%%%%%

\section{Probability Distributions for Wind–Wave Conditions}
\label{A:fx}
This appendix outlines the probability distributions used in the \textbf{\textit{Wind–Wave Sampler}} described in~\autoref{sec:windwave_samp}, considering mean wind speed ($U$), significant wave height ($H_s$), wave peak period ($T_p$), and wind–wave misalignment angle ($M_{ww}$). The distributions for $U$, $H_s$, and $T_p$ are defined via their cumulative distribution functions (CDFs), which are then differentiated to obtain probability density functions (PDFs), while $M_{ww}$ follows a von Mises distribution. The resulting distributions, based on \citet{Papi_2022}, are shown in \autoref{fig:pdfs}.

\subsection{Mean Wind Speed}
The mean wind speed $U$ is modeled as a random variable following an exponentiated Weibull distribution:
\begin{equation}
    F_U(U) = \left( 1 - \exp\left[ -\left( \frac{U}{\alpha} \right)^{\beta} \right] \right)^{\delta}
\end{equation}

\noindent Differentiating the CDF yields the PDF:
\begin{equation}
f_U(U) = \delta \frac{\beta}{\alpha} \left( \frac{U}{\alpha} \right)^{\beta - 1}
\left( 1 - \exp\left[ -\left( \frac{U}{\alpha} \right)^\beta \right] \right)^{\delta - 1}
\exp\left[ -\left( \frac{U}{\alpha} \right)^\beta \right]
\label{eq:fu}
\end{equation}

\noindent with $\alpha=12.773$, $\beta=2.345$ and $\delta=0.880$. To account for wind turbulence, an IEC category (A, B, or C) is selected, with corresponding reference intensities \( I_{\text{ref}} \in \{0.16, 0.14, 0.12\} \). The standard deviation of the wind speed is then calculated as:
\begin{equation}
    \sigma_w = I_{\text{ref}}\left(0.75 \cdot U + 5.6\right)
    \label{eq:sigmaw}
\end{equation}

\subsection{Significant Wave Height}
The significant wave height $H_s$ is modeled as a random variable conditioned on the mean wind speed $U$, following an exponentiated Weibull distribution:

\begin{equation}
F_{H_s}(H_s \mid U) = \left( 1 - \exp\left[ -\left( \frac{H_s}{\alpha_{H_s}} \right)^{\beta_{H_s}} \right] \right)^5
\label{eq:FHs}
\end{equation}

\begin{figure*}[t!]
    \centering
    \hspace{0.03\textwidth}
    \subfloat[Mean Wind Speed $U$]{%
        \label{fig:pdfs_a}
        \includegraphics[width=0.34\textwidth]{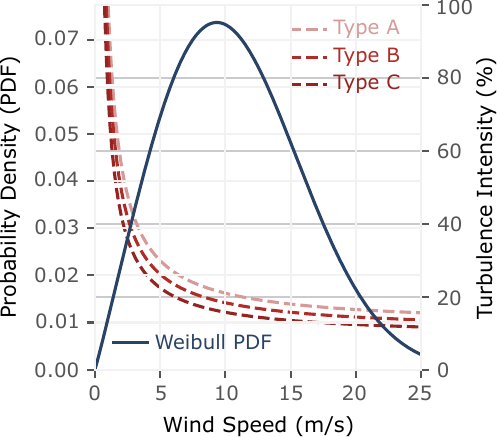}
    }\hspace{0.03\textwidth}
    \subfloat[Significant Wave Height $H_s$]{%
        \label{fig:pdfs_b}
        \includegraphics[width=0.34\textwidth]{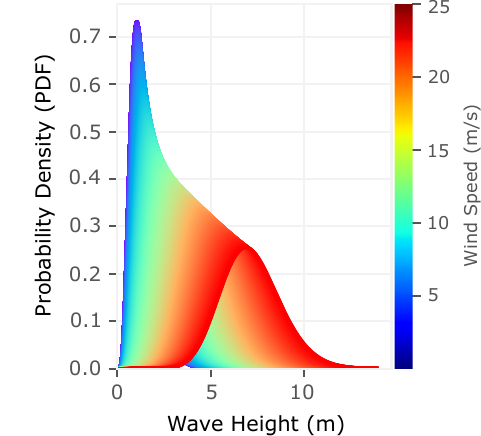}
    }\hspace{0.03\textwidth}\\[3ex]
    \subfloat[Wave Peak Period $T_p$]{%
        \label{fig:pdfs_c}
        \includegraphics[width=0.34\textwidth]{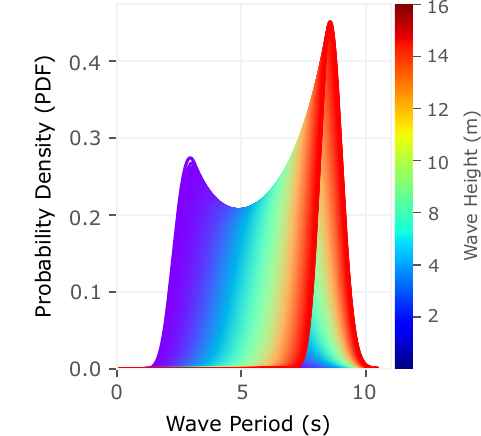}
    }\hspace{0.03\textwidth}
    \subfloat[Mean Wave Direction $M_{ww}$]{%
        \label{fig:pdfs_d}
        \includegraphics[width=0.34\textwidth]{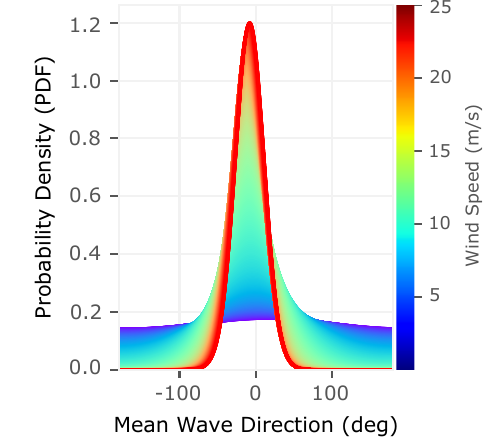}
    }
    \caption{PDFs of the environmental parameters used in the \textbf{FLOAT} \textbf{\textit{Wind–Wave Sampler}}.}
    \label{fig:pdfs}
\end{figure*}

\noindent Differentiating the CDF yields the conditional PDF:
\begin{equation}
\begin{aligned}
f_{H_s}(H_s \mid U) = 5 \cdot \frac{\beta_{H_s}}{\alpha_{H_s}} \left( \frac{H_s}{\alpha_{H_s}} \right)^{\beta_{H_s} - 1}
\cdot & \left( 1 - \exp\left[ -\left( \frac{H_s}{\alpha_{H_s}} \right)^{\beta_{H_s}} \right] \right)^4 \\
\cdot \exp\left[ -\left( \frac{H_s}{\alpha_{H_s}} \right)^{\beta_{H_s}} \right]
\end{aligned}
\end{equation}
\noindent where the parameters $\alpha_{H_s}$ and $\beta_{H_s}$ depend on $U$ as:
\begin{align*}
\alpha_{H_s} &= \frac{1.25 + 0.01 U^{1.98}}{2.0445^{1/\beta_{H_s}}} \\
\beta_{H_s} &= 1.1 + \frac{1.37}{1 + \exp\left[ -0.27(U - 15.86) \right]}
\end{align*}

\subsection{Wave Peak Period}
The wave peak period $T_p$, is modeled as a log-normal distribution conditioned on the significant wave height $H_s$. Its CDF is given by:
\begin{equation}
F_{T_p}(T_p \mid H_s) = \frac{1}{2} \left[ 1 + \text{erf} \left( \frac{\ln T_p - \mu_{T_p}}{\sqrt{2} \sigma_{T_p}} \right) \right]
\label{eq:Ftp}
\end{equation}
\noindent Differentiating the CDF yields the corresponding PDF:
\begin{equation}
f_{T_p}(T_p \mid H_s) = \frac{1}{T_p \sigma_{T_p} \sqrt{2\pi}} \exp\left[ -\frac{(\ln T_p - \mu_{T_p})^2}{2 \sigma_{T_p}^2} \right]
\end{equation}
\noindent where  parameters $\mu_{T_p}$ and $\sigma_{T_p}$ depend on $H_s$ as:
\begin{align*}
\mu_{T_p} &= \ln\left(5.94 + 9.42 \sqrt{\frac{H_s}{g}} \right) \\
\sigma_{T_p} &= 0.24 \exp(-0.11 H_s)
\end{align*}

\subsection{Wave–wind Misalignment Angle}
The mean wave–wind misalignment angle $M_{ww}$ is modeled using a von Mises distribution conditioned on the mean wind speed $U$. The corresponding PDF is given by:
\begin{equation}
f_{M_{ww}}(M_{ww} \mid U) = \frac{\exp\left[k \cos(M_{ww} - \mu_w)\right]}{2\pi I_0(k)}
\end{equation}

\noindent where $I_0(k)$ is the modified Bessel function of the first kind and order zero. The parameters $k$ and $mu_w$ are defined as functions of $U$:
\begin{subequations}
\begin{align}
k &= \frac{10.04}{1 + \exp\left[-0.28(U - 15.89)\right]} \\
\mu_w &= 0.24 - 0.05 U + 0.0014 U^2
\end{align}
\end{subequations}

\section{FLOAT Workflow Details} \label{A:worflows}

This appendix presents the detailed procedures and algorithms of the core modules integrated into \textbf{FLOAT}. Cross-references to the corresponding equations and sections are included to ensure clarity and reproducibility. Each algorithm specifies the primary inputs, procedural steps, and resulting outputs of its respective module. By formalizing these processes, the appendix provides a transparent and modular perspective of the overall methodology.

\subsection{Wind–Wave Sampler} \label{Aa:FLOWTS}
To generate representative environmental conditions for fatigue simulations, it follows the key steps outlined in \autoref{alg:FLOWTS}:

%%%%%%%%%%%%%%%%%%%%%%%%%%%%%%%%%%%%%%%%%%%%%%%%%%%%%%%%%%%%%%%%%%%%%%%%%%%%%%%%%%%%%%%%%%%%%%%%%%%%%%
%%%%%%%%%%%%%%%%%%%%%%%%%%%%%%%%%%%%%%%%%%%%%%%%%%%%%%%%%%%%%%%%%%%%%%%%%%%%%%%%%%%%%%%%%%%%%%%%%%%%%%
\begin{algorithm}[h!]
\caption{\textbf{\textit{Wind–Wave Sampler}} Workflow.}
\label{alg:FLOWTS}
\begin{algorithmic}[1]

\Statex \textbf{Inputs:}
\Statex \quad \textbf{Distributions:}
\Statex \quad \quad $f_U(U)$ \hfill (PDF of mean wind speed, \hyperref[eq:fu]{Eq.~\ref*{eq:fu}})
\Statex \quad \quad $F_{H_s}(H_s \mid U)$ \hfill (CDF of significant wave height,  \hyperref[eq:FHs]{Eq.~\ref*{eq:FHs}})
\Statex \quad \quad $F_{T_p}(T_p \mid H_s)$ \hfill (CDF of wave peak period, \hyperref[eq:Ftp]{Eq.~\ref*{eq:Ftp}})
\Statex \quad \textbf{Parameters:}
\Statex \quad \quad $n_U$ \hfill (number of wind speed samples)
\Statex \quad \quad $n_{H_s}$ \hfill (number of wave height samples)
\Statex \quad \quad $n_{T_p}$ \hfill (number of wave period samples)

\Statex \quad \quad $M_{ww} = 0$ \hfill (fixed mean wave direction)
\Statex \quad \quad $V_{\text{in}}, V_{\text{out}}$ \hfill (cut-in and cut-out wind speeds)
\Statex \quad \quad $\text{Turbulent}$ \hfill (true: turbulent wind; false: steady wind)
\Statex \quad \quad $\text{Class}_{\text{IEC}}$ \hfill (IEC turbulence class, if turbulent)
\Statex \quad \quad $n_{\text{U}_{\text{seeds}}}$ \hfill (number of seeds per wind speed, if turbulent)

\State \textbf{Step 1: Wind Speed Sampling}
\State $id = 0$
\State $\Delta_{\text{bin}} = \frac{V_{\text{out}} - V_{\text{in}}}{n_U}$ \hfill (\hyperref[eq:bin]{Eq.~\ref*{eq:bin}})
\For{$i = 1$ to $n_U$}
    \State $U_i = V_{\text{in}} + \left(i - \frac{1}{2}\right) \cdot \Delta_{\text{bin}}$ \hfill (\hyperref[eq:ui]{Eq.~\ref*{eq:ui}})
    \State $p_{U_i} = \int_{U_i - \Delta_{\text{bin}}/2}^{U_i + \Delta_{\text{bin}}/2} f_U(U)\, dU$ \hfill (\hyperref[eq:pui]{Eq.~\ref*{eq:pui}})

    \State \textbf{Step 2: Wave Height Sampling}
    \For{$j = 1$ to $n_{H_s}$}
        \State $H_{s_{ij}} = F_{H_s}^{-1} \left( \frac{j - 0.5}{n_{H_s}} \,\middle|\, U_i \right)$ \hfill (\hyperref[eq:Hsij]{Eq.~\ref*{eq:Hsij}})

        \State \textbf{Step 3: Wave Period Sampling}
        \For{$k = 1$ to $n_{T_p}$}
            \State $T_{p_{ijk}} = F_{T_p}^{-1} \left( \frac{k - 0.5}{n_{T_p}} \,\middle|\, H_{s_{ij}} \right)$ \hfill (\hyperref[eq:Tpijk]{Eq.~\ref*{eq:Tpijk}})

            \State \textbf{Step 4: Mean Wave Direction Assignment}
            \State $id \gets id + 1$
            \State $Sim_{id} \gets (U_i, H_{s_{ij}}, T_{p_{ijk}}, M_{ww})$
            \State $p_{id} \gets p_{U_i}$
        \EndFor
    \EndFor
\EndFor

\State $N_{\text{sim}} = id$

\State \textbf{Step 5: Turbulence Expansion (if applicable)}
\If{Turbulent}
    \For{$id = 1$ to $N_{\text{sim}}$}
\State $(U_i, H_{s_{ij}}, T_{p_{ijk}}, M_{ww}) \gets \text{Sim}_{id}$
        \For{$s = 1$ to $n_{\text{U}_{\text{seeds}}}$}
                        \State $Sim_{id_{\text{turb}}} \gets (U_i, s, \text{Class}_{\text{IEC}}, H_{s_{ij}}, T_{p_{ijk}}, M_{ww})$

        \EndFor
    \EndFor
\EndIf
\State \textbf{Step 6: Probability Normalization}
\State $Z = \sum_{id = 1}^{N_{\text{sim}}} p_{id}$
\For{$id = 1$ to $N_{\text{sim}}$}
    \State $p_{id} \gets \frac{p_{id}}{Z}$ \hfill (\hyperref[eq:normpid]{Eq.~\ref*{eq:normpid}})
\EndFor
\Statex \textbf{Output:}
\Statex \quad \textbf{Simulation Samples} (for all $id$):
\Statex \quad \quad ${Sim}_{id}$ \hfill (simulation sample for ID $id$) \Statex \quad \quad $p_{id}$ \hfill (probability of $Sim_{id}$) join this two
\end{algorithmic}
\end{algorithm}
%%%%%%%%%%%%%%%%%%%%%%%%%%%%%%%%%%%%%%%%%%%%%%%%%%%%%%%%%%%%%%%%%%%%%%%%%%%%%%%%%%%%%%%%%%%%%%%%%%%%%%
%%%%%%%%%%%%%%%%%%%%%%%%%%%%%%%%%%%%%%%%%%%%%%%%%%%%%%%%%%%%%%%%%%%%%%%%%%%%%%%%%%%%%%%%%%%%%%%%%%%%%%

\begin{enumerate}
\item \textbf{Wind Speed Sampling.} Following IEC 61400-3~\cite{IEC61400-3-1, IEC61400-3-2}, the operational range from cut-in ($V_{\text{in}}$) to cut-out ($V_{\text{out}}$), typically 3–25m/s, is divided into $n_U$ uniform bins of width:
\begin{equation}
\Delta_{\text{bin}} = \frac{V_{\text{out}} - V_{\text{in}}}{n_U}.
\label{eq:bin}
\end{equation}
Mean wind speed is sampled at each bin center:
\begin{equation}
U_i = V_{\text{in}} + \left(i - \frac{1}{2}\right) \cdot \Delta_{\text{bin}}.
\label{eq:ui}
\end{equation}

\noindent
Since environmental states are conditioned on $U$, their probability corresponds to that of the associated wind speed, computed from the PDF $f_U(U)$ (\autoref{eq:fu}):
\begin{equation}
p_{U_i} = \int_{U_i - \Delta_{\text{bin}}/2}^{U_i + \Delta_{\text{bin}}/2} f_U(U)\, dU
\label{eq:pui}
\end{equation}

\item \textbf{Wave Height Sampling.}  For each wind speed $U_i$, $n_{H_s}$ values are sampled by drawing from a uniform distribution in $(0,1)$ and mapping through the inverse conditional CDF $F_{H_s}(H_s \mid U)$ (\autoref{eq:FHs}). For $U_i$, the $j$-th sample is:
\begin{equation}
H_{s_{ij}} = F_{H_s}^{-1} \left( \frac{j - 0.5}{N_{H_s}} \,\middle|\, U_i \right)
\label{eq:Hsij}
\end{equation}

\item \textbf{Wave Period Sampling.}  
For each $H_{s_{ij}}$, $n_{T_p}$ uniform values in $(0,1)$ are mapped through the inverse conditional CDF $F_{T_p}(T_p \mid H_s)$ (\autoref{eq:Ftp}). For $H_{s_{ij}}$, the $k$-th sample is:
\begin{equation}
T_{p_{ijk}} = F_{T_p}^{-1} \left( \frac{k - 0.5}{N_{T_p}} \,\middle|\, H_{s_
{ij}} \right)
\label{eq:Tpijk}
\end{equation}

\item \textbf{Mean Wave Direction Assignment.} As \autoref{fig:pdfs} shows, mean wave direction is concentrated near zero and therefore set as $M_{ww} = 0$.

\item \textbf{Turbulence Expansion.}  
For turbulent wind, turbulence intensity \( \sigma_w \) is computed from the IEC class (A, B, or C) for each \( U_i \) (\autoref{eq:sigmaw}). Stochastic wind fields are generated using \( n_{\text{U}_{\text{seeds}}} \) seeds. Not applied in steady wind.

\item \textbf{Probability Normalization.}  
Each sample is assigned the probability of its corresponding wind speed, \( p_{id} = p_{U_i} \), and normalized across all \( N_{\text{sim}} \) samples to ensure \( \sum p_{id} = 1 \):
\begin{equation}
p_{id} = \frac{p_{id}}{\sum_{id = 1}^{N_{\text{sim}}} p_{id}}
\label{eq:normpid}
\end{equation}

\end{enumerate}

\noindent The total number of simulations becomes:
\begin{equation}
N_{\text{sim}} = 
\begin{cases}
n_U \cdot n_{H_s} \cdot n_{T_p} \cdot n_{\text{U}_{\text{seeds}}}, & \text{turbulent wind} \\
n_U \cdot n_{H_s} \cdot n_{T_p}, & \text{steady wind}
\end{cases}
\end{equation}

\subsection{Numerical Simulator} \label{sec:FastFOWT}
The following steps are performed to execute the fatigue simulations, as detailed in \autoref{alg:FastFOWT}:

\begin{enumerate}
\item \textbf{Generate Turbulent Wind.} For turbulent wind conditions, TurbSim generates stochastic wind fields based on the mean wind speed \( U_i \), seed \( s \), and IEC turbulence class. These fields are then used as inputs to OpenFAST.

\item \textbf{Compute Initial Values.} For each wind speed, a steady-wind OpenFAST run under still water with platform pitch DOF disabled yields time-averaged values, after discarding initial seconds, of blade pitch, rotor speed, platform surge, and platform heave, used as initial conditions to accelerate convergence. It also provides the mean rotor force and moment for pitch calibration.

\item \textbf{Pitch–Platform Calibration.} For semi-submersible FOWTs, the platform pitch is calibrated using the pre-simulation \( F_\text{rotor} \) and \( M_\text{rotor} \). Given a safety factor, the required water height in selected columns (\autoref{eq:hcolwater}) and total platform mass (\autoref{eq:mplatform}) are computed.

\item \textbf{Heave–Platform Calibration (Optional).} For semi-submersible FOWTs, and when the tower geometry is modified, the platform heave is recalibrated to preserve static equilibrium. The adjustment is performed by redistributing platform mass to offset the tower mass variation.

\item \textbf{Run the Simulations.} With all initial values and calibrated platform parameters defined, the simulations are run in OpenFAST on HPC via Inductiva. The simulation time and number of virtual machines must be specified.
\end{enumerate}

%%%%%%%%%%%%%%%%%%%%%%%%%%%%%%%%%%%%%%%%%%%%%%%%%%%%%%%%%%%%%%%%%%%%%%%%%%%%%%%%%%%%%%%%%%%%%%%%%%%%%%
%%%%%%%%%%%%%%%%%%%%%%%%%%%%%%%%%%%%%%%%%%%%%%%%%%%%%%%%%%%%%%%%%%%%%%%%%%%%%%%%%%%%%%%%%%%%%%%%%%%%%%
\begin{algorithm}[t!]
\caption{\textbf{\textit{Numerical Simulator}} Workflow.}
\label{alg:FastFOWT}
\begin{algorithmic}[1]

\Statex \textbf{Inputs:}
\Statex \quad \textbf{Simulation Samples} (for all $id$):
\Statex \quad \quad $Sim_{id}$ \hfill (includes wind speed \( U \), seed \( s \), and turbulence class \( \text{Class}_{\text{IEC}} \))

\Statex \quad \textbf{Parameters:}
\Statex \quad \quad $t_{\text{sim}_0},\, t_{\text{init}_0}$ \hfill (simulation and trim times, steady-wind)
\Statex \quad \quad $t_{\text{sim}}$ \hfill (simulation time, turbulent-wind)
\Statex \quad \quad $\text{SF}$ \hfill (safety factor for platform pitch calibration)
\Statex \quad \quad $N_\text{vm}$ \hfill (number of virtual machines for HPC)

\State \textbf{Step 1: Generate Turbulent Wind}
\State Define wind field parameters: $U$, $s$, and $\text{Class}_{\text{IEC}}$
\State Generate wind field using TurbSim~\cite{turbsim}

\State \textbf{Step 2: Compute Initial Values}
\State Run steady-wind OpenFAST~\cite{Jonkman_2013} simulation for $t_{\text{sim}_0}$ under still water with platform pitch DOF disabled
\State Discard the first $t_{\text{init}_0}$ seconds to remove startup transients
\State Compute time-averaged blade pitch, rotor speed, platform surge, platform heave, $F_\text{rotor}$, and $M_\text{rotor}$

\State \textbf{Step 3: Pitch–Platform Calibration}
\State Using $F_\text{rotor}$, $M_\text{rotor}$, and $\text{SF}$:
\Statex \quad Select the columns used for pitch balancing
\Statex \quad Compute column water height (\hyperref[eq:hcolwater]{Eq.~\ref*{eq:hcolwater}})
\Statex \quad Compute total platform mass (\hyperref[eq:mplatform]{Eq.~\ref*{eq:mplatform}})

\State \textbf{Step 4: Heave–Platform Calibration (Optional)}
\Statex \quad If tower mass varies, adjust platform mass  
\Statex \quad Enforce equilibrium: $\Delta m_\text{tower} + \Delta m_\text{platform} = 0$

\State \textbf{Step 5: Run the Simulations}
\State Use wind fields from Step 1
\State Use initial values from Step 2
\State Use platform calibration parameters from Step 3
\State Run OpenFAST~\cite{Jonkman_2013} for each $Sim_{id}$ with duration $t_{\text{sim}}$
\State Execute on $N_\text{vm}$ virtual machines via Inductiva~\cite{InductivaOpenFAST}

\Statex \textbf{Output:}
\Statex \quad \textbf{Simulation Outputs} (for all $id$):
\Statex \quad \quad $\text{Response}_{id}$ \hfill (time series outputs for simulation ID $id$)
\end{algorithmic}
\end{algorithm}

%%%%%%%%%%%%%%%%%%%%%%%%%%%%%%%%%%%%%%%%%%%%%%%%%%%%%%%%%%%%%%%%%%%%%%%%%%%%%%%%%%%%%%%%%%%%%%%%%%%%%%
%%%%%%%%%%%%%%%%%%%%%%%%%%%%%%%%%%%%%%%%%%%%%%%%%%%%%%%%%%%%%%%%%%%%%%%%%%%%%%%%%%%%%%%%%%%%%%%%%%%%%%

\subsection{Frequency Response Analyser} \label{sec:FRA}

The following steps, summarized in \autoref{alg:FRATower}, define the frequency-domain evaluation of the tower-base fore–aft moment via PSD:

%%%%%%%%%%%%%%%%%%%%%%%%%%%%%%%%%%%%%%%%%%%%%%%%%%%%%%%%%%%%%%%%%%%%%%%%%%%%%%%%%%%%%%%%%%%%%%%%%%%%%%
%%%%%%%%%%%%%%%%%%%%%%%%%%%%%%%%%%%%%%%%%%%%%%%%%%%%%%%%%%%%%%%%%%%%%%%%%%%%%%%%%%%%%%%%%%%%%%%%%%%%%%
\begin{algorithm}[t!]
\caption{\textbf{\textit{Frequency Response Analyser}} Workflow.}
\label{alg:FRATower}
\begin{algorithmic}[1]

\Statex \quad ($N_{\text{sim}}$ simulations indexed by $id = 1, \dots, N_{\text{sim}}$)
\Statex \quad  ($N_{U_b}$ wind speed bins indexed by $b = 1\dots, N_{U_b}$)
\Statex \quad ($N_{U_j}$ mean wind speed groups indexed by $j = 1\dots, N_{U_j}$)

\Statex \textbf{Inputs:}
\Statex \quad \textbf{Simulation Outputs} (for all $id$):
\Statex \quad \quad $\text{Response}_{id}$ \hfill (time series outputs for simulation $id$)
\Statex \quad \quad $U_{id}$ \hfill (mean wind speed of simulation $id$)

\Statex \quad \textbf{Parameters}:
\Statex \quad \quad $t_{\text{sim}_0},\, t_{\text{init}_0}$ \hfill (simulation and trim times, steady-wind)

\Statex \quad \quad $t_{\text{sim}},\, t_{\text{init}}$ \hfill (simulation and trim times, turbulent-wind)

\Statex \quad \quad $f_s$  \hfill (sampling frequency)
\Statex \quad \quad $L$  \hfill (Welch segment length)
\Statex \quad \quad $(f_{\min}, f_{\max}), \; (U_{\min}, U_{\max})$ \hfill (plot ranges)

\State \textbf{Step 1: Simulation PSD Estimation}
\For{$id = 1$ to $N_{\text{sim}}$}

    \State \textbf{Step 1.1: Remove Startup Transients}:     
    \State $\text{Response}_{id} \gets \text{Response}_{id}[t_{\text{init}}, t_{\text{sim}}]$

    \State \textbf{Step 1.2: Compute PSD}:  
    \State Compute the one-sided PSD using Welch’s method~\cite{1161901} using $f_s$ and $L$

\EndFor

\State \textbf{Step 2: Harmonic Frequency Identification}

\For{$b = 1$ to $N_{U_b}$}
\State \textbf{Step 2.1: Compute Mean Rotor Speed}
\State Run steady-wind OpenFAST~\cite{Jonkman_2013} simulation for $t_{\text{sim}_0}$ under still water with pitch/heave–platform calibration
\State Discard first $t_{\text{init}_0}$ seconds to remove transients
\State Compute time-averaged rotor speed, $\overline{\mathrm{RPM}}(U_b)$

\State \textbf{Step 2.2: Compute Harmonic Frequencies}
    \State $f_{1\mathrm{P}}(U_b) = \overline{\mathrm{RPM}}(U_b)/60$
    \State $f_{n\mathrm{P}}(U_b) = n\, f_{1\mathrm{P}}(U_b)$ \; for $n \in \{3,6,9\}$ \hfill  (\hyperref[eq:f_rpm]{Eq.~\ref*{eq:f_rpm}})
\EndFor

\State \textbf{Step 3: PSD Heatmap Assembly}
\State \textbf{Step 3.1: PSD Aggregation}
\For{$j = 1$ to $N_{U_j}$}
\State $\mathbf{S}(f,U_j) = \dfrac{1}{|\mathcal{J}(U_j)|}\sum_{i \in \mathcal{J}(U_j)} \widehat{S}_{xx}^{(i)}(f)$ \hfill  (\hyperref[eq:sf_psd]{Eq.~\ref*{eq:sf_psd}})
\EndFor

\State \textbf{Step 3.2: Heatmap Visualization}  
\State $\text{Heatmap}(f,{U}_j) = \log_{10}\mathbf{S}(f, {U}_j)$  
\State Plotted over $(f_{\min},f_{\max}) \times (U_{\min},U_{\max})$ with harmonic overlays.

\Statex \textbf{Output:}
\Statex \quad \textbf{PSD Heatmap} \hfill ($\log_{10}\mathbf{S}(f,{U}_j)$ with $n$P overlays)

\end{algorithmic}
\end{algorithm}
%%%%%%%%%%%%%%%%%%%%%%%%%%%%%%%%%%%%%%%%%%%%%%%%%%%%%%%%%%%%%%%%%%%%%%%%%%%%%%%%%%%%%%%%%%%%%%%%%%%%%%
%%%%%%%%%%%%%%%%%%%%%%%%%%%%%%%%%%%%%%%%%%%%%%%%%%%%%%%%%%%%%%%%%%%%%%%%%%%%%%%%%%%%%%%%%%%%%%%%%%%%%%

\begin{enumerate}
\item \textbf{Simulation PSD Estimation.}  
For each simulation, the tower-base fore–aft bending moment time series is processed as follows:

\begin{enumerate}[label*=\arabic*.] 
    \item \textbf{Remove Startup Transients.}  
    Discard the initial portion of the time series to exclude transient effects.

    \item \textbf{Compute PSD.}  
    Estimate the one-sided PSD using Welch’s method~\cite{1161901}, with prescribed sampling frequency, segment length, Hann window, and 50\% overlap.
\end{enumerate}

\item \textbf{Harmonic Frequency Identification.} Identify the harmonic components associated with rotor-induced periodicity:
\begin{enumerate}[label*=\arabic*.] 
    \item \textbf{Compute Mean Rotor Speed.}  
    For each wind speed bin, perform a steady-wind OpenFAST simulation under still-water conditions with pitch/heave–platform calibration. After trimming transients, compute the time-averaged rotor speed.
 
    \item \textbf{Compute Harmonic Frequencies.}  
    Obtain the 1P frequency and higher-order harmonics from the mean rotor speed (\autoref{eq:f_rpm}).
\end{enumerate}

\item \textbf{PSD Heatmap Assembly.}  Assemble a unified spectral representation across wind conditions:
\begin{enumerate}[label*=\arabic*.]
    \item \textbf{PSD Aggregation.}  
    Group simulations by their mean wind speed, average the corresponding spectra (\autoref{eq:sf_psd}), and assemble them into the PSD matrix.

    \item \textbf{Heatmap Visualization.}  
    Plot $\log_{10}$ of the PSD over frequencies in $[f_{\min}, f_{\max}]$ and wind speeds in $[U_{\min}, U_{\max}]$, showing frequency (rows), wind speed (columns), and overlay the harmonic curves (1P, 3P, 6P, 9P).
\end{enumerate}
\end{enumerate}

\subsection{Fatigue Analyser} \label{sec:FAITower}
The following steps, summarized in \autoref{alg:FAITower}, describe the fatigue analysis procedure:

\begin{enumerate}

\item \textbf{Simulation Fatigue Evaluation.}  
For each simulation, and for each tower section \( i \), damage is computed as follows:

\begin{enumerate}[label*=\arabic*.] 
    \item \textbf{Remove Startup Transients.} Trim the initial portion of the moment time series to exclude transient effects.

    \item \textbf{Interpolate Moments.}  
    Since the fore-aft bending moments are recorded only at discrete tower heights, the moment time series at each section midpoint is interpolated from the closest available data.

\item \textbf{Compute Rainflow Cycles.}  Apply rainflow counting to the interpolated moment series \( M_{\text{FA}} \), yielding cycles indexed by \( j \), each with count \( n_{ij} \) and moment range \( \Delta M_{\text{FA}_{ij}} \). These are converted to stress ranges via elastic bending theory for circular sections:

\begin{equation}
\Delta \sigma_{ij} = \frac{\Delta M_{\text{FA}_{ij}} \cdot r_i}{I_i}
\label{eq:stress_from_moment}
\end{equation}
\noindent where \( r_i \) is the outer radius and \( I_i = \frac{\pi}{4} \left(r_i^4 - (r_i - t_i)^4\right) \) is the second moment of area at section \( i \), with \( t_i \) denoting the wall thickness.

\item \textbf{Compute Cycles to Failure.}  
Compute the number of cycles to failure \( N_{ij} \) for each stress range \( \Delta \sigma_{ij} \) using the S-N curve in~\autoref{eq:ni_explicit}.

\item \textbf{Compute Sample Damage.}  
Compute damage using \( n_{ij} \) and \( N_{ij} \) via the Palmgren–Miner rule in~\autoref{eq:damage}.
\end{enumerate}

\item \textbf{Lifetime Fatigue Estimation.}  
For each simulation \( id \), the damage is weighted by its expected number of occurrences \( n_{id} \) (see~\autoref{eq:nj_t}) over the design lifetime. The total accumulated damage is then computed using~\autoref{eq:damage_t}.
\end{enumerate}

%%%%%%%%%%%%%%%%%%%%%%%%%%%%%%%%%%%%%%%%%%%%%%%%%%%%%%%%%%%%%%%%%%%%%%%%%%%%%%%%%%%%%%%%%%%%%%%%%%%%%%
%%%%%%%%%%%%%%%%%%%%%%%%%%%%%%%%%%%%%%%%%%%%%%%%%%%%%%%%%%%%%%%%%%%%%%%%%%%%%%%%%%%%%%%%%%%%%%%%%%%%%%
\begin{algorithm}[h!]
\caption{\textbf{\textit{Fatigue Analyzer}} Workflow.}
\label{alg:FAITower}
\begin{algorithmic}[1]
\Statex \quad ($n$ selected tower cross sections indexed by $i = 1, \dots, n$)
\Statex \quad ($N_{\text{sim}}$ simulations indexed by $id = 1, \dots, N_{\text{sim}}$)

\Statex \textbf{Inputs:}
\Statex \quad \textbf{Tower} (for all $i$):
\Statex \quad \quad $r_i$ \hfill (outer radius at section $i$) \Statex \quad \quad $t_i$ \hfill (wall thickness at section $i$)
\Statex \quad \textbf{S-N Curve Parameters}:
\Statex \quad \quad $m$ \hfill (slope) \Statex \quad \quad $k$ \hfill (thickness exponent) \Statex \quad \quad $t_{\text{ref}}$ \hfill (reference thickness)
\Statex \quad \textbf{Simulation Outputs} (for all $id$):
\Statex \quad \quad $\text{Response}_{id}$ \hfill (time series outputs for simulation $id$)
\Statex \quad \quad $p_{id}$ \hfill (probability of simulation $id$)
\Statex \quad \textbf{Parameters}:
\Statex \quad \quad$t_{\text{sim}}$ \hfill (simulation time)
\Statex \quad \quad$t_{\text{init}}$ \hfill (startup trim time)
\Statex \quad \quad $LT$ \hfill (wind turbine lifetime)

\State \textbf{Step 1: Simulation Fatigue Evaluation}
\For{$id = 1$ to $N_{\text{sim}}$}
    \State \textbf{Step 1.1: Remove Startup Transients}
    \State $\text{Response}_{id} \gets \text{Response}_{id}[t_{\text{init}}, t_{\text{sim}}]$
\For{$i = 1$ to $n$}
    \State \textbf{Step 1.2: Interpolate Moments}
    
        \State Interpolate $M_{\text{FA}}(z_i, t)$ using \text{Response}$_{id}$

\State \textbf{Step 1.3: Compute Rainflow Cycles}
        \State Apply rainflow method to $M_{\text{FA}}(z_i, t) \gets         \Delta M_{ij}$, $n_{ij}$
        \State $\Delta \sigma_{ij} = \dfrac{4 \cdot \Delta M_{ij} \cdot r_i}{\pi \left( r_i^4 - (r_i - t_i)^4 \right)}$ \hfill (\hyperref[eq:stress_from_moment]{Eq.~\ref*{eq:stress_from_moment}})
 
    \State \textbf{Step 1.4: Compute Cycles to Failure}
        \State $N_{ij}(\Delta\sigma_{ij}) = \bar{a} \left[ \Delta \sigma_{ij} \left( \dfrac{t_i}{t_{\text{ref}}} \right)^k \right]^{-m}
   $ \hfill (\hyperref[eq:ni_explicit]{Eq.~\ref*{eq:ni_explicit}})

    \State \textbf{Step 1.5: Compute Sample Damage}
        \State $D_{id,i} = \sum_j \dfrac{n_{ij}(\Delta\sigma_{ij})}{N_{ij}(\Delta\sigma_{ij})}$\hfill (\hyperref[eq:damage]{Eq.~\ref*{eq:damage}})
    \EndFor
\EndFor

\State \textbf{Step 2: Lifetime Fatigue Estimation}
\For{$i = 1$ to $n$}
    \State   $t_{id} = t_{\text{sim}} - t_{\text{init}}$
    \State  $n_{id} = \dfrac{LT}{t_{id}}\cdot p_{id}$ \hfill (\hyperref[eq:nj_t]{Eq.~\ref*{eq:nj_t}})
    \State $D^{\text{t}}_i = \sum_{id=1}^{N_{\text{sim}}} D_{id, i} \cdot n_{id}$\hfill (\hyperref[eq:damage_t]{Eq.~\ref*{eq:damage_t}})
\EndFor

\Statex \textbf{Output:}
\Statex \quad \textbf{Fatigue Damage Tower} (for all $i$):
\Statex \quad  \quad $D^{\text{t}}_i$ \hfill (expected fatigue damage at each tower section)

\end{algorithmic}
\end{algorithm}
%%%%%%%%%%%%%%%%%%%%%%%%%%%%%%%%%%%%%%%%%%%%%%%%%%%%%%%%%%%%%%%%%%%%%%%%%%%%%%%%%%%%%%%%%%%%%%%%%%%%%%
%%%%%%%%%%%%%%%%%%%%%%%%%%%%%%%%%%%%%%%%%%%%%%%%%%%%%%%%%%%%%%%%%%%%%%%%%%%%%%%%%%%%%%%%%%%%%%%%%%%%%%

\subsection{Design Optimizer} \label{sec:OPTITower}

The following steps, summarized in \autoref{alg:OPTITower}, describe the design optimization procedure:

%%%%%%%%%%%%%%%%%%%%%%%%%%%%%%%%%%%%%%%%%%%%%%%%%%%%%%%%%%%%%%%%%%%%%%%%%%%%%%%%%%%%%%%%%%%%%%%%%%%%%%
%%%%%%%%%%%%%%%%%%%%%%%%%%%%%%%%%%%%%%%%%%%%%%%%%%%%%%%%%%%%%%%%%%%%%%%%%%%%%%%%%%%%%%%%%%%%%%%%%%%%%%
\begin{algorithm}[h!]
\caption{\textbf{\textit{Design Optimizer}} Workflow.}
\label{alg:OPTITower}
\begin{algorithmic}[1]
\Statex \quad (Optimization uses a land-based tower model; the floating frequency limit is mapped using \hyperref[eq:float_to_land]{Eq.~\ref*{eq:float_to_land}}.)
\Statex \quad ($n$ selected tower cross sections indexed by $i = 1, \dots, n$)

\Statex \textbf{Inputs:}

\Statex \quad \textbf{Reference Tower} (for all $i$):
\Statex \quad \quad $d^{\text{ref}}_i$ \hfill (outer diameter at section $i$)
\Statex \quad \quad $t^{\text{ref}}_i$ \hfill (thickness at section $i$)
\Statex \quad \quad $D^{\text{ref}}_i$ \hfill (fatigue damage at section $i$)

\Statex \quad \textbf{Safety Factors:}
\Statex \quad \quad $\gamma_f$, $\gamma_m$, $\gamma_n$, $\gamma_d$ 

\Statex \quad \textbf{Design Variables:}
\Statex \quad \quad $d_i \in [d_{\min}, d_{\max}]$ \hfill (outer diameter, \text{for } i = 0, \dots, n)
\Statex \quad \quad $t_j \in [t_{\min}, t_{\max}]$ \hfill (wall thickness, \text{for } j = 1, \dots, n)

\Statex \quad \textbf{Design Constraints:}
\Statex \quad \quad $\frac{\gamma_f \cdot \gamma_m \cdot \gamma_n \cdot \sigma_{\text{vM}}(\boldsymbol{x})}{\sigma_y} \leq 1.0
$ \hfill (stress, \hyperref[eq:stress1]{Eq.~\ref*{eq:stress1}})
\Statex \quad \quad $\text{LSF}, \text{GF} \leq 1.0$ \hfill (buckling, \hyperref[eq:buck1]{Eq.~\ref*{eq:buck1}})
\Statex \quad \quad $f_1^{\min} \leq f_1 \leq f_1^{\max}$ \hfill (frequency, \hyperref[eq:freq1]{Eq.~\ref*{eq:freq1}})
\Statex \quad \quad $D_i \cdot \gamma_d \leq 1.0$ \hfill (fatigue, \hyperref[eq:damage1]{Eq.~\ref*{eq:damage1}})
\Statex \quad \quad $d_{i+1} \leq d_i, \quad t_{i+1} \leq t_i$ \hfill (monotonicity, \hyperref[eq:mono1]{Eq.~\ref*{eq:mono1}})
\Statex \quad \quad $\left( \frac{d}{t} \right)_{\text{min}} \leq \frac{d_i}{t_i} \leq \left( \frac{d}{t} \right)_{\text{max}}$ \hfill (diameter-to-thickness, \hyperref[eq:d_to_p1]{Eq.~\ref*{eq:d_to_p1}})
\Statex \quad \quad $\text{taper}_{\min} \leq \frac{d_{i+1}}{d_i} \leq \text{taper}_{\max}$ \hfill (taper, \hyperref[eq:taper1]{Eq.~\ref*{eq:taper1}})

\Statex \quad \textbf{Optimization Settings:}
\Statex \quad \quad \text{SLSQP} \hfill (optimization method)
\Statex \quad \quad \text{Central finite differences} \hfill (gradient estimation)
\Statex \quad \quad $\delta$ \hfill (finite difference step size)
\Statex \quad \quad $N_{\text{max}}$ \hfill (maximum number of iterations)
\Statex \quad \quad $\epsilon_{\text{opt}}$ \hfill (optimality tolerance)

\Statex \quad \textbf{Loading:}
\Statex \quad \quad $\mathbf{F}_\text{top}$, $\mathbf{M}_\text{top}$ \hfill (applied force and moment at tower top)
\Statex \quad \quad $m_\text{RNA}$, $\mathbf{r}_\text{CoM}$, $\mathbf{I}_\text{RNA}$ \hfill (RNA: mass, CoM, inertia)

\State  \textbf{Step 1: Initialize Design Vector}
\State \quad $\boldsymbol{x}^{(0)} = \{r_i^{\text{ref}}, t_i^{\text{ref}}\}$ \hfill (initial design)

\For{$k = 0$ to $N_{\text{max}}$}
    \State \textbf{Step 2: Compute Tower Response}
\Statex \quad \quad Compute response using $\mathbf{F}_\text{top}$, $\mathbf{M}_\text{top}$, $m_\text{RNA}$, $\mathbf{r}_\text{CoM}$, $\mathbf{I}_\text{RNA}$
    \Statex \quad \quad Compute stress $\sigma_i$, LSF, GF, $f_1$
    
    \State \textbf{Step 3: Estimate Fatigue Damage}
    \Statex \quad \quad Estimate $D_i$ via \textit{Fatigue Estimator} \hfill (\autoref{alg:FASTower})

    \State \textbf{Step 4: Check Constraints}
    \Statex \quad \quad Check all defined constraints.

    \State \textbf{Step 5: Compute Objective}
    \Statex \quad \quad Compute tower mass $M(\boldsymbol{x}^{(k)})$ \hfill (\hyperref[eq:mass]{Eq.~\ref*{eq:mass}})
    
    \State \textbf{Step 6: Update Design}
\Statex \quad \quad $\boldsymbol{x}^{(k+1)} \leftarrow \text{SLSQP}(\boldsymbol{x}^{(k)})$ using finite differences ($\delta$)

    \State \textbf{Step 7: Check Convergence}
\If{$\|\nabla \mathcal{L}(\boldsymbol{x}^{(k+1)})\| < \epsilon_{\text{opt}}$}
        \State \textbf{break}
    \EndIf
\EndFor

\Statex \textbf{Output:}
\Statex \quad \textbf{Optimization Outputs} (for all $i$ and $j$):
\Statex \quad \quad $d_i^*$, $t_j^*$ \hfill (optimized diameters and thicknesses)
\end{algorithmic}
\end{algorithm}
%%%%%%%%%%%%%%%%%%%%%%%%%%%%%%%%%%%%%%%%%%%%%%%%%%%%%%%%%%%%%%%%%%%%%%%%%%%%%%%%%%%%%%%%%%%%%%%%%%%%%%
%%%%%%%%%%%%%%%%%%%%%%%%%%%%%%%%%%%%%%%%%%%%%%%%%%%%%%%%%%%%%%%%%%%%%%%%%%%%%%%%%%%%%%%%%%%%%%%%%%%%%%

\begin{enumerate}

\item \textbf{Initialize Design Vector.}
Set the initial values for the outer diameters and wall thicknesses at each tower section, based on the reference tower geometry.

\item \textbf{Compute Tower Response.}
Evaluate stress, buckling safety factors, and the first natural frequency using the structural model for the selected DLC.

\item \textbf{Estimate Fatigue Damage.}
Use the \textbf{\textit{Fatigue Estimator}} to compute the fatigue damage at each tower section for the current design, using the reference tower geometry and its corresponding damage as a reference.

\item \textbf{Check Constraints.}
Verify whether the current design satisfies all imposed constraints.

\item \textbf{Compute Objective.}
Compute the objective function, defined as the total tower mass.

\item \textbf{Update Design.}
Apply the SLSQP algorithm with central finite differencing and a fixed step size to update the design vector within the prescribed bounds.

\item \textbf{Check Convergence.}
If the convergence criterion, defined by the optimality tolerance, is met, the optimization terminates. Otherwise, the process returns to Step 2 and continues until convergence is achieved or the maximum number of iterations is reached.
\end{enumerate}

\subsection{Fatigue Estimator} \label{sec:FASTower}

The following steps, summarized in \autoref{alg:FASTower}, describe the fatigue estimation procedure:

\begin{enumerate}
\item \textbf{{Calibration Phase (Calibration Tower).}} For each selected cross section $i$, the calibration tower geometry $(r^{\text{cal}}_i$ and $t^{\text{cal}}_i)$, the corresponding fatigue damage $(D^{\text{cal}}_i)$, and the S-N curve parameters are used to compute the constant $C_i$:
    \begin{equation}
    C_i = D^{\text{cal}}_i \cdot \left(r^{\text{cal}}_i\right)^{2m} \cdot \left(t^{\text{cal}}_i\right)^{m} \cdot \left( \frac{t_{\text{ref}}}{t^{\text{cal}}_i} \right)^{k \cdot m}
    \label{eq:C_ref}
\end{equation}

\item \textbf{{Prediction Phase (New Tower).}} With each $C_i$ calibrated, the model estimates the fatigue damage $D^{\text{new}}_i$ for the new tower design at the same cross section $i$, using its geometry $(r^{\text{new}}_i$ and $t^{\text{new}}_i)$ and the same S-N curve parameters:
    \begin{equation}
    D^{\text{new}}_i = C_i \cdot \left(r^{\text{new}}_i\right)^{-2m} \cdot \left(t^{\text{new}}_i\right)^{-m} \cdot \left( \frac{t^{\text{new}}_i}{t_{\text{ref}}} \right)^{k \cdot m}
    \label{eq:D_new}
    \end{equation}
\end{enumerate}

%%%%%%%%%%%%%%%%%%%%%%%%%%%%%%%%%%%%%%%%%%%%%%%%%%%%%%%%%%%%%%%%%%%%%%%%%%%%%%%%%%%%%%%%%%%%%%%%%%%%%%
%%%%%%%%%%%%%%%%%%%%%%%%%%%%%%%%%%%%%%%%%%%%%%%%%%%%%%%%%%%%%%%%%%%%%%%%%%%%%%%%%%%%%%%%%%%%%%%%%%%%%%
\begin{algorithm}[h!]
\caption{\textbf{\textit{Fatigue Estimator}} Workflow.}
\label{alg:FASTower}
\begin{algorithmic}[1]
\Statex \quad ($n$ selected tower cross sections indexed by $i = 1, \dots, n$)
\Statex \textbf{Inputs:}
\Statex \quad \textbf{Calibration Tower} (for all $i$):
\Statex \quad \quad $r^{\text{cal}}_i$ \hfill (outer radius at section $i$)
\Statex \quad \quad $t^{\text{cal}}_i$ \hfill (wall thickness at section $i$)
\Statex \quad \quad $D^{\text{cal}}_i$ \hfill (fatigue damage at section $i$)
\Statex \quad \textbf{S-N Curve Parameters}:
\Statex \quad \quad $m$ \hfill (slope) \Statex \quad \quad $k$ \hfill (thickness exponent) \Statex \quad \quad $t_{\text{ref}}$ \hfill (reference thickness)
\Statex \quad \textbf{New Tower} (for all $i$):
\Statex \quad \quad $r^{\text{new}}_i$ \hfill (outer radius at section $i$)
\Statex \quad \quad $t^{\text{new}}_i$ \hfill (wall thickness at section $i$)

\State \textbf{Step 1: Calibration Phase}
\For{$i = 1$ to $n$}
    \State $C_i = D^{\text{cal}}_i \cdot \left(r^{\text{cal}}_i\right)^{2m} \cdot \left(t^{\text{cal}}_i\right)^m \cdot \left( \dfrac{t_{\text{ref}}}{t^{\text{cal}}_i} \right)^{k \cdot m}$ \hfill (\hyperref[eq:C_ref]{Eq.~\ref*{eq:C_ref}})
\EndFor

\State \textbf{Step 2: Prediction Phase}
\For{$i = 1$ to $n$}
    \State $D^{\text{new}}_i = C_i \cdot \left(r^{\text{new}}_i\right)^{-2m} \cdot \left(t^{\text{new}}_i\right)^{-m} \cdot \left( \dfrac{t^{\text{new}}_i}{t_{\text{ref}}} \right)^{k \cdot m}$ \hfill (\hyperref[eq:D_new]{Eq.~\ref*{eq:D_new}})
\EndFor

\Statex \textbf{Output:}
\Statex \quad \textbf{Fatigue Damage Tower} (for all $i$):
\Statex \quad \quad $D^{\text{new}}_i$ \hfill (fatigue damage at section $i$)
\end{algorithmic}
\end{algorithm}
%%%%%%%%%%%%%%%%%%%%%%%%%%%%%%%%%%%%%%%%%%%%%%%%%%%%%%%%%%%%%%%%%%%%%%%%%%%%%%%%%%%%%%%%%%%%%%%%%%%%%%
%%%%%%%%%%%%%%%%%%%%%%%%%%%%%%%%%%%%%%%%%%%%%%%%%%%%%%%%%%%%%%%%%%%%%%%%%%%%%%%%%%%%%%%%%%%%%%%%%%%%%%

\section{Equivalence of Monte Carlo Weights and Lifetime Fatigue Probabilities} \label{A:monte_carlo_val}

Each environmental state $j$ is simulated for a duration $t_j$, yielding total damage $D_j$. The damage rate can thus be written as:
\begin{equation}
    D(U_j, H_{s,j}, T_{p,j}, M_{ww,j}) = \frac{D_j}{t_j}.
\end{equation}

Substituting into the Monte Carlo approximation (\autoref{eq:monte}) gives:
\begin{equation}
    D_t \approx \sum_{j=1}^{N} D_j \cdot n_j,
\end{equation}
with the effective number of repetitions of each state defined as:
\begin{equation}
    n_j = \frac{LT}{t_j} \cdot w_j.
\end{equation}

Comparing this result with the expected count formulation in \autoref{eq:nj_t} 
shows that the Monte Carlo weights $w_j$ are equivalent to the probabilities $p_j$ 
of occurrence for each environmental state, ensuring consistency between the 
probabilistic model and the sampled simulations.

\section{Implementation Details of the Pitch–Platform Calibration} \label{A:pitch_calib}

\begin{figure}[h!]
    \centering
    \includegraphics[width=\columnwidth]{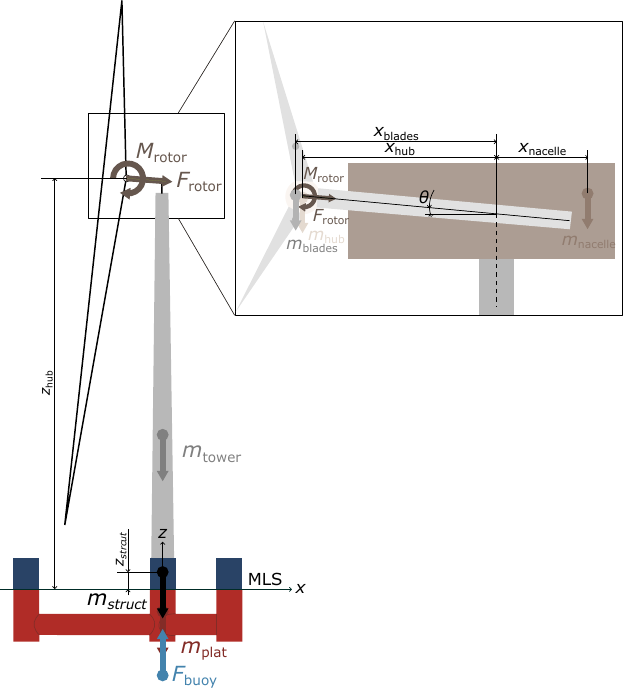}
    \caption{Applied loads and geometric references used to compute the structural moments for pitch–platform calibration.}
    \label{fig:moment_balance}
\end{figure}

The total structural moment results from gravitational and aerodynamic contributions, which are defined below and illustrated in \autoref{fig:moment_balance}.

The gravitational moment is:
\begin{equation}
M_\text{weight} = g \cdot (m_\text{nacelle} \cdot x_\text{nacelle} + m_\text{hub} \cdot x_\text{hub} + m_\text{blades} \cdot x_\text{blades})
\end{equation}
where \( g \) is gravity, \( m \) the mass of each component, and \( x \) its horizontal position. The aerodynamic moment is:
\begin{equation}
M_\text{aero} = F_\text{rotor}  \cdot \cos(\theta) \cdot (z_\text{hub} - z_\text{struct}) + F_\text{rotor}  \cdot \sin(\theta) \cdot x_\text{hub} + M_\text{rotor} \end{equation}

\noindent
where \( \theta \) is the shaft tilt angle, and \( F_\text{rotor} \) and \( M_\text{rotor} \) are the mean rotor thrust and moment, computed with fixed platform pitch under steady wind and still water conditions.

To implement pitch–platform calibration in the simulation, the required ballast mass is computed from the restoring moment:
\begin{equation}
m_\text{colwater} = \frac{M_\text{colwater}}{L \cdot g}
\end{equation}
where \( L \) is the horizontal distance from the column to the platform center, as illustrated in \autoref{fig:x} (\( L \) for upwind; \( L \cdot \cos(60^\circ) \cdot 2 = L \) for port/starboard). The corresponding water height is:

\begin{equation}
h_\text{colwater} = \frac{m_\text{colwater}}{\rho \cdot \pi \cdot \frac{d_\text{colwater}^2}{4}} \cdot SF
\label{eq:hcolwater}
\end{equation}
where \( \rho \) is the water density, \( d_\text{colwater} \) is the column diameter, and \( SF \) is a safety factor. To preserve the total platform mass, the equivalent structural mass is reduced to compensate for the added water:
\begin{equation}
m_\text{platform} = m_\text{platform,init} - n_\text{colwater} \cdot m_\text{colwater} \cdot SF
\label{eq:mplatform}
\end{equation}
where \( n_\text{colwater} \) is 1 (upwind) or 2 (port/starboard).

\section{Geometric Parameters of the IEA~22~MW Semi-Submersible FOWT Reference Tower}
\label{A:table_ref}

\autoref{tab:geom_tower} summarizes the geometric parameters of the IEA~22~MW semi-submersible FOWT reference tower, including the 31 outer diameters ($d_i$), 30 section heights ($h_i$), and 30 wall thicknesses ($t_i$), defined from bottom to top across the 30 tower sections.

%%%%%%%%%%%%%%%%%%%%%%%%%%%%%%%%%%%%%%%%%%%%%%%%%%%%%%%%%%%%%%%%%%%%%%%%%%%%%%%%%%%%%%%%%%%%%%%%%%%%%%
%%%%%%%%%%%%%%%%%%%%%%%%%%%%%%%%%%%%%%%%%%%%%%%%%%%%%%%%%%%%%%%%%%%%%%%%%%%%%%%%%%%%%%%%%%%%%%%%%%%%%%
\begin{table}[h!]
\footnotesize
\centering
\caption{Geometric parameters of the IEA~22~MW semi-submersible FOWT reference tower~\cite{1cd3e417f8854b808c5372670588d3d0}: 31 diameters $d_i$, 30 heights $h_i$, and 30 thicknesses $t_i$ across the 30 sections (from bottom to top).}
\begin{tabular}{@{}p{0.5cm}p{1cm}p{1cm}p{1cm}@{}}
\toprule
\textbf{i} & \textbf{d$_i$ (m)} & \textbf{h$_i$  (m)} & \textbf{t$_i$   (mm)} \\ \midrule
0  & 10.000 & - & -  \\
1  & 10.000 & 3.1885 & 66.329 \\
2  & 10.000 & 5.0410 & 64.618 \\
3  & 9.912  & 5.0420 & 62.569 \\
4  & 9.799  & 5.0410 & 61.597 \\
5  & 9.683  & 4.0400 & 60.882 \\
6  & 9.565  & 5.0410 & 60.151 \\
7  & 9.444  & 5.0420 & 59.404 \\
8  & 9.320  & 5.0410 & 58.639 \\
9  & 9.194  & 5.0410 & 57.857 \\
10 & 9.064  & 5.0420 & 57.056 \\
11 & 8.931  & 5.0410 & 56.234 \\
12 & 8.794  & 5.0410 & 55.391 \\
13 & 8.654  & 5.0420 & 54.526 \\
14 & 8.510  & 5.0410 & 53.636 \\
15 & 8.361  & 5.0410 & 52.720 \\ 
16 & 8.207  & 5.0420 & 51.775 \\ 
17 & 8.049  & 5.0410 & 50.800 \\ 
18 & 7.885  & 5.0410 & 49.792 \\ 
19 & 7.715  & 5.0420 & 48.748 \\ 
20 & 7.538  & 5.0410 & 47.663 \\ 
21 & 7.353  & 5.0410 & 46.533 \\ 
22 & 7.161  & 5.0410 & 45.354 \\ 
23 & 6.954  & 5.0420 & 44.109 \\ 
24 & 6.747  & 5.0410 & 42.818 \\ 
25 & 6.494  & 5.0410 & 41.380 \\ 
26 & 6.171  & 5.0420 & 40.387 \\ 
27 & 6.000  & 5.0410 & 39.479 \\ 
28 & 6.000  & 5.0410 & 38.444 \\ 
29 & 6.000  & 5.0420 & 38.444 \\
30 & 6.000  & 5.0405 & 38.444 \\
\bottomrule
\end{tabular}
\label{tab:geom_tower}
\end{table}
%%%%%%%%%%%%%%%%%%%%%%%%%%%%%%%%%%%%%%%%%%%%%%%%%%%%%%%%%%%%%%%%%%%%%%%%%%%%%%%%%%%%%%%%%%%%%%%%%%%%%%
%%%%%%%%%%%%%%%%%%%%%%%%%%%%%%%%%%%%%%%%%%%%%%%%%%%%%%%%%%%%%%%%%%%%%%%%%%%%%%%%%%%%%%%%%%%%%%%%%%%%%%

\section{Wind–Wave Sampling Example for the Case Study}
\label{A:wave_samples}

\begin{figure*}[h!]
    \centering
    \includegraphics[width=1\textwidth]{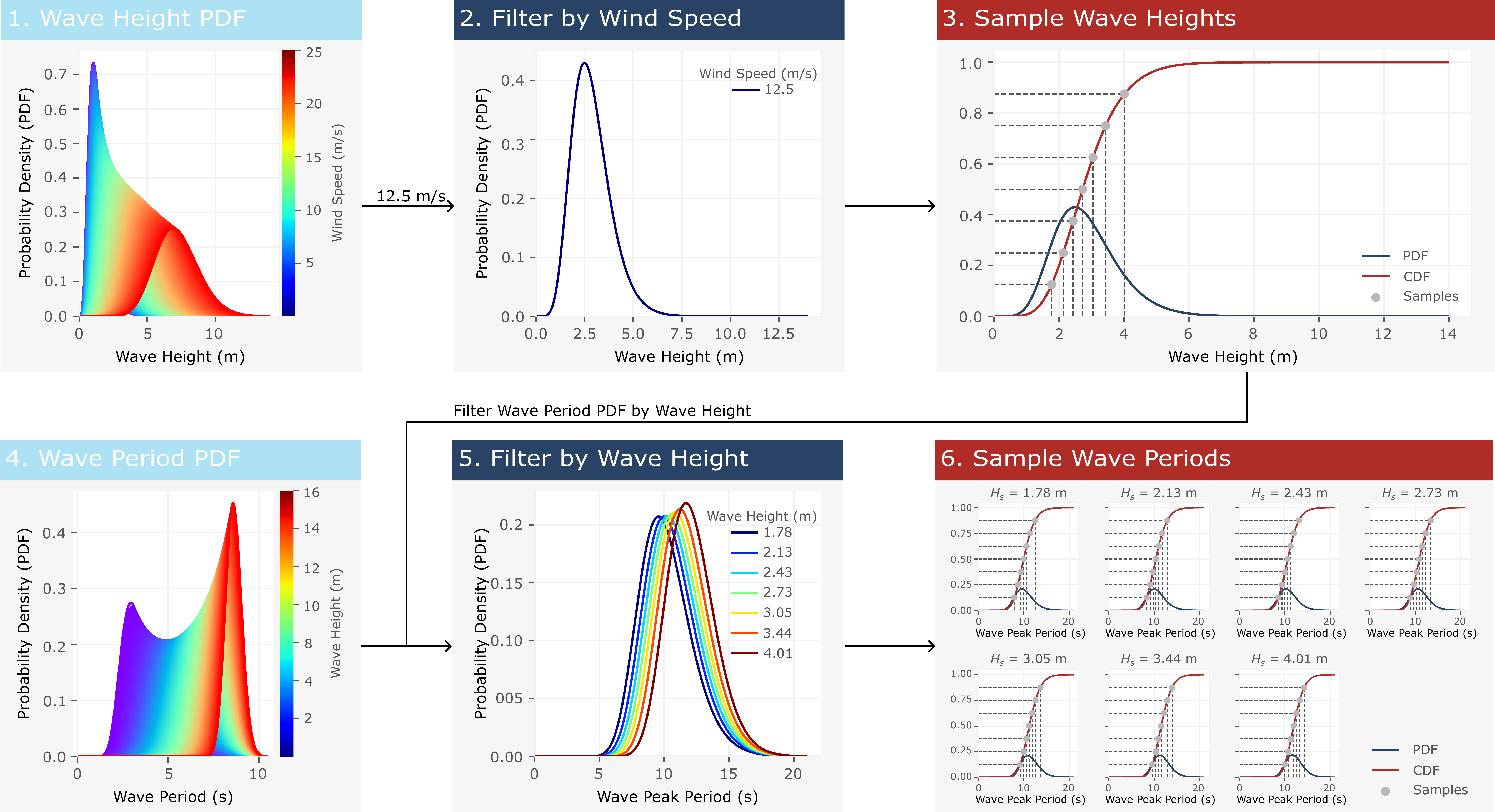}
\caption{Sampling by the \textbf{FLOAT} \textbf{\textit{Wind–Wave Sampler}} for a wind speed of 12.5~m/s: seven significant wave heights ($H_s$) were selected, and for each, seven peak periods ($T_p$), yielding 49 $(H_s, T_p)$ combinations. The same procedure was applied to all wind speeds in the case study.}
    \label{fig:wind}
\end{figure*}

The \textbf{\textit{Wind–Wave Sampler}} module of \textbf{FLOAT} (see~\autoref{sec:windwave_samp}) 
was used to define the environmental conditions for the numerical simulations. 
In the case study (see~\autoref{sec:case_study}), a total of 22 discrete wind speeds were considered. 
The sampling strategy is illustrated in \autoref{fig:wind} for a representative wind speed of 12.5~m/s 
and follows the six-step procedure outlined below:

\begin{enumerate}
\item \textbf{Wave Height PDF.} 
The PDF of the significant wave height ($H_s$) was obtained 
from the joint wind–wave distribution.

\item \textbf{Filter by Wind Speed.} 
The $H_s$ distribution was conditioned on the selected wind speed.

\item \textbf{Sample Wave Heights.} 
Seven representative $H_s$ values were selected using stratified sampling 
based on cumulative probability.

\item \textbf{Wave Period PDF.} 
For each sampled $H_s$, the conditional PDF of the peak wave period ($T_p$) was retrieved.

\item \textbf{Filter by Wave Height.} 
The $T_p$ distribution was conditioned on the corresponding wave height.

\item \textbf{Sample Wave Periods.} 
Seven representative $T_p$ values were sampled from each conditional distribution, 
resulting in 49 $(H_s, T_p)$ combinations for the selected wind speed.
\end{enumerate}

The same six-step sampling procedure was systematically applied to all wind speeds considered in the case study. 
Considering six independent turbulence seeds per wind speed, 
this resulted in a total of 6{,}468 environmental scenarios per tower design.

%%%%%%%%%%%%%%%%%%%%%%%%%%%%%%%%%%%%%%%%%%%%%%%%%%%%%%%%%%%%%%%%%%%%%%%%%%%%%%%%%%%%%%%%%%%%%%%%%%%%%%
%%%%%%%%%%%%%%%%%%%%%%%%%%%%%%%%%%%%%%%%%%%%%%%%%%%%%%%%%%%%%%%%%%%%%%%%%%%%%%%%%%%%%%%%%%%%%%%%%%%%%%
%%%%%%%%%%%%%%%%%%%%%%%%%%%%%%%%%%%%%%%%%%%%%%%%%%%%%%%%%%%%%%%%%%%%%%%%%%%%%%%%%%%%%%%%%%%%%%%%%%%%%%
%%%%%%%%%%%%%%%%%%%%%%%%%%%%%%%%%%%%%%%%%%%%%%%%%%%%%%%%%%%%%%%%%%%%%%%%%%%%%%%%%%%%%%%%%%%%%%%%%%%%%%

\section{Benchmarking the FLOAT Numerical Simulator with the IEA 22 MW Reference}
\label{A:benchmarking_fado_iea}

This appendix presents a benchmark study to assess the accuracy of the \textbf{\textit{Numerical Simulator}} module in \textbf{FLOAT}. The results are compared against the IEA 22 MW Reference Wind Turbine model~\cite{1cd3e417f8854b808c5372670588d3d0}.

\subsection{Simulation Setup} Simulations were conducted for the 22 wind speeds defined in the case study (see~\autoref{sec:case_study}), under steady wind and still water conditions. Each simulation ran for 200 seconds, with the initial 100 seconds excluded to eliminate transient effects. 

\subsection{Results and Comparison}
\begin{figure*}[h!]
    \centering
    \subfloat[Rotor performance]{%
        \includegraphics[width=0.775\textwidth]{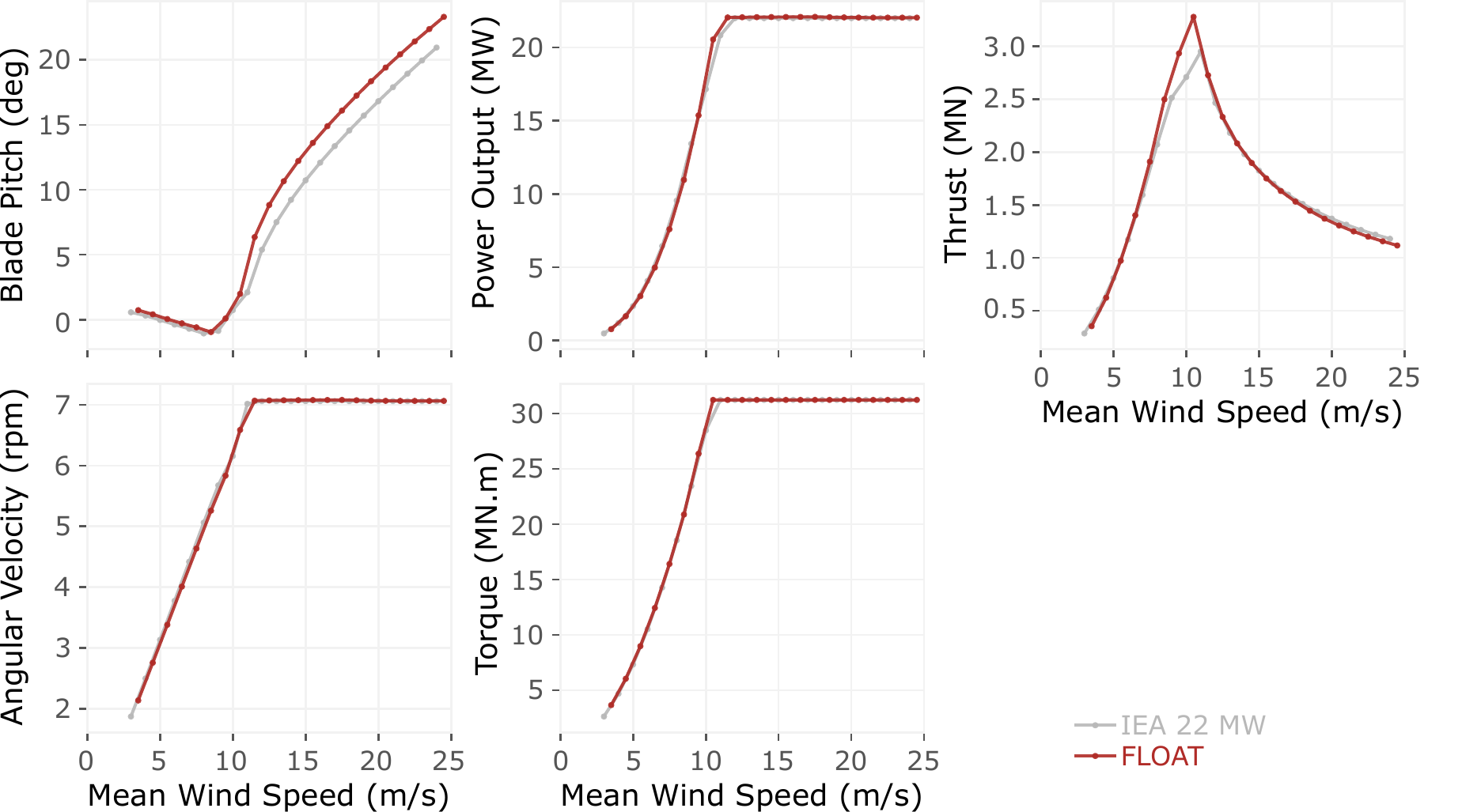}
        \label{fig:rotor_perf}
    }
    \vspace{0.5cm}
    \hfill
    \subfloat[Frequency response]{%
        \includegraphics[width=0.8\textwidth]{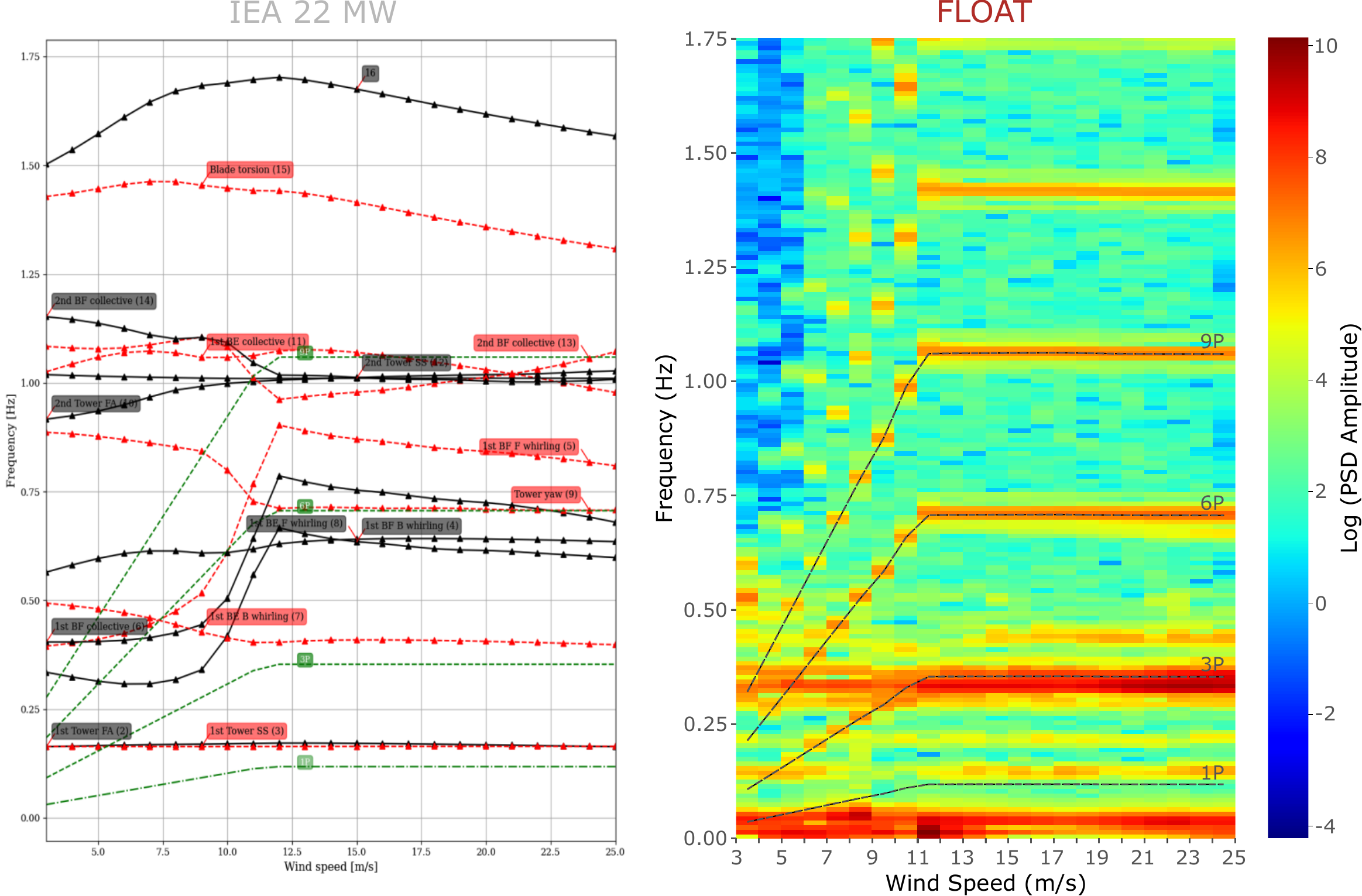}
        \label{fig:freq_resp}
    }
    
    \caption{Benchmarking the \textbf{FLOAT} \textbf{\textit{Numerical Simulator}} with the IEA 22 MW Reference~\cite{1cd3e417f8854b808c5372670588d3d0} for 22 wind speeds under steady wind and still water conditions.}
    \label{fig:rotor_benchmark}
\end{figure*}

The comparison focuses on two key aspects, rotor performance and frequency response, both shown in \autoref{fig:rotor_benchmark}.

\textbf{\textit{Rotor Performance.}} The curves for blade pitch, power output, thrust, angular velocity, and torque produced by \textbf{FLOAT} closely follow those from the IEA 22 MW reference model. Minor deviations in pitch and thrust are observed and attributed to platform tilt effects, as the reference turbine is modeled as a fixed-bottom system, whereas the benchmark was conducted under floating conditions.

\textbf{\textit{Frequency Response.}} The spectrogram of the rotor’s rotational speed across wind speeds shows harmonic trends for 1P, 3P, 6P, and 9P, which align with those reported in the IEA 22 MW reference model. These harmonics, associated with rotor-induced excitations, display consistent scaling throughout the operational.

\subsection{Validation Summary}
The good agreement with the IEA 22~MW reference model in both rotor performance and frequency response validates the accuracy of the simulator. These results support its use to generate the fatigue simulations presented in the case study (see~\autoref{sec:case_study}).

\section{Pitch/Heave–Platform Calibration Analysis of the FLOAT Numerical Simulator}
\label{A:pitch_fado}

This appendix evaluates the pitch/heave–platform calibration implemented in the \textbf{\textit{Numerical Simulator}} module of the \textbf{FLOAT} framework 
by comparing the platform’s rotational (pitch) and translational (heave) responses with and without calibration, highlighting improved platform stability.

\subsection{Simulation Setup} 
Simulations were carried out for the 22 wind speeds defined in the case study (see~\autoref{sec:case_study}), under steady-wind and still-water conditions. Each run lasted 200 s, with the first 100 s discarded to remove transients. Pitch calibration was performed using the IEA~22~MW reference tower, while heave calibration was evaluated using the tower design obtained after the first optimization cycle of the IEA~22~MW reference tower (Optimized~1).

\subsection{Results and Comparison}

\begin{figure*}[h!]
    \centering
    \subfloat[Without Calibration]{%
        \includegraphics[width=0.4955\textwidth]{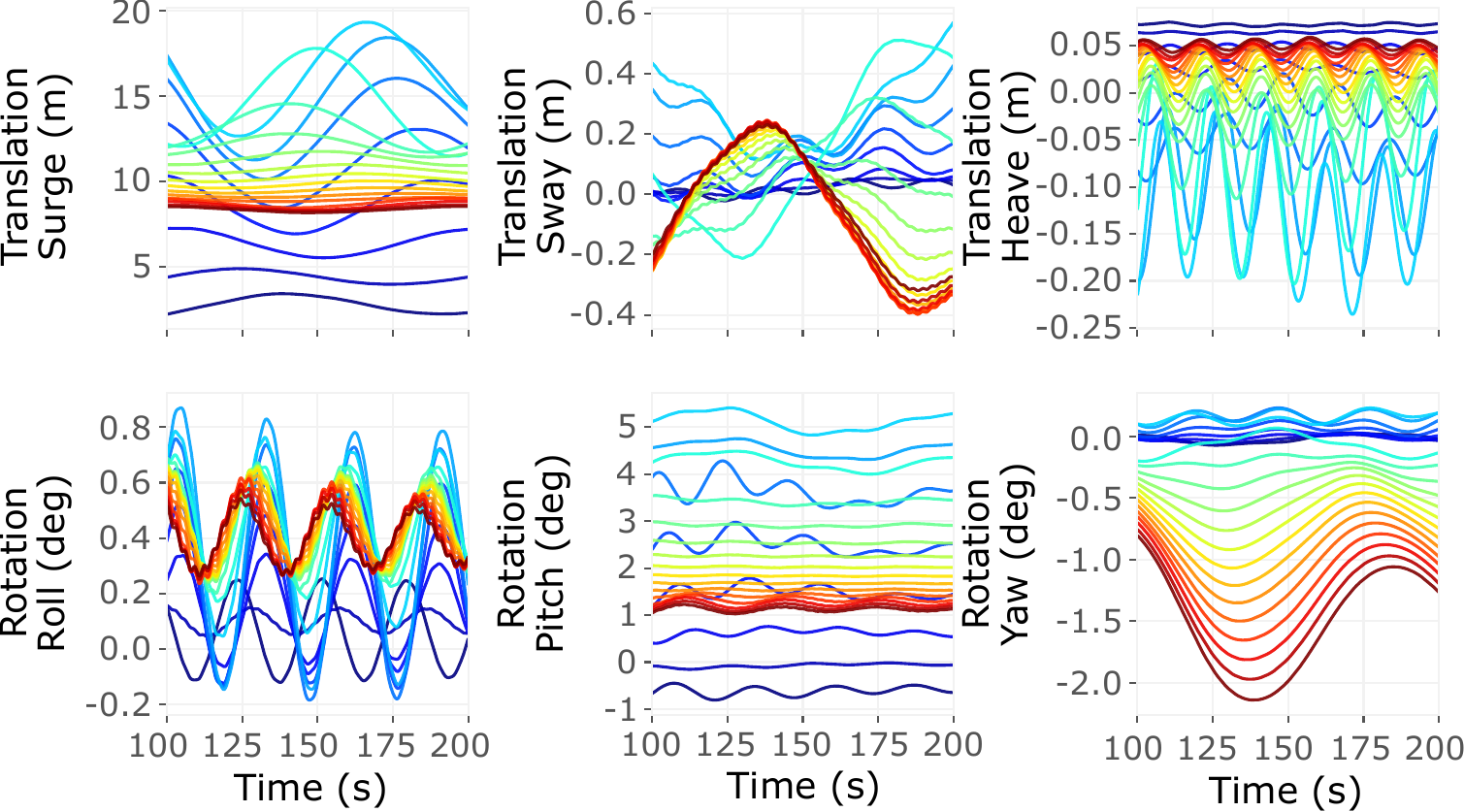}
    }
    \subfloat[With Calibration]{%
        \includegraphics[width=0.4955\textwidth]{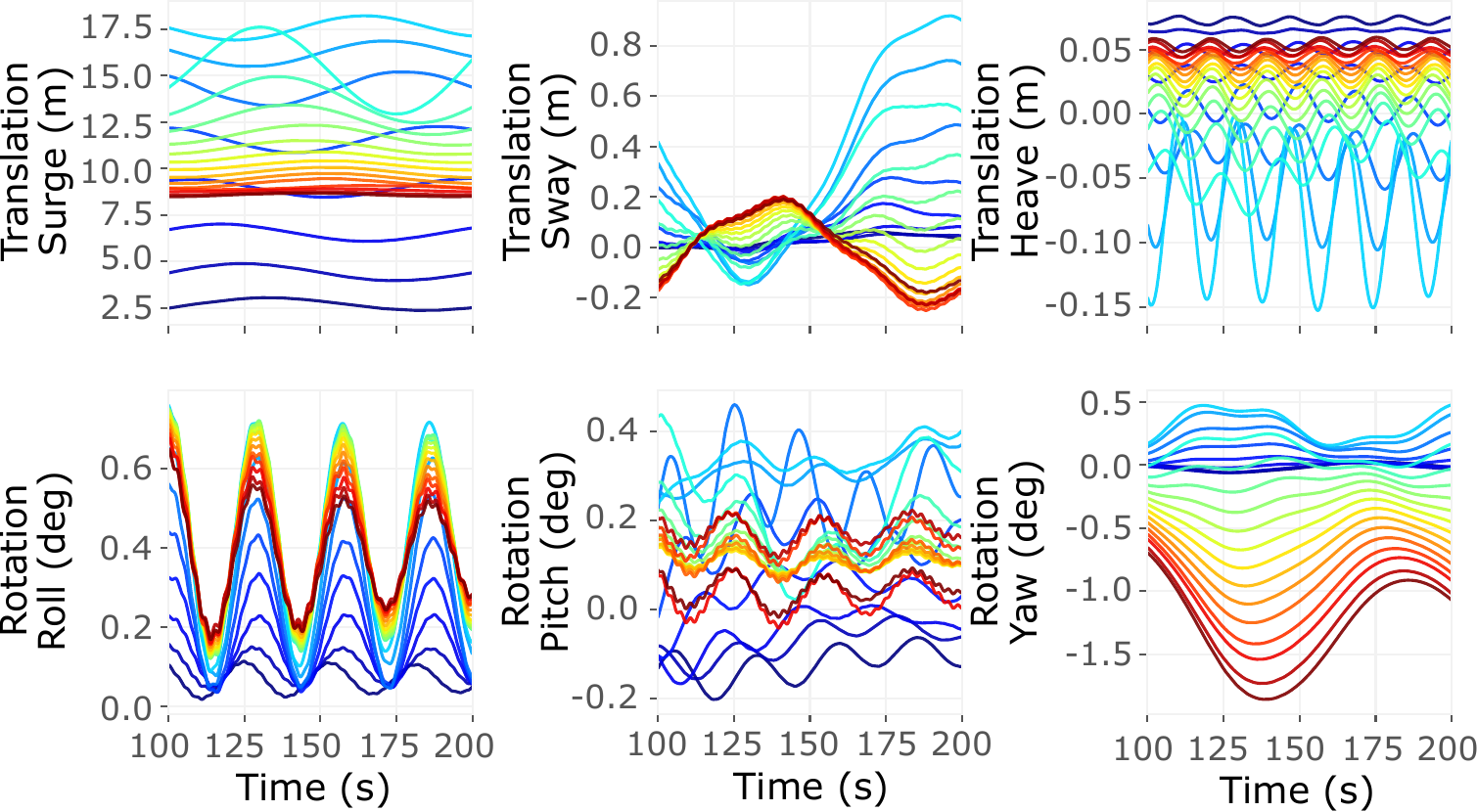}
    }
    \vspace{0.3cm}
    \hfill
    \includegraphics[width=1\textwidth]{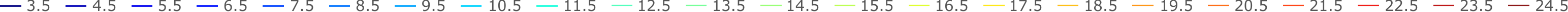}

    \caption{Pitch–platform calibration: platform motion responses across 22 wind speeds (3.5–24.5 m/s) under steady wind and still water using the \textbf{FLOAT} \textbf{\textit{Numerical Simulator}} for the IEA 22 MW Reference tower.}
    \label{fig:pitch_calibration_appendix}
\end{figure*}

\autoref{fig:pitch_calibration_appendix} compares the platform’s translational and rotational responses with and without pitch–platform calibration, 
with particular emphasis on the pitch rotation subplot (in degrees), which is the most relevant metric for this analysis.  
Without calibration, the pitch response ranges from $-1^{\circ}$ to $5^{\circ}$, with most cases clustered between $1^{\circ}$ and $3^{\circ}$.  
With calibration, this range is narrowed to $-0.2^{\circ}$–$0.4^{\circ}$, with most values concentrated between $0^{\circ}$ and $0.2^{\circ}$.

\begin{figure*}[h!]
    \centering
    \subfloat[Without Calibration]{%
        \includegraphics[width=0.4955\textwidth]{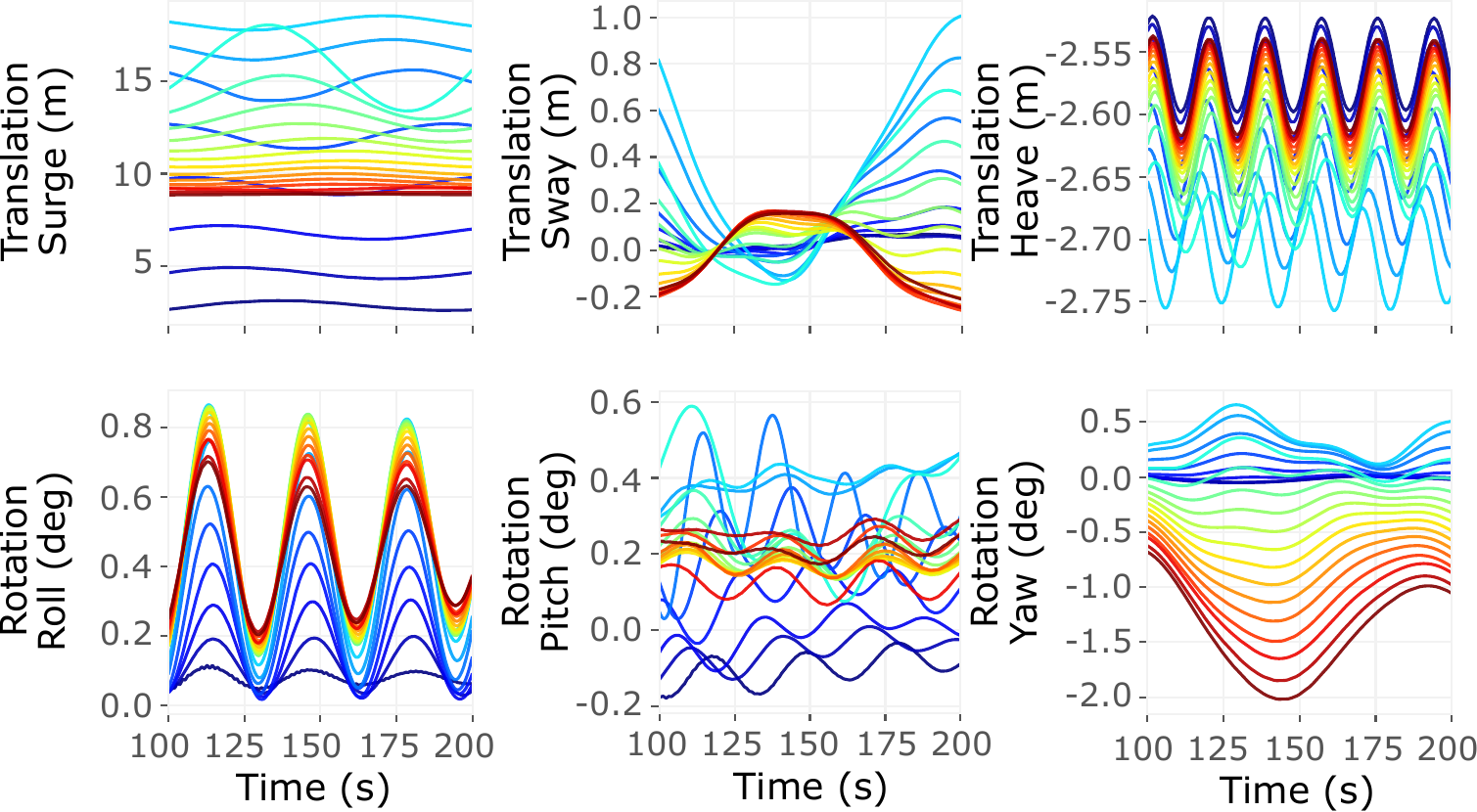}
    }
    \subfloat[With Calibration]{%
        \includegraphics[width=0.4955\textwidth]{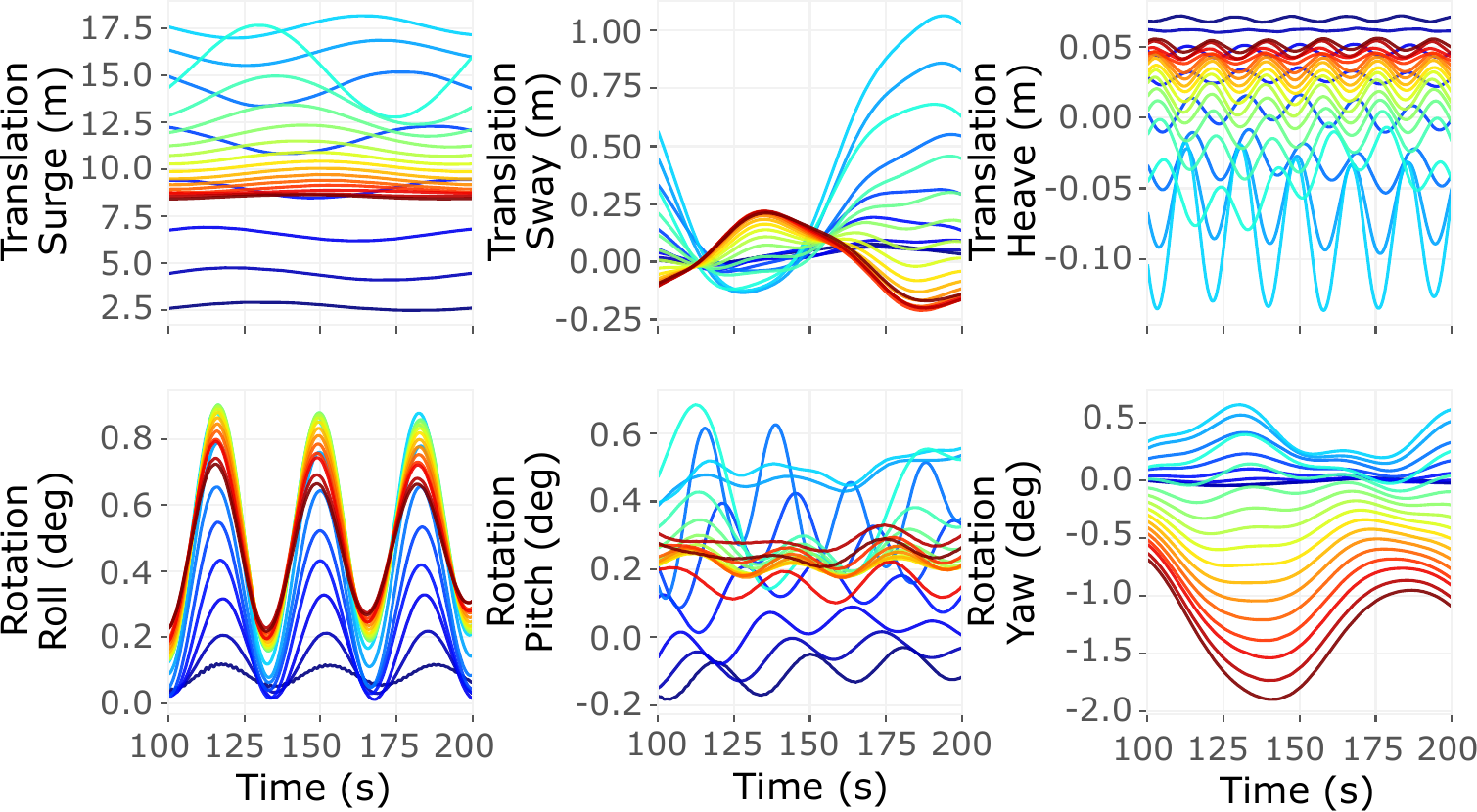}
    }
    \vspace{0.3cm}
    \hfill
    \includegraphics[width=1\textwidth]{figs/flabel.pdf}

    \caption{Heave–platform calibration: platform motion responses across 22 wind speeds (3.5–24.5 m/s) under steady wind and still water using the \textbf{FLOAT} \textbf{\textit{Numerical Simulator}} for the 22 MW Optimized 1 tower.}
    \label{fig:heave_calibration_appendix}
\end{figure*}

\autoref{fig:heave_calibration_appendix} compares the platform’s responses with and without heave–platform calibration, 
with particular emphasis on the translation heave subplot (in meters), which is the most relevant metric for this analysis.  
Without calibration, the heave response varies between $-2.75$ m and $-2.55$ m.  
With calibration, the range is reduced to $-0.15$ m–$0.05$ m, with most cases clustered near $0.05$ m.

\subsection{Validation Summary} 
The comparison confirms that pitch/heave–platform calibration improves the fidelity of the simulator by stabilizing both the rotational (pitch) and translational (heave) platform responses, driving them to values close to zero after calibration and ensuring physically realistic behavior. 
These results validate the numerical simulator for use in the fatigue and performance simulations presented in the case study (see~\autoref{sec:case_study}).

\section{HPC Benchmarking of the FLOAT Numerical Simulator}
\label{A:benchmark_hpc}

This appendix presents the benchmarking of the \textbf{FLOAT} \textbf{\textit{Numerical Simulator}} on the Inductiva platform~\cite{InductivaOpenFAST} using Google Cloud VMs. Each VM follows the convention \texttt{family-configuration-vCPUs} (e.g., \texttt{c2d-highcpu-2}). 
Cost estimates in this section, unless otherwise noted, are based on preemptible (spot) VM pricing, where spot instances are virtual machines offered at a reduced price but may be interrupted by the provider when resources are needed elsewhere, typically 60–91\% lower than on-demand rates.

\subsection{Top VM Types}
\begin{figure*}[t!]
    \centering
    \includegraphics[width=1.0\textwidth]{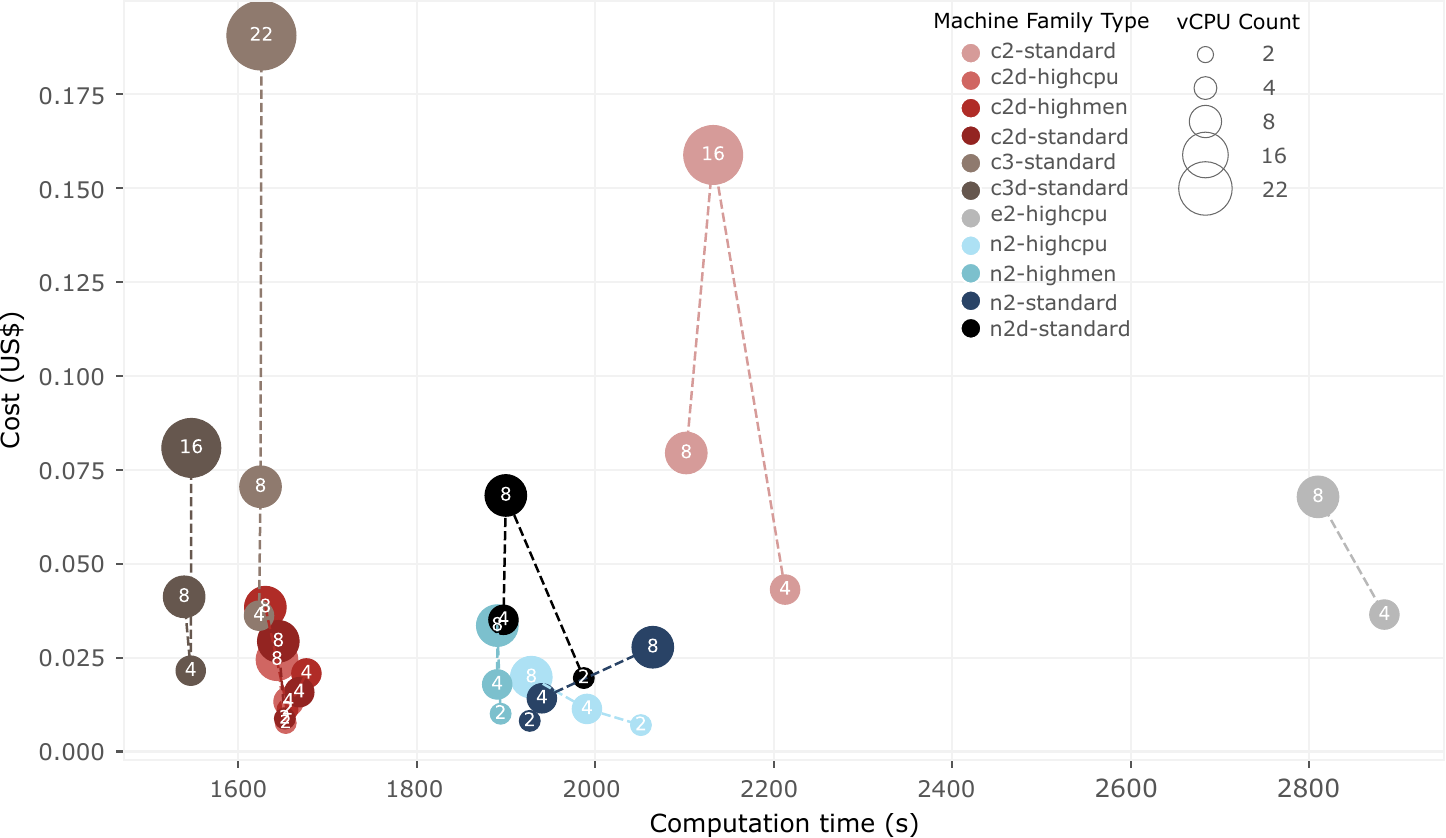}
    \caption{Benchmarking of the \textbf{FLOAT} \textbf{\textit{Numerical Simulator}} using the Inductiva platform~\cite{InductivaOpenFAST}. 
Computation time (s) versus hourly cost (US\$) for 36 Google Cloud VM types. 
    Bubble size represents the number of vCPUs, and color indicates the VM family. 
    Each point corresponds to the average of two simulation runs.}
    \label{fig:benchmark_hpc}
\end{figure*}

To identify the most efficient VM types, a benchmark was conducted using 36 Google Cloud instances from the 
\texttt{c2}, \texttt{c2d}, \texttt{c3}, \texttt{c3d}, \texttt{n2}, \texttt{n2d}, and \texttt{e2} families. 
These include \texttt{standard}, \texttt{highcpu}, and \texttt{highmem} configurations, 
ranging from 2 to 22 virtual CPUs (vCPUs). 
Each VM type was evaluated through two repeated simulations to account for variability, 
using identical environmental conditions: a turbulent wind speed of 12~m/s, 
wave height of 0.83~m, and wave period of 6.9~s. 
Each simulation was run for 1000~s with a time step of 0.01~s. 
The results were averaged, and mean values are reported in \autoref{fig:benchmark_hpc}.

Based on these results, \autoref{tab:machine_benchmark_100} highlights the two most efficient VM types. Both ran in $\sim$1650 s per simulation, but the \texttt{c2d-highcpu-2} was slightly cheaper, making it the preferred choice for large-scale \textbf{FLOAT} simulations.

\begin{table}[h!]
\footnotesize
\centering
\caption{Computation time and cost per simulation for the top two Google Cloud VM types (single machine).}
\begin{tabular}{@{}C{1.9cm}C{1.5cm}C{1.1cm}@{}}
\toprule
\textbf{Machine \break Type} & \textbf{Computation Time (s)} & \textbf{Cost/h (US\$)} \\
\midrule
\textbf{\texttt{c2d-standard-2}} & 1652.1 & 0.0088 \\
\rowcolor{firebrick!20}
\textbf{\texttt{c2d-highcpu-2}} & 1653.1 & 0.0076 \\
\bottomrule
\end{tabular}
\label{tab:machine_benchmark_100}
\end{table}

\subsection{Scalability Evaluation}
To assess the scalability of the top-performing VM type (\texttt{c2d-highcpu-2}), the estimated cost and runtime were extrapolated from a single simulation (1653~s, $\sim$27.5~min) to the full set of 6,468 simulations in the case study (see~\autoref{sec:case_study}). If executed sequentially on one machine with 2 vCPUs, this would require 10.7 million seconds of wall-clock time (5939 CPU-h), or 123.8 days, making large-scale fatigue analysis practically infeasible, as shown in \autoref{tab:machine_benchmark_500}.

\begin{table}[h!]
\footnotesize
\centering
\caption{Estimated wall-clock runtime, CPU hours, and cost for 6,468 simulations on \texttt{c2d-highcpu-2}, comparing sequential and parallel execution. 
Values are extrapolated from the per-simulation runtime (1653 s, 2 vCPUs).}
\begin{tabular}{@{}C{1cm}C{1.6cm}C{1.5cm}C{1cm}C{1.5cm}@{}}
\toprule
\textbf{Execution Mode} & \textbf{Wall-Clock Time} & \textbf{CPU \break Hours} & \textbf{Machine Number} & \textbf{Total Cost (US\$)} \\
\midrule
Sequential & 123.8 days & 5939 & 1 & 22.61\\
\rowcolor{firebrick!20}
Parallel & 11.9 hours & 5939 & 250 & 22.61 \\
\bottomrule
\end{tabular}
\label{tab:machine_benchmark_500}
\end{table}

This is precisely where the benefit of HPC becomes clear: by distributing the workload across 250 machines (500 vCPUs), the estimated wall-clock runtime was reduced to $\sim$11.9~h, with an estimated total cost of \$22.61. Importantly, while the wall-clock time decreased by more than two orders of magnitude, the total CPU-hours and cost remained essentially unchanged, as both are determined by the overall computational workload.

This corresponds to a $\sim$250× reduction in runtime relative to sequential execution, consistent with the number of machines deployed in parallel, and demonstrates scalability close to the ideal linear speedup. This confirms that large-scale fatigue datasets can be generated efficiently through HPC deployment within the \textbf{FLOAT} framework while maintaining cost-effectiveness.

\subsection{Benchmark Validation with Public Pricing}

To ensure transparency, the estimated cost of \$22.61 for the 6,468 simulations executed within the \textbf{FLOAT} framework is compared in \autoref{tab:public_pricing} with public pricing from both Inductiva and Google Cloud. As the \textbf{FLOAT} HPC deployment was carried out via Inductiva using spot instances, the estimate is primarily validated against Inductiva spot pricing. The comparison with Google Cloud confirms that the estimated cost is consistent with prevailing market prices across providers. Overall, the results show that spot pricing, offering substantial reductions relative to on-demand rates, significantly lowers execution costs, with the estimated \$22.61 falling within the expected spot pricing ranges of both providers.

\begin{table}[h!]
\footnotesize
\centering
\caption{Per-machine cost comparison for \texttt{c2d-highcpu-2} VMs, with total costs extrapolated to 6,468 simulations.}

\begin{tabular}{@{}C{1.9cm}C{1.9cm}C{1.8cm}C{1.7cm}@{}}
\toprule
\textbf{Cloud \break Provider} & \textbf{Price \break Model} & \textbf{Cost/h \break (US\$)} & \textbf{Total Cost (US\$)} \\
\midrule
Google Cloud & On-demand               & 0.075  & $\sim$223.13 \\
Google Cloud & Spot  & 0.0067 – 0.0300 & $\sim$19.89 – 89.25 \\
Inductiva & On-demand                  & 0.083     & $\sim$246.93 \\
Inductiva & Spot    & 0.0075 – 0.0332 & $\sim$22.31 – 98.86 \\
\rowcolor{firebrick!20}
\textbf{FLOAT} & Inductiva spot  & 0.0076 & 22.61 \\
\bottomrule
\end{tabular}
\label{tab:public_pricing}
\end{table}

\subsection{Real-Case Comparison}
The estimation for the best-performing VM type, \texttt{c2d-highcpu-2}, was validated against a real HPC deployment of the 6,468 simulations, as shown in~\autoref{tab:machine_benchmark_real}. 
The real-case confirmed the extrapolation, completing all simulations in 12.58 hours, only 43 minutes longer than the estimated 11.87 hours. 
The total cost, however, increased from the estimated \$22.61 to \$47. 
This difference reflects the billing model of cloud providers, which charges wall-clock allocation per machine, together with overheads from scheduling and parallel execution across 250 nodes.

\begin{table}[h!]
\footnotesize
\centering
\caption{Estimated vs. real runtime and cost for 6,468 simulations on 250 \texttt{c2d-highcpu-2} machines.}
\begin{tabular}{@{}C{1cm}C{1.9cm}C{1.5cm}C{1cm}C{1.1cm}@{}}
\toprule
 & \textbf{Wall-Clock Time (h)} & \textbf{Machine \break Number} & \textbf{Cost (US\$)} \\
\midrule
\rowcolor{firebrick!15}
Estimation & 11.87 & 250 & 22.61 \\
\rowcolor{midnightblue!20}
Real & 12.58 & 250 & 47.00 \\
\bottomrule
\end{tabular}
\label{tab:machine_benchmark_real}
\end{table}

\subsection{Validation Summary} 
The \texttt{c2d-highcpu-2} instance was identified as the best option, offering an optimal trade-off between cost and speed. 
Combined with HPC deployment, it enabled execution of the 6,468 FOWT simulations in $\sim$13~h on 250 machines at a total cost of \$47, 
compared to more than 100 days on a single machine 
(speed-up $\sim$205$\times$).

\section{Geometric Parameters of the FLOAT~22~MW Semi-Submersible FOWT Tower} \label{A:table_opt}

\autoref{tab:geom_tower_opt2} summarizes the geometric parameters of the FLOAT~22~MW semi-submersible FOWT tower, 
namely the final design optimized from the IEA~22~MW semi-submersible reference tower using the \textbf{FLOAT} framework. The table includes the 31 outer diameters ($d_i$), 30 section heights ($h_i$), and 30 wall thicknesses ($t_i$), 
defined from bottom to top along the tower.

%%%%%%%%%%%%%%%%%%%%%%%%%%%%%%%%%%%%%%%%%%%%%%%%%%%%%%%%%%%%%%%%%%%%%%%%%%%%%%%%%%%%%%%%%%%%%%%%%%%%%%
%%%%%%%%%%%%%%%%%%%%%%%%%%%%%%%%%%%%%%%%%%%%%%%%%%%%%%%%%%%%%%%%%%%%%%%%%%%%%%%%%%%%%%%%%%%%%%%%%%%%%%
\begin{table}[t!]
\footnotesize
\centering
\caption{Geometric parameters of the FLOAT~22~MW semi-submersible FOWT tower: 31 diameters $d_i$, 30 heights $h_i$, and 30 thicknesses $t_i$ across the 30 sections (from bottom to top).}
\begin{tabular}{@{}p{0.5cm}p{1cm}p{1cm}p{1cm}@{}}
\toprule
\textbf{i} & \textbf{d$_i$ (m)} & \textbf{h$_i$ (m)} & \textbf{t$_i$ (mm)} \\ \midrule
0  & 12.000  & -       & -       \\
1  & 12.000  & 3.1885  & 118.225 \\
2  & 12.000  & 5.0410  & 113.191 \\
3  & 12.000  & 5.0420  & 107.086 \\
4  & 12.000  & 5.0410  & 101.219 \\
5  & 12.000  & 4.0400  &  95.529 \\
6  & 12.000  & 5.0410  &  89.934 \\
7  & 12.000  & 5.0420  &  84.557 \\
8  & 11.973  & 5.0410  &  79.579 \\
9  & 11.960  & 5.0410  &  74.789 \\
10 & 11.516  & 5.0420  &  73.361 \\
11 & 11.497  & 5.0410  &  71.914 \\
12 & 11.048  & 5.0410  &  70.489 \\
13 & 11.039  & 5.0420  &  69.020 \\
14 & 10.581  & 5.0410  &  67.560 \\
15 & 10.563  & 5.0410  &  66.073 \\
16 & 10.104  & 5.0420  &  64.599 \\
17 & 10.080  & 5.0410  &  63.076 \\
18 &  9.608  & 5.0410  &  61.535 \\
19 &  9.582  & 5.0420  &  59.966 \\
20 &  9.093  & 5.0410  &  58.392 \\
21 &  9.060  & 5.0410  &  56.728 \\
22 &  8.563  & 5.0410  &  55.083 \\
23 &  8.523  & 5.0420  &  53.392 \\
24 &  8.020  & 5.0410  &  51.695 \\
25 &  7.980  & 5.0410  &  50.080 \\
26 &  7.462  & 5.0420  &  48.672 \\
27 &  7.452  & 5.0410  &  47.376 \\
28 &  7.080  & 5.0410  &  45.860 \\
29 &  7.080  & 5.0420  &  44.524 \\
30 &  6.741  & 5.0405  &  44.524 \\
\bottomrule
\end{tabular}
\label{tab:geom_tower_opt2}
\end{table}

\clearpage

\bibliographystyle{elsarticle-num-names} 

\bibliography{references}

@Article{en14092484,
AUTHOR = {Rinaldi, Giovanni and Thies, Philipp R. and Johanning, Lars},
TITLE = {Current Status and Future Trends in the Operation and Maintenance of Offshore Wind Turbines: A Review},
JOURNAL = {Energies},
VOLUME = {14},
YEAR = {2021},
NUMBER = {9},
ARTICLE-NUMBER = {2484},
ISSN = {1996-1073},
DOI = {10.3390/en14092484}
}

@article{MCMORLAND2022112499,
title = {Operation and maintenance for floating wind turbines: A review},
journal = {Renewable and Sustainable Energy Reviews},
volume = {163},
pages = {112499},
year = {2022},
issn = {1364-0321},
doi = {10.1016/j.rser.2022.112499},
author = {J. McMorland and M. Collu and D. McMillan and J. Carroll}
}

@article{Borg_2016,
doi = {10.1088/1742-6596/753/8/082024},
year = {2016},
month = {sep},
publisher = {IOP Publishing},
volume = {753},
number = {8},
pages = {082024},
author = {Borg, Michael and Hansen, Anders Melchior and Bredmose, Henrik},
title = {Floating substructure flexibility of large-volume 10{MW} offshore wind turbine platforms in dynamic calculations},
journal = {Journal of Physics: Conference Series}
}

@article{ALVESRIBEIRO2025126294,
title = {Offshore wind turbine tower design and optimization: A review and AI-driven future directions},
journal = {Applied Energy},
volume = {397},
pages = {126294},
year = {2025},
issn = {0306-2619},
doi = {10.1016/j.apenergy.2025.126294},
author = {João {Alves Ribeiro} and Bruno {Alves Ribeiro} and Francisco Pimenta and Sérgio {M.O. Tavares} and Jie Zhang and Faez Ahmed}
}

@article{Zahle_2024,
doi = {10.1088/1742-6596/2767/2/022065},
year = {2024},
month = {jun},
publisher = {IOP Publishing},
volume = {2767},
number = {2},
pages = {022065},
author = {Zahle, Frederik and Li, Ang and Lønbæk, Kenneth and Sørensen, Niels N. and Riva, Riccardo},
title = {Multi-fidelity, steady-state aeroelastic modelling of a 22-megawatt wind turbine},
journal = {Journal of Physics: Conference Series}
}

@article{Collier_2024,
doi = {10.1088/1742-6596/2767/5/052042},
year = {2024},
month = {jun},
publisher = {IOP Publishing},
volume = {2767},
number = {5},
pages = {052042},
author = {Collier, W and Ors, D and Barlas, T and Zahle, F and Bortolotti, P and Marten, D and Jensen, C S L and Branlard, E and Zalkind, D and Lønbæk, K},
title = {Aeroelastic code comparison using the IEA 22MW reference turbine},
journal = {Journal of Physics: Conference Series}
}

@article{DANGI2025122248,
title = {The effect of turbulent coherent structures in atmospheric flow on wind turbine loads},
journal = {Renewable Energy},
volume = {241},
pages = {122248},
year = {2025},
issn = {0960-1481},
doi = {10.1016/j.renene.2024.122248},
author = {Nirav Dangi and Jurij Sodja and Carlos Simão Ferreira and Wei Yu}
}

@article{PACHECO2023115913,
title = {Experimental evaluation of strategies for wind turbine farm-wide fatigue damage estimation},
journal = {Engineering Structures},
volume = {285},
pages = {115913},
year = {2023},
issn = {0141-0296},
doi = {10.1016/j.engstruct.2023.115913},
author = {Pacheco, Jo{\~a}o and Pimenta, Francisco and Pereira, S{\'e}rgio and Cunha, {\'A}lvaro and Magalh{\~a}es, Filipe}
}

@article{Papi_2022,
doi = {10.1088/1742-6596/2385/1/012117},
year = {2022},
month = {dec},
publisher = {IOP Publishing},
volume = {2385},
number = {1},
pages = {012117},
author = {Papi, F. and Perignon, Y. and Bianchini, A.},
title = {Derivation of Met-Ocean Conditions for the Simulation of Floating Wind Turbines: a European case study},
journal = {Journal of Physics: Conference Series}
}

@article{PIMENTA2024117367,
title = {Modal properties of floating wind turbines: Analytical study and operational modal analysis of an utility-scale wind turbine},
journal = {Engineering Structures},
volume = {301},
pages = {117367},
year = {2024},
issn = {0141-0296},
doi = {10.1016/j.engstruct.2023.117367},
author = {Pimenta, Francisco and Ribeiro, Daniel and Román, Adela and Magalhães, Filipe}
}

@article{Vlachogiannis_2025,
title = {Redefining fatigue predictions: A multi-sea state HPC framework for FOWTs},
journal = {Ocean Engineering},
volume = {338},
pages = {121961},
year = {2025},
issn = {0029-8018},
doi = {10.1016/j.oceaneng.2025.121961},
author = {Prokopios Vlachogiannis and Christophe Peyrard and Ajit C. Pillai and David Ingram and Pierre Bousseau and Maurizio Collu}
}

@article{matsuishi_1968,
    title = {Fatigue of metals subjected to varying stress},
    year = {1968},
    journal = {Proceedings of the Kyushu Branch of Japan Society of Mechanics Engineering},
    author = {Matsuishi, M and Endo, T},
    pages = {37--40}
}

@article{ZHAO2021106075,
title = {Calibration of design fatigue factors for offshore structures based on fatigue test database},
journal = {International Journal of Fatigue},
volume = {145},
pages = {106075},
year = {2021},
issn = {0142-1123},
doi = {10.1016/j.ijfatigue.2020.106075},
author = {Wangwen Zhao}
}

@article{dowling_1972,
    title = {Fatigue Failure Predictions for Complicated Stress-Strain Histories},
    year = {1972},
    author = {Dowling, N},
    journal = {Journal of Materials},
}

@article{YANG2023119111,
title = {Performance and fatigue analysis of an integrated floating wind-current energy system considering the aero-hydro-servo-elastic coupling effects},
journal = {Renewable Energy},
volume = {216},
pages = {119111},
year = {2023},
issn = {0960-1481},
doi = {10.1016/j.renene.2023.119111},
author = {Yang Yang and Jianbin Fu and Zhaobin Shi and Lu Ma and Jie Yu and Fang Fang and Shunhua Chen and Zaibin Lin and Chun Li}
}

@article{YUAN2024119667,
title = {Deep analysis of power regulation on fatigue loads and platform motion in floating wind turbines},
journal = {Ocean Engineering},
volume = {313},
pages = {119667},
year = {2024},
issn = {0029-8018},
doi = {10.1016/j.oceaneng.2024.119667},
author = {Xinyu Yuan and Dongran Song and Sifan Chen and Jian Yang and Mi Dong and Renyong Wei and M. Talaat and Young Hoon Joo}
}

@Article{en16010002,
AUTHOR = {Barooni, Mohammad and Ashuri, Turaj and Velioglu Sogut, Deniz and Wood, Stephen and Ghaderpour Taleghani, Shiva},
TITLE = {Floating Offshore Wind Turbines: Current Status and Future Prospects},
JOURNAL = {Energies},
VOLUME = {16},
YEAR = {2023},
NUMBER = {1},
ARTICLE-NUMBER = {2},
URL = {https://www.mdpi.com/1996-1073/16/1/2},
ISSN = {1996-1073},
DOI = {10.3390/en16010002}
}

@ARTICLE{1161901,
  author={Welch, P.},
  journal={IEEE Transactions on Audio and Electroacoustics}, 
  title={The use of fast Fourier transform for the estimation of power spectra: A method based on time averaging over short, modified periodograms}, 
  year={1967},
  volume={15},
  number={2},
  pages={70-73},
  doi={10.1109/TAU.1967.1161901}}

@article{LIU2024120238,
title = {On long-term fatigue damage estimation for a floating offshore wind turbine using a surrogate model},
journal = {Renewable Energy},
volume = {225},
pages = {120238},
year = {2024},
issn = {0960-1481},
doi = {10.1016/j.renene.2024.120238},
author = {Ding Peng Liu and Giulio Ferri and Taemin Heo and Enzo Marino and Lance Manuel}
}

@article{LI2020570,
title = {Long-term fatigue damage assessment for a floating offshore wind turbine under realistic environmental conditions},
journal = {Renewable Energy},
volume = {159},
pages = {570-584},
year = {2020},
issn = {0960-1481},
doi = {10.1016/j.renene.2020.06.043},
author = {Xuan Li and Wei Zhang}
}

@techreport{IEC61400-1,
  type = {Standard},
  title        = {Wind Energy Generation Systems - Part 1: Design Requirements},
  year         = {2019},
  Author       = {{IEC 61400-1:2019}},
  institution  = {International Electrotechnical Commission},
  url          = {https://webstore.iec.ch/publication/26423},
  address ={Geneva, CH}
}

@techreport{IEC61400-3-1,
  type = {Standard},
  title        = {Wind Energy Generation Systems - Part 3-1: Design Requirements for Fixed Offshore Wind Turbines},
  year         = {2019},
  Author       = {{IEC 61400-3-1:2019}},
  institution  = {International Electrotechnical Commission},
  url          = {https://webstore.iec.ch/publication/29360},
  address = {Geneva, CH}
}

@techreport{IEC61400-3-2,
  type = {Standard},
  title        = {Wind Energy Generation Systems - Part 3-2: Design Requirements for Floating Offshore Wind Turbines},
  year         = {2019},
  Author       = {{IEC 61400-3-2:2019}},
  institution  = {International Electrotechnical Commission},
  url          = {https://webstore.iec.ch/en/publication/29244},
  address = {Geneva, CH}
}

@techreport{Eurocode3-Part1-9,
  type = {Standard},
  title        = {Eurocode 3: Design of Steel Structures - Part 1-9: Fatigue},
  institution = {European Committee for Standardization},
  Author       = {{EN 1993-1-9}},
    year         = {2005},
  url          = {https://www.en-standard.eu/csn-en-1993-1-9-eurocode-3-design-of-steel-structures-part-1-9-fatigue/},
  address = {Brussels, BE}
}

@techreport{DNV-RP-C202,
type = {Standard},
  title        = {Buckling strength of shells},
  institution = {DNV},
  year         = {2021},
  Author       = {{DNV-RP-C202}},
  url          = {https://www.dnv.com/energy/standards-guidelines/dnv-rp-c202-buckling-strength-of-shells/},
  address   = {H{\o}vik, NO}
}

@techreport{DNV-RP-C203,
type = {Standard},
  title        = {Fatigue Design of Offshore Steel Structures},
  institution = {DNV},
  year         = {2024},
  Author       = {{DNV-RP-C203}},
  url          = {https://www.dnv.com/oilgas/download/dnv-rp-c203-fatigue-design-of-offshore-steel-structures/},
  address   = {H{\o}vik, NO}
}

@techreport{osti_1660012,
  author       = {Allen, Christopher and Viscelli, Anthony and Dagher, Habib and Goupee, Andrew and Gaertner, Evan and Abbas, Nikhar and Hall, Matthew and Barter, Garrett},
  title        = {Definition of the UMaine VolturnUS-S Reference Platform Developed for the IEA Wind 15-Megawatt Offshore Reference Wind Turbine},
  institution  = {National Renewable Energy Laboratory},
  doi          = {10.2172/1660012},
  address        = {Golden, CO, US},
  year         = {2020},
  month        = {07}}

@techreport{osti_1155123,
  author       = {Robertson, A. and Jonkman, J. and Masciola, M. and Song, H. and Goupee, A. and Coulling, A. and Luan, C.},
  title        = {Definition of the Semisubmersible Floating System for Phase II of OC4},
  institution  = {National Renewable Energy Laboratory},
  doi          = {10.2172/1155123},
  address        = {Golden, CO, US},
  year         = {2014},
  month        = {09}}

@techreport{osti_979456,
  author       = {Jonkman, J},
  title        = {Definition of the Floating System for Phase IV of OC3},
  institution  = {National Renewable Energy Laboratory},
  doi          = {10.2172/979456},
  address        = {Golden, CO, US},
  year         = {2010},
  month        = {05}}

@techreport{osti_921803,
  author       = {Jonkman, J M},
  title        = {Dynamics Modeling and Loads Analysis of an Offshore Floating Wind Turbine},
  institution  = {National Renewable Energy Laboratory},
  doi          = {10.2172/921803},
  address        = {Golden, CO, US},
  year         = {2007},
  month        = {12}}

@techreport{1cd3e417f8854b808c5372670588d3d0,
  author       = {Zahle, Frederik and Barlas, Thanasis and Lonbaek, Kenneth and Bortolotti, Pietro and Zalkind, Daniel and Wang, Lu and Labuschagne, Casper and Sethuraman, Latha and Barter, Garrett},
  title        = {Definition of the IEA Wind 22-Megawatt Offshore Reference Wind Turbine},
  institution  = {National Renewable Energy Laboratory \& Technical University of Denmark},
  doi          = {10.11581/DTU.00000317},
  address = {Golden, CO, USA \& Lyngby, DNK},
  publisher    = {Technical University of Denmark},
  year         = {2024},
  month        = {04}}

@techreport{osti_2409185,
  author       = {Zahle, Frederik and Barlas, Thanasis and Bortolotti, Pietro and Zalkind, Daniel and Collier, William},
  title        = {International Energy Agency 22 MW Offshore Reference Wind Turbine},
  institution  = {National Renewable Energy Laboratory},
  doi          = {10.2172/2409185},
  address      = {Golden, CO, US},
  year         = {2024},
  month        = {06}}

@techreport{turbsim,
  author       = {B.J. Jonkman and M.L. Buhl, Jr.},
  title        = {TurbSim User’s Guide},
  institution  = {National Renewable Energy Laboratory},
  url          = {https://docs.nrel.gov/docs/fy06osti/39797.pdf},
  address      = {Golden, CO, US},
  year         = {2006},
  month        = {09}}

@inproceedings{Jonkman_2013,
author = {Jonkman, Jason},
title = {The New Modularization Framework for the {FAST} Wind Turbine {CAE} Tool},
booktitle = {51st AIAA Aerospace Sciences Meeting including the New Horizons Forum and Aerospace Exposition},
doi = {10.2514/6.2013-202},
pages = {},
year = "2013",
}

@proceedings{10.1115/IOWTC2021-3533,
    author = {Jonkman, Jason and Wright, Alan and Barter, Garrett and Hall, Matthew and Allison, James and Herber, Daniel R.},
    title = {Functional Requirements for the WEIS Toolset to Enable Controls Co-Design of Floating Offshore Wind Turbines},
    volume = {ASME 2021 3rd International Offshore Wind Technical Conference},
    series = {International Conference on Offshore Mechanics and Arctic Engineering},
    pages = {V001T01A007},
    year = {2021},
    month = {02},
    doi = {10.1115/IOWTC2021-3533},
}

@inproceedings{wang2005fatigue,
    author = {Wang, Xiaozhi and Cheng, Zhan and Wirsching, Paul H. and Sun, Haihong},
    title = {Fatigue Design Factors and Safety Level Implied in Fatigue Design of Offshore Structures},
    booktitle = {24th International Conference on Offshore Mechanics and Arctic Engineering: Volume 3},
    pages = {231-236},
    year = {2005},
    month = {06},
    doi = {10.1115/OMAE2005-67488}
}

@inproceedings{bak2013dtu,
  title = {The {DTU} 10-{MW} Reference Wind Turbine},
author = {Bak, Christian and Zahle, Frederik and Bitsche, Robert and Kim, Taeseong and Yde, Anders and Henriksen, Lars Christian and Hansen, Morten Hartvig and Blasques, Jos{\'e} Pedro Albergaria Amaral and Mac Gaunaa and Natarajan, Anand},
  year = {2013},
pages = {},
  booktitle = {Proceedings of the Danish Wind Power Research Conference},
  address = {Copenhagen, DNK},
  url = {https://orbit.dtu.dk/en/publications/the-dtu-10-mw-reference-wind-turbine},
}

@inproceedings{Sarmento2024,
  author    = {Sarmento, Luis and Penedones, Hugo and Santos, Sérgio and Barbosa, Paulo},
  title     = {Running Hydraulics Simulations at Scale Using {Inductiva} Python {API}},
  booktitle = {The First International Conference on Technologies for Marine and Coastal Ecosystems},
  year      = {2024},
  month     = {November},
  pages     = {1--2},
  publisher = {{IARIA}},
  address   = {Valencia, ESP},
  isbn      = {978-1-68558-325-5},
  url = {https://www.thinkmind.org/library/COCE/COCE_2024/coce_2024_1_10_30003.html},
}

@misc{15mw_new,
	title = {{Project to Demonstrate 15+ MW Turbine on Ocergy’s Floating Platform Kicks Off}},
	url = {https://www.offshorewind.biz/2025/06/23/project-to-demonstrate-15-mw-turbine-on-ocergys-floating-platform-kicks-off/},
	author = {{Adnan Memija}},
        note = {Accessed: 2025-21-08},
year={2011}
}

@misc{20mw_new,
	title = {{20 MW: World’s largest floating wind turbine spanning 7 soccer fields tested by China}},
	url = {https://interestingengineering.com/energy/china-largest-floating-wind-turbine-qihang},
	author = {{
Kaif Shaikh}},
        note = {Accessed: 2025-21-08},
year={2011}
}

@misc{HywindDemo,
  title = {{Hywind Demo (Statoil)}},
  author = {{Statoil (now Equinor)}},
  year = {2009},
  note = {Accessed: 2025-08-21},
  url = {https://questfwe.com/wp-content/uploads/2018/02/Windfloat-Presentation-PPI.pdf}
}

@misc{WindFloat1,
	title = {{WindFloat 1}},
	url = {https://www.principlepower.com/projects/windfloat1},
	author = {{Principle Power}},
        note = {Accessed: 2025-21-08},
year={2011}
}

@misc{WindFloatAtlantic,
	title = {{WindFloat Atlantic}},
	url = {https://www.principlepower.com/projects/windfloat-atlantic},
	author = {{Principle Power}},
        note = {Accessed: 2025-21-08},
year={2020}
}

@misc{Kincardine,
	title = {{Kincardine}},
	url = {https://www.principlepower.com/projects/kincardine-offshore-wind-farm},
	author = {{Principle Power}},
        note = {Accessed: 2025-21-08},
year={2021}
}

@misc{hywind-tampen,
	title = {{Hywind Tampen}},
	url = {https://www.equinor.com/energy/hywind-tampen},
	author = {{equinor}},
        note = {Accessed: 2025-21-08},
year={2022}
}

@misc{hywind-scotland,
	title = {{Hywind Scotland}},
	url = {https://www.equinor.com/energy/hywind-scotland},
	author = {{equinor}},
        note = {Accessed: 2025-21-08},
year={2017}
}

@misc{Bladed, url={https://www.dnv.com/software/services/bladed/}, title={{Bladed}},  note = {Accessed: 2025-07-19}, year={2025}, author={DNV}}

@misc{InductivaOpenFAST,
  author = {{Inductiva}},
  title = {{OpenFAST Tutorial}},
  year = {2024},
  url = {https://tutorials.inductiva.ai/simulators/OpenFAST.html},
  note = {Accessed: 2024-11-12}
}

@misc{AWSOpenFAST,
  author = {{Amazon Web Services}},
  title = {Physics on {AWS}: Optimizing Wind Turbine Performance Using {OpenFAST} in a Digital Twin},
  year = {2024},
  url = {https://aws.amazon.com/blogs/architecture/physics-on-aws-optimizing-wind-turbine-performance-using-openfast-in-a-digital-twin/},
  note = {Accessed: 2024-11-12}
}

@phdthesis{Stewart16,
  title={Design Load Analysis of Two Floating Offshore Wind Turbine Concepts},
  author={Stewart, Gordon M},
  year={2016},
  school={University of Massachusetts Amherst},
  doi = {10.7275/7627466.0},
}

@misc{aws,
  author       = {{Amazon Web Services, Inc.}},
  title        = {{AWS}},
  year         = 2024,
  url          = {https://aws.amazon.com/},
        note = {Accessed: 2024-11-05},
}

@book{92172f5db61f409db7a831d414c019bd,
title = "FAST.Farm User{\textquoteright}s Guide and Theory Manual",
author = "Jason Jonkman and Kelsey Shaler",
year = "2021",
url    = {https://research-hub.nrel.gov/en/publications/fastfarm-users-guide-and-theory-manual},
publisher =  {{National Renewable Energy Laboratory}},
}

@book{osti_2345922,
  author       = {Dykes, Katherine and Ning, S. Andrew and Scott, George and Graf, Peter and Barter, Garrett and Bortolotti, Pietro and Key, Alicia and Abbas, Nikhar and Gaetner, Evan and Quick, Julian and others},
  title        = {WISDEM® v3.15.2 2024 (Wind-Plant Integrated System Design and Engineering Model) [SWR-14-05]},
  doi          = {10.11578/dc.20240502.1},
  place        = {United States},
publisher =  {{National Renewable Energy Laboratory}},
  year         = {2024},
  month        = {04}}

@book{18aac95355e641309b54a6830618c5ca,
title = "How 2 HAWC2, the User's Manual",
author = "Larsen, {Torben J.} and Hansen, {Anders Melchior}",
year = "2007",
isbn = "978-87-550-3583-6",
series = "Denmark. Forskningscenter Risoe. Risoe-R",
number = "1597(ver. 3-1)(EN)",
url    = {https://orbit.dtu.dk/en/publications/how-2-hawc2-the-users-manual
},
publisher = "Ris{\o} National Laboratory",
}

%% else use the following coding to input the bibitems directly in the
%% TeX file.

% \begin{thebibliography}{00}

% %% \bibitem{label}
% %% Text of bibliographic item

% \bibitem{}

% \end{thebibliography}
\end{document}